\documentclass[14pt]{article}
\usepackage{amsmath}
\usepackage{bm}
\usepackage{color}
\usepackage{amssymb}
\usepackage{latexsym}
\usepackage[greek,english]{babel}
\usepackage[utf8x]{inputenc}

\usepackage{graphicx}
\usepackage{amsfonts}
\usepackage{yfonts}
\usepackage{physics}
\def\notcong{\relax{\cong \kern-.90em /}}
\def\ksl{\relax{k \kern-.50em /}}
\def\qsl{\relax{q \kern-.50em /}}
\def\psl{\relax{p \kern-.45em /}}
\def\I1{\relax{\rm 1\kern-.40em 1}}
\def\dsl{\relax{\partial \kern-.50em /}}
\def\Dsl{\relax{D \kern-.50em /}}
\def\Esl{\relax{E \kern-.50em /}}
\def\Asl{\relax{A \kern-.50em /}}
\def\Ksl{\relax{K \kern-.50em /}}
\def\esl{\relax{e \kern-.50em /}}
\def\epsilonsl{\relax{\rm \epsilon \kern-.50em /}}
\def\partialsl{\relax{\rm \partial \kern-.50em /}}
\def\1/2{\frac{1}{2}}
\def\3/2{\frac{3}{2}}
\def\p{\partial}
\def\d{{\rm d}}
\def\e{{\rm e}}
\def\im{{\rm i}}

\def\bmm{\begin{displaymath}}
\def\em{\end{displaymath}}

\begin{document}
\setcounter{page}{1}
\def\IZ{\relax{\rm Z\kern-.40em Z}}
\def\nn{\nonumber}

\def\h{\hbar}

\def\hbar{{\mathchar'26\mkern-9muh}}

\def\be{\begin{equation}}

\def\ee{\end{equation}}

\def\bq{\begin{eqnarray}}

\def\eq{\end{eqnarray}}

\def\1/2{\frac{1}{2}}

\def\sqr#1#2{{\vcenter{\vbox{\hrule

height.#2pt\hbox{\vrule width.#2pt height#1pt

\kern#1pt \vrule width.#2pt}\hrule
 
height.#2pt}}}}
\thispagestyle{empty}

\begin{center}
{\Large \bf From Many Models to ONE THEORY}
\vskip 0.5cm
{\it From the Dark Ages to Enlightenment}
\vskip 2cm
{\bf{J. Iliopoulos}}
\vskip 0.5cm
 Laboratoire de Physique de l'Ecole Normale Supérieure\\
 ENS-PSL, CNRS, Sorbonne Univ., Univ. Paris Cité, Paris, France \\
(ilio@phys.ens.fr)

\vskip 2cm
{\bf ABSTRACT}
\end{center}
The year 2024 marks two anniversaries: The 70th anniversary of CERN and the 50th anniversary of the $J/\Psi$ discovery. At this occasion I have been asked to give review talks on the significance of these anniversaries. This article is an expanded version of these talks. 

\eject
\tableofcontents
\eject
\section{Introduction}

By a worth noting coincidence, the year 2024 marks two anniversaries: The 70th anniversary of CERN\cite{CERN-hist} and the 50th anniversary of the $J/\Psi$ discovery\cite{J/Psidisc}. 

CERN's founding fathers were among Europe's leading scientists. CERN became the most successful International Institution. Scientists built the Scientific European Union long before politicians thought of the Economic European Union. If the Scientific Institution appears to be more solid than the Economic one, it is because its foundations are made with ideas.

Regarding the second anniversary, I want to argue that $J/\Psi$ was not just a new resonance; we knew already a large number of them. It  was not even only the first indirect evidence for the existence of a new quark flavor. It was all that, but it was also much more. It was the final proof which convinced the large majority of our community that we were witnessing a radical change of paradigm in our understanding of microscopic physics. From phenomenological models and specific theories, each one applied to a restricted set of experimental data, we had to think in terms of a fundamental theory of universal validity. From many models to the STANDARD THEORY of particle physics. For most of us this transition was a revelation, for some others it was a painful experience. It is appropriate to combine it with CERN's anniversary because the experimental verification of the Standard Model started at CERN with the neutral currents and was completed also at CERN with the BEH scalar boson. It is this story that I will attempt to narrate in this note, although I do not consider myself to be an unbiased observer.

In order to avoid any misunderstanding, I want to start by emphasising what this article is not meant to be. It is not a review article. I will not cite all important papers which were published in the last one hundred years. Although it tells a story, it is not a scholarly article on the history of high energy physics. I do not feel capable of writing such a history. My purpose is to present a personal view of some ideas among those which contributed to the establishment of the Standard Theory. It is neither exhaustive nor objective.

\section{Elementary Particles: The origins}
\label{Orig}
Our field is called “The Physics of Elementary Particles”. Although the discontinuous structure of matter\footnote{The atomic hypothesis is attributed to Leucippus and his student Democritus of Abdera (c.460 – c.370 BC)   for whom the basic constituents of matter are the “atoms" and the “vacuum",  i.e. empty space between the atoms.\textgreek{“Νόμω γάρ χροιή, νόμω γλυκύ, νόμω πικρόν, ετεή δ' άτομα καί κενόν"}, which in free translation says that “Laws determine the tone, the sweetness or the bitterness, but everything consists of atoms and empty space".} has been the subject of great scientific debates for many years, by the beginning of last century the question was settled\cite{hist1}. However, the nature of the elementary constituents kept on changing  following the increasing resolution power of our microscopes. We thus discovered the chain {\it molecules $\rightarrow$ atoms $\rightarrow$ nuclei} + {\it electrons $\rightarrow$ protons} + {\it neutrons} + {\it electrons} $\rightarrow$ {\it quarks} + {\it leptons} $\rightarrow$ ??  We may believe intuitively that there exists such a thing as “an innermost layer", but there is no  proof for such belief, and even less, for the claim that we have already reached these hypothetical “truly elementary” particles. Obviously, this question cannot be addressed independently of  that concerning the nature of the second Democritean  concept, that of the vacuum, a question  related to the microscopic structure of space and time, for which we know practically nothing. It follows that, for the moment, the only “definition” we can give is that an elementary particle is an object for which we have not yet been able to detect any internal structure. Although this definition is logically correct, it describes only a small part of the objects we study in our research which include in addition all hadrons and hadronic resonances. So, the practical definition is a cyclic one, namely that an elementary particle is an object which appears as an entry in the “Table of Elementary Particles”. 
\vskip 0.3cm

If I had to assign a year to the birth of experimental particle physics I would choose 1950, the discovery of the neutral pion  by H.J. Steinberger, W.K.H. Panofsky and J.S. Steller at the Berkeley electron synchrotron\cite{Steinbetal}. It is the first “elementary” particle discovered with an accelerator (the charged pion was discovered in 1947 in cosmic rays\footnote{The first cyclotron was built at Berkeley  by E. O. Laurence and his student M. S. Livingstone in 1930. It was a toy accelerator with a diameter of only 4 inches, but in the following years Laurence built in his laboratory a series of larger and larger accelerators reaching a diameter of 184 inches. They were capable of producing pions abundantly but the experimentalists had not yet developed detection techniques suitable to the new environment. It is not enough to acquire a new toy, you must also learn how to play with it. It is also true that Laurence was more interested in building accelerators than in making experiments and  did not allocate sufficient resources to detectors.}). The existence of $\pi^0$ was predicted in 1938 by N. Kemmer\cite{Kemmer} who wrote the first isospin invariant theory for the nuclear forces\footnote{Kemmer has not received the appropriate recognition for his groundbreaking work. His equations for the pion-nucleon interaction can be found in every textbook, but his name is rarely mentioned.}. So, $\pi^0$ is the first particle whose existence was predicted by an argument based on an internal symmetry and also the first particle to be discovered in an accelerator. Since that time accelerators became the main engines of discovery in particle physics\cite{hist2}. The first consequence was the separation of the two communities, particle and cosmic ray physicists\footnote{The last conference which was common to both was held in France, at Bagnères-de-Bigorre, in July 1953. It was organised by the French L. Leprince-Ringuet and the cosmic ray physicists complained claiming that he allocated too much time to the still scarce results coming from accelerators. C.F. Powell, the discoverer of the charged pion and a leading figure in cosmic ray research, said in his closing lecture: “Gentlemen, we have been invaded \dots The accelerators are here”. An immediate measure of this “invasion”: In the same conference the Australian R.H. Dalitz presented a simple method to study the properties of the newly discovered particles which exhibited three-body decays. It is the famous  “Dalitz plot” in which every event is represented by a point in a graph. In 1953 Dalitz had 13 points in his plot, all coming from cosmic rays. At the 1955 Rochester Conference he had 53, with 42 still coming from cosmic rays. The following year he had more than 600, mostly coming from accelerators. Note however, that in recent years the two communities often joined forces again with the emergence of a new discipline, the “Astroparticles”.}. With the use of accelerators the field expanded very rapidly and the nature of experimental work changed\footnote{We should also acknowledge the fact that this rapid expansion was to a certain extent due to the success of the Manhattan project during the war.}. Dedicated research  centers were built -- Brookhaven in the USA, CERN in Europe -- which were not associated with a particular University. The drive to build larger and more complex detectors, and the rate with which data were collected, led to the creation of large international collaborations. We entered the era of “Big Science”. A measure of this exponential expansion is given by the number of elementary particles. In 1940 we knew the proton and the neutron, the electron and the neutrino (not yet detected), the photon and the particle which was believed to be the Yukawa meson (it was in fact the muon). In 1960 the Table of Elementary Particles had several dozen entries and that of today several hundred, although we know that very few among them are “elementary”.
\vskip 0.3cm

If experimental particle physics followed a monotonically rising trajectory, that of its theoretical counterpart was more circuitous. Modern theoretical physics has a precise date of birth: June 2 1947, the date of the Shelter Island Conference. It is the first postwar meeting on the Foundations of Quantum Mechanics, sponsored by  the USA National Academy of Sciences. It was held at Long Island's Shelter Island at Ram's Head Inn and  gathered 24 participants around J.R. Oppenheimer. With the aura of the  Manhattan Project, he was considered as the founding father of the American School of theoretical physics. Regarding the Conference, Oppenheimer later declared it was the most successful scientific meeting he had ever attended. Indeed, in retrospect Shelter Island is a landmark of theoretical physics. The most important contributions which were presented in that conference were not theoretical breakthroughs but two experimental results: the non-zero values of the Lamb shift and of the electron anomalous magnetic moment. Both these results were important because, for the first time, they showed, beyond any possible doubt, that the predictions based on the Dirac equation for the electron are only approximately correct. This in turn motivated a serious study of quantum field theory. As S. Weinberg  has put it\cite{W1}: {\it “The great thing accomplished by the discovery of the Lamb shift was not so much that it forced us to change our physical theories, as that it forced us to take them seriously”.} Indeed, the formalism of quantum field theory existed already for both bosons and fermions.  But the enormous prestige of the Dirac theory on the one hand, and the absence of a clear physical motivation on the other, discouraged theorists to face seriously the problem of the divergences of perturbation theory. The fact that the theoretical ideas existed  in a subconscious form is witnessed by the fact that it took practically no time to develop them. The first estimation of the Lamb shift in a non-relativistic approximation was done by H.A. Bethe in the train which brought him back from New York to Ithaca. In the following months R.P. Feynman and J.S. Schwinger, independently, using  apparently different  formulations, set up the program for the renormalised perturbation expansion of  Quantum Electrodynamics  and Schwinger gave the first calculation of $g-2$. As it turned out, similar results were obtained independently in Japan by Sin-Itiro Tomonaga, who obtained also the first complete calculation of the Lamb shift. The equivalence of all these approaches  was formally shown by F. Dyson in 1948. Rare are the examples in physics in which so much progress was accomplished in such a short time.

The Shelter Island conference was meant to be the first in a series on the same subject. Indeed a follow up  conference was organised in Pocono (Pennsylvania) in 1948 and a third one in Oldstone (New York) in 1949. After that the programme of the Foundations of Quantum Mechanics was declared complete. Today we know that this was a rather optimistic view, but at the time it seemed to be justified. The results obtained by applying to quantum electrodynamics the newly established principles of renormalised perturbation theory were in spectacular agreement with experiment. 

The first hint that everything was not perfect came from a seemingly mathematical problem. With the exception of some very simple toy models, we have no exact solutions in quantum field theories. Every result is obtained in the form of a formal power series in a parameter, known as the coupling constant, whose value reflects the strength of the interaction. For QED it is the fine structure constant $\alpha \approx 1/137$. The theory of renormalisation guarantees that every term in this expansion is calculable but, in practice, the actual calculations are so complicated that, even today, only very few terms have been computed. However, since $\alpha <<1$, the higher order terms are believed   to give negligible contributions. Nevertheless, from the mathematical point of view, there remains the obvious question concerning the convergence properties of the expansion. In his fundamental papers, which formulated the theory of quantum electrodynamics, Dyson repeatedly emphasised that the theory was mathematically incomplete because the convergence properties of the series were not known. He addressed specifically this question in the years around 1950 and, in the prevailing euphoria, he was convinced he would prove that the series was convergent, at least for the part which corresponds to high frequencies{\footnote{It seems that W. Pauli had warned him against such expectation.}. However pretty soon he found a very simple argument showing that this was not true; the perturbation series diverges\cite{Dyson}. The argument is based on the obvious physical remark that, for $\alpha <0$,  a state with an arbitrarily large number of electron-positron pairs will have negative energy which is  unbounded from below. This proves that the series in powers of $\alpha$ cannot be absolutely convergent. Although this result does not exclude weaker forms of convergence, even today we have no rigorous proofs for a physical theory like QED. To quote Dyson\cite{DysonCom}, {\it “So my program for making quantum field theory consistent, by purely mathematical manipulations  without any new physics, came to an ignominious end. The hard-earned lesson, \dots is that new physics is unavoidable.”} 

Going beyond perturbation,  A.S. Wightman formulated in 1955 a rigorous mathematical framework for a quantum field theory\cite{Wight}.  It consists of a set of six axioms that every physically acceptable quantum field theory is supposed to satisfy at the non-perturbative level. Although they are very general -- Lorentz invariance, locality, positivity of the energy etc --  they have two important physical consequences: they imply the invariance of any field theory under the $CPT$ transformation\footnote{The product of the three discrete transformations “charge conjugation” ($C$), “parity” ($P$) and “time reversal” ($T$).}, as well as the connection between spin and statistics. Both are in excellent agreement with observations, but the problem is that we do not know whether any among the field theories we use in particle physics admits non-trivial solutions satisfying the Wightman axioms. This was certainly a drawback for a mathematician like Dyson, but I do not think that the high energy physics community worries too much about it. Since the low order results agree so well with experiment, we can leave the study of the mathematical properties for later. I will come back to this remark in a later section. 

The natural next step was to apply this approach to the other two interactions, to wit the strong and the weak ones. The former were represented by the isospin invariant pion-nucleon interaction and it was shown that the renormalisation program applied to it. However, the results were useless because the effective pion-nucleon coupling constant turned out to be very large, on the order of 10, making the power series expansion meaningless. The dynamics appears to be dominated by phenomena such as resonance production, which cannot be described by ordinary perturbation theory. There remained the weak interactions. Naturally, they are much weaker, but now a new problem appeared: the renormalisation program is a quite complex process and applies only to a small number of quantum field theories. The Fermi theory, which describes very well the low energy weak interactions, is not one of them. 

This double failure soon tarnished the glory of quantum field theory. The disappointment was such that the subject was not even taught in many universities. By the late fifties the theoretical high energy physics landscape was fragmented in many disconnected domains, having no common trends and often ignoring each other. For strong interaction processes the main approach was based on the assumed analytic properties of the $S$-matrix elements\footnote{This approach, which from exaggerated heights of almost religious faith has fallen into equally unjustified depths of oblivion, has produced many fundamental concepts in particle physics, such as Regge theory,  bootstrap methods, dual models, the Veneziano amplitude and string theory.}, but we had also several simple models, none with any solid theoretical basis, each one applied to a particular corner of phase space. For weak interactions the Fermi theory proved to be a very good  phenomenological model, but it had no logical justification and no obvious connection with anything else. The most important progress  in our understanding of Nature's fundamental laws came mainly from the application of symmetry principles and not from dynamical calculations. Quantum field theory was noticeable essentially by its absence. A totally marginal subject confined to very few precision calculations in quantum electrodynamics. Many physicists had only vague and often erroneous ideas about it and, to a certain extent, this misunderstanding has survived even today.

\section{The secret road to the New Theory}

The title of this section could have been: {\it Quantum Field Theory strikes back}, but it would have been misleading. The new theory did not emerge out of the blue. I want to show that it is built on results coming from all lines of research and each one contributed its fair share to it. 
The construction of the Standard Model, which became gradually the
Standard Theory of elementary particle physics, is, probably,  the most
remarkable achievement of modern theoretical physics\footnote{For a historical account of the Standard Model see, for example\cite{SM-hist1} and \cite{SM-hist2}.}. It started with the weak interactions. It may sound strange that a revolution in the physics of elementary particles was initiated by the study of the weakest among their interactions (the effects of the gravitational interactions are not measurable in high energy physics), but through the history of our field, the weak interactions have triggered many such revolutions and we often had  the occasion to meditate on the fundamental significance of “tiny” effects. 

It is usually said that progress in science occurs when an unexpected
experimental result contradicts the current theoretical beliefs. This forces
scientists to change their ideas and leads to a new theory. This has often
been the case in the past, but the revolution we are going to describe here
had a theoretical, rather an aesthetic motivation. It was a triumph of
abstract theoretical thought which brought geometry into physics. The road has
been long and circuitous and many a time it gave the impression of leading to
a dead end. In fact it 
was a long series of isolated and mostly confidential contributions.  Essentially all the milestones went
unnoticed when they were first proposed. Many important ideas had to be rediscovered
again and again. This complicates the task of the narrator because a chronological order would have been hard to follow. 
I choose instead to group together all contributions into a few central topics: I start with a series of seemingly uncorrelated discoveries, most of which had motivations unrelated to their final application. They form what I call “the background”. Then I move to the two main components of the Standard Model, namely the electroweak theory, which paved the way to the new era, followed by the strong interactions and QCD. 

\subsection{The background}
In the secret road there was no well defined direction. Several milestones did not seem to point to a single path. The pioneers were often unaware of each other's work and it is only now that we can see a coherent picture. I will mention only very few contributions, particularly some that are not so well-known, but a complete study should include many more\footnote{The way we learn physics we often get the impression that it progresses towards a clear and well defined goal. This is almost unavoidable because we learn only those attempts which have succeeded. However, scientific research resembles more a monte-carlo algorithm. Many directions in the phase space of ideas are explored and we pursue only the most promising ones. Occasionally we find later that they lead to a dead end and we have to come back and start a new direction. Text books never present this real story, thus creating the illusion of a unique line of thought.}.

\subsubsection{Gauge theories}
\label{GT}

$\bullet$ {\it Classical electrodynamics.} The first concept of gauge invariance goes back to classical electrodynamics\cite{RevGauge} but its history and evolution is quite complicated. I am not able to present a full story and I will mention only a few steps which, to my understanding, have played an important role in shaping our present ideas on gauge invariance. They are not necessarily among the most important contributions to classical electrodynamics.   

By the end of the 18th century the field of electrostatics was quite well understood. In France the main actors were C. Coulomb, and his famous experiments that established the $1/r^2$ law, and S. D. Poisson who wrote the differential equation for the electrostatic potential $\Delta V+4\pi \rho=0$, in analogy with Newtonian dynamics. Similar results, although starting from apparently different assumptions,  were obtained in England by B. Franklin. The question of action at a distance puzzled  people for many years, as it had puzzled Newton\footnote{In a letter addressed to Richard Bentley in 1687, Newton writes {\it “That one body may act upon another at a distance through a vacuum, without the mediation of something else, by and through which their action and force may be conveyed from one to another, is to me so great an absurdity, that I believe no man, who has in philosophical matters a competent faculty of thinking, can ever fall into it...”}.} more than one century earlier. On the other hand magnetic phenomena were believed to have a different origin and were not considered as equally important. This belief was shuttered during the early years of the 19th century following various experiments performed not only with static charges, but  also with electric currents and magnets. Of particular interest for the concept of gauge invariance turned out to be the experiments of H.C. Oersted and  A.-M.  Ampère, in particular those measuring the interactions among closed electric circuits carrying steady currents.  

 I do not know who was the first to remark that the dynamical system described by the components of the electric and magnetic fields ${\bm E}$ and ${\bm B}$ -- that is, six degrees of freedom in our counting -- was in fact redundant because some of the equations do not involve any time derivatives and should be considered as constraints. It seems that the first person who attempted to reduce the redundancy was  C.F. Gauss who, in some manuscript notes in 1835, introduced the concept of the “vector potential” ${\bm A}$. It was further developed by several authors and was fully written by G. Kirchoff in 1857, following earlier work, in particular by F. Neumann. The components of the electric and magnetic fields could be expressed in terms of the vector and scalar potentials, thus reducing the number of degrees of freedom from six to four. It was soon
noticed that it still carried redundant variables and several ``gauge conditions''
were used. The condition, which in modern notation is written as 
$\partial_{\mu}A^{\mu}=0$, was proposed by the Danish mathematical physicist
L.V. Lorenz in 1867. The funny remark I often make is that  most physics books misspell
Lorenz's name as {\it Lorentz}, adding a “t”, thus erroneously attributing the condition to
the famous Dutch H.A. Lorentz, of the Lorentz transformations\footnote{In
  French: On ne pr\^ete qu'aux riches.}. It seems that Maxwell favored the condition ${\bm \nabla}\cdot {\bm A}(x)\!=\!0$ which today we call {\it “the Coulomb gauge”}\footnote{It is obvious that Coulomb never wrote such a condition. The name comes from the fact that, in this gauge, the scalar potential satisfies the Poisson equation $\Delta V(x)+\rho(x)=0$ which, for $\rho$ given by a static point source, reproduces the Coulomb potential.}. Using Maxwell's equations we can immediately see the redundancy of the system $(\Phi, {\bm A})$ because the equation for $\Phi$ does not involve any time derivative. Lorenz arrived to this conclusion because he had a formulation of classical electrodynamics equivalent to Maxwell's. 

In this context, an interesting story is the following\cite{RevGauge}:  Around the years 1840 F.E.  Neumann and, independently, W.E. Weber, studied the interaction between two closed electric circuits carrying currents $I$ and $I'$, respectively. Their methods and physical assumptions were different and they present only a historical interest today, so I give only their final expressions for the magnetic interaction energy $W$ between the two circuits, using the same modern notation for both. 
\be
\label{Gauge1}
\d W_N=\frac{II'}{c^2}\frac{{\bm n}\cdot{\bm n'}}{r}\d s \d s' ~~~~~\d W_W=\frac{II'}{c^2}\frac{({\bm n}\cdot {\hat {{\bm r}}})({\bm n'}\cdot {\hat {{\bm r}}})}{r}\d s \d s'
\ee
where the subscripts $N$ and $W$ stand for Neumann and Weber, respectively. The notation is the following: ${\bm x}$ and ${\bm x'}$ denote two points on the circuits $C$ and $C'$, ${\bm r}={\bm x}-{\bm x'}$, ${\hat {{\bm r}}}$ is the unit vector in the direction of ${\bm r}$, and the line elements $\d s$ and  $\d s'$ are parametrised as $\d {\bm s}={\bm n}\d s$ and $\d {\bm s'}={\bm n'}\d s'$. The total interaction energy $W$ can be obtained by integrating the differential expressions (\ref{Gauge1}) along the two circuits. Neither Neumann nor Weber wrote explicitly  the corresponding vector potentials but, had they done so, they would have arrived at expressions of the form:
\be
\label{Gauge2}
{\bm A}_N({\bm x},t) = \frac{1}{c} \int \d^3 x' \frac{1}{r} {\bm J}({\bm x'},t)~~~~~{\bm A}_W({\bm x},t) = \frac{1}{c} \int \d^3 x' \frac{1}{r} {\hat {{\bm r}}}({\hat {{\bm r}}}\cdot {\bm J}({\bm x'},t))
\ee
with ${\bm J}({\bm x'},t)$ a general current density. The interesting part of the story is that, in the years after 1870, H.L.F. von Helmholtz criticised and compared these expressions. In particular, he noticed that the two formulae for the elementary magnetic interaction energy differ by a quantity which can be expressed as a multiple of the perfect differential 
\be
\label{Gauge3}
\d s \d s' \frac{\p ^2 r}{\p s \p s'} = \d s \d s' \frac {({\bm n}\cdot {\hat {{\bm r}}})({\bm n'}\cdot {\hat {{\bm r}}})-({\bm n}\cdot {\bm n'})}{r}
\ee
so, they give the same result when integrated over the closed circuits. In our present terminology he showed that the two expressions are {\it gauge equivalent.} He even went a step further: he generalised Neumann's and Weber's results by exhibiting a one-parameter family of expressions for the magnetic energy which interpolate between Neumann and Weber:
\be
\label{Gauge4}
\d W_{\alpha} =\frac{II'}{2c^2 r}[(1+\alpha) ({\bm n}\cdot{\bm n'})+(1-\alpha)({\bm n}\cdot {\hat {{\bm r}}})({\bm n'}\cdot {\hat {{\bm r}}})]\d s \d s'
\ee
This linear combination differs from either one of the expressions in eq. (\ref{Gauge1}) by a multiple of the perfect differential of eq. (\ref{Gauge3}) and, therefore, it is consistent with Ampère’s observations for any value of the parameter $\alpha$. The analogue for the vector potentials of eq. (\ref{Gauge2}), is the expression
\be
\label{Gauge5}
{\bm A}_{\alpha}=\1/2 (1+\alpha){\bm A}_N + \1/2 (1-\alpha){\bm A}_W
\ee
Helmholtz made the gauge equivalence clear by writing 
\be
\label{Gauge6}
{\bm A}_{\alpha}={\bm A}_N +\frac{1-\alpha}{2} {\bm \nabla}\Psi ~~~~~\Psi=-\frac{1}{c}\int {\hat {{\bm r}}}\cdot {\bm J}({\bm x'},t)\d^3 x'
\ee
He then went on to show that, if $\Phi$ denotes the electrostatic potential,  $\Psi$, ${\bm A}_{\alpha}$ and $\Phi$  satisfy the equations
\be
\label{Gauge7}
\Delta \Psi =\frac{2}{c} \frac{\p \Phi}{\p t} ~~~~~ {\bm \nabla}\cdot {\bm A}_{\alpha}=-\frac{\alpha}{c} \frac{\p \Phi}{\p t}
\ee
Notice that the second equation is Lorenz's gauge condition for the potentials used by Helmholtz. He even remarked that the choice $\alpha=0$ gives the Coulomb gauge used by Maxwell. 
It was the first example of a {\it family of gauges.}

By the end of the century H.A. Lorentz published a book and some encyclopedia articles with the full classical electromagnetic theory. It took more than a century to obtain it and the  invariance under  gauge transformations of the vector and scalar potentials is an integral part of it.
\vskip 0.3cm

$\bullet$ {\it Gauge invariance in General Relativity and the attempts to unify electromagnetism and gravitation.} The development of the general theory of relativity offered a new paradigm for
a gauge theory. I believe that its mathematical formulation as a theory  invariant under
local translations (more precisely  diffeomorphisms) was first understood by D. Hilbert\cite{Hilbert}.  But apart from this point, the gauge theories we use in the Standard Model are logically unrelated to general relativity and gravitation because they deal with internal symmetries and not with those of space and time. Therefore I will not present these developments in any detail, although  for many  decades the two were entangled and general relativity was the
starting point for many studies of theories invariant under local
transformations. 

Since the two classically known interactions, electromagnetism and gravitation, were found to obey a gauge principle, it was normal to search for a unified description. The attempt which has survived today carries the names of 
T. Kaluza and O. B. Klein\cite{KK}. Their approach consists in writing general relativity in a five dimensional space-time. They show that, at least  in some simple cases, the solution for the metric tensor splits into a four dimensional space, for example Minkowski, and a compact fifth dimension representing a circle of radius $R$. At the limit when $R$ goes to zero, a four-dimensional observer will see the components of the metric along the fifth dimension as a four-dimensional vector endowed with a gauge principle, thus mimicking a theory of gravitation and electromagnetism. It is the approach which is mostly used today in supergravity and superstring theories.  What is less known is that
the idea was introduced earlier by the Finnish Gunnar Nordstr\"{o}m\cite{Nord}
who had 
constructed a scalar theory of gravitation. In 1914 he wrote a
five-dimensional theory of electromagnetism and showed that, if one assumes
that the fields are independent of the fifth coordinate, the assumption made
later by Kaluza, the electromagnetic vector potential splits into a four
dimensional one and a scalar field identified to his scalar graviton.  In some sense, Kaluza and Klein wrote the mirror image of Nordstr\"{o}m's theory.

The first attempt  to an e.m. -- gravitation unification in the framework of general relativity is due to H.K.H. Weyl in 1919\cite{Weyl1}. He proposed to enlarge the invariance of general relativity by including local scale transformations in which the metric transforms as $g\rightarrow \e^{2\lambda(x)}g$, with $\lambda(x)$ an arbitrary function of the space-time point $x$. Weyl wanted to associate this new gauge transformation to electromagnetism. Geometrically, it corresponds to a change of the unit of length  from point to point, therefore, it was natural to call the resulting invariance {\it eichinvarianz} in German, which was translated in English as {\it gauge invariance}. Before looking at the physics of this theory, let me stress that, from the mathematical point of view, it is very interesting. Weyl was an accomplished mathematician and in this work he develops and introduces many concepts from differential geometry which took decades to reach the theoretical physics community. He was probably the first person to understand the underlying geometry of gauge invariance. In this article he defends his point of view by noting {\it “\dots Riemannian geometry \dots} [contains] {\it an element of “remote geometry”  (ferngeometrisches Element)—with no good reason \dots The metric allows the two magnitudes of two vectors to be compared, not only at the same point, but at any arbitrarily separated points\dots”} For Weyl a true infinitesimal geometry should not allow to compare the lengths of two vectors  at a distance, the same way it does not allow to compare their directions. Concerning the physics, he was  very enthusiastic. In a letter to Einstein he wrote: {\it “\dots I succeeded, as I believe, to derive electricity and gravitation from a common source\dots”}. The reaction of the leading physicists of the time was more reserved. Among them Einstein, who liked the mathematical part of the theory, spotted immediately a physical inconsistency. In a postcard he sent to Weyl he pointed out that, if we accept a nonintegrable length connection, then the behavior of clocks would depend on their history, in clear contradiction with the existence of stable atomic spectra. Pauli, in his usual very aggressive style, wrote to Weyl {\it “\dots you have been given a chair in ‘‘Physics’’ in America. I admire your courage; since the conclusion is inevitable that you wish to be judged, not for success in pure mathematics, but for your true but unhappy love for physics.”} In retrospect, I think that, in his 1919 paper, Weyl asked the right mathematical question but gave the wrong physical answer\footnote{I realise that I am making the mistake a real historian should not make: to interpret old papers in the light of today's knowledge. But, first I am not a real historian and second, every time I make this mistake I will try to make it clear.}.  Take the example of general relativity: we start from a theory invariant under a group of global transformations, in this case the Poincaré transformations of special relativity. Promoting the invariance from global to local introduces new geometrical quantities -- called “connections” -- which are interpreted physically as giving rise to the forces of gravitation. Weyl understood that in this respect, the classical electromagnetic theory was mathematically incomplete. People had discovered by trial and error its property of gauge invariance, but the underlying global symmetry was missing. This was the right mathematical question. The wrong physical answer was to identify it with scale transformations. The laws of physics are in no way invariant under global scale transformations. As it turned out, the correct answer was found in the framework of quantum mechanics to which we turn next.

\vskip 0.3cm

$\bullet$ {\it Gauge invariance and quantum mechanics.} The discovery of quantum mechanics in the 1920s offered a new arena to study gauge theories. In the Schr{\"o}dinger picture, the wave function $\Psi({\bm x}, t)$ takes complex values and, since the equation is linear and homogenious, the theory is invariant under the transformation $\Psi \rightarrow C\Psi$, with $C$ any complex number. If we add the normalisation condition, we obtain $|C|^2=1$, so the invariance is reduced to a group of global $U(1)$ transformations. The fact that this abelian symmetry can be the missing symmetry of electromagnetism was first noticed by V. A. Fock in 1926\cite{Fock}, just after Schr{\"o}dinger wrote his equation. Fock writes the invariance of Schr{\"o}dinger's equation for an electron in an electromagnetic field as we know it today (equations (5) and (9) in his paper, I only modernise slightly the notation and set $c=\hbar=1$.)
\be
\label{Gauge8}
{\bm A}={\bm A}_1 + {\bm \nabla} f~;~~~\Phi=\Phi_1-\frac{\p f}{\p t}~;~~~\theta=\theta_1-e f~;~~~\Psi=\Psi_1\e^{-\im e f}
\ee
where ${\bm A}$ and $\Phi$ are the vector and scalar potentials, $e$ is the electric charge and $f=f({\bm x},t)$ an arbitrary function of space and time. In a second section of his article Fock extends the analysis to include general relativity. I do not know whether he was aware of Weyl's 1919 article, but he does not refer to it\footnote{He refers instead to a preliminary version of a later paper, presumably ref\cite{Weyl2}, which Weyl had communicated to him in manuscript form.}. 

The same conclusion was reached also by Fritz London in 1927\cite{London}. He started from Weyl's scale transformation $g\rightarrow \e^{2\lambda(x)}g$ and noted a formal analogy with the phase transformation in quantum mechanics provided $\lambda$ takes complex values. The titles of his articles are revealing: {\it Die Theorie von Weyl und die Quantenmechanik} and {\it Quantenmechanische Deutung der Theorie von Weyl}. He cites both Weyl and Fock. Apparently Weyl shared London's point of view. In the 1928 edition of his book on group theory and quantum mechanics\cite{Weyl1} he writes: {\it “\dots I now believe that this gauge invariance does not tie together electricity and gravitation, but rather electricity and matter\dots”}. The following year he published a remarkable article\cite{Weyl2} in which he extends the principle of gauge invariance to the Dirac electron, in both the special and the general theory of relativity. It is one of the classic papers of theoretical physics introducing many new concepts, such as the Weyl two-component spinors and the vierbein and spin-connection formalism\footnote{It seems that after the publication of this article, Pauli withdrew his sarcastic comments regarding Weyl's contributions to physics.}. Although the theory is no more scale invariant, he still used the term {\it eichinvarianz} --gauge invariance -- a term which has survived ever since.
\vskip 0.3cm

$\bullet$ {\it Non-abelian internal symmetries.} The concept of non-abelian internal symmetries in particle physics is attributed to W. Heisenberg with his introduction of the isospin formalism in 1932\cite{Heisenberg1}, but the real story is more complicated. Heisenberg's 1932 papers are an incredible mixture of the old and the new. The neutron had just been discovered, but for many people it was a new bound state of a proton and an electron, like a “small" hydrogen atom. Heisenberg does not reject this idea. Although for his work he considers the neutron as a spin one-half Dirac fermion, something incompatible with a proton-electron bound state, he notes that {\it “\dots under suitable circumstances the neutron will break up into a proton and an electron in which case the conservation laws of energy and momentum probably do not apply."} On the $\beta$-decay controversy and the existence of the neutrino he does not take any clear stand, but he sides more with his master Bohr than with his friend  Pauli: {\it “\dots The admittedly hypothetical validity of Fermi statistics for neutrons as well as the failure of the energy law in $\beta$-decay proves the inapplicability of present Quantum Mechanics to the structure of the neutron."} In fact, Heisenberg's fundamental contribution should be appreciated not {\it despite}  these shortcomings, but precisely {\it because} of them. We should remember that in 1932 
experimental data on nuclear forces were almost entirely absent.  Heisenberg had to guess the values of the nuclear attractive forces between nucleon pairs by using a strange analogy with molecular forces. He postulated a  $p-n$ and an $n-n$ nuclear force, but not a $p-p$ one, so his theory was not really isospin invariant. Nevertheless, he made the   conceptual step  to describe the nucleon wave functions in terms of a new  two-component object  
\be
\label{Isospin1}
 \Psi (x)=\begin{pmatrix} \Psi_{p}(x) \cr \Psi_{n} (x) \cr \end{pmatrix}
\ee
and the nuclear potential in the form of two-by-two Pauli matrices. The components in equation (\ref{Isospin1}) denote the wave functions of a proton and a neutron. $\Psi (x)$ has the form of a two-component spinor, but, here is the new element, it is NOT a spinor in our three dimensional space.  We shall assume that there exists a second such space, isomorphic to, but distinct from the one in which we live. A rotation in this new space mixes  $\Psi_{p}$ and $\Psi_{n}$ the same way that a rotation in our space mixes the up and down components of a spinor. It was the first introduction of a multi-dimensional internal space\footnote{Fock, in his 1926 paper\cite{Fock}, had described the parameter space of the quantum mechanical phase as a fifth dimension.}. In 1937 E. P. Wigner proposed the term {\it isotopic spin} which, simplified to {\it isospin}, has been used until now. 
 
In the following years three important developments allowed Heisenberg's initial suggestion to become a complete isospin invariant theory. The first, and probably the most important, was the progress in experimental techniques, which brought more detailed and more precise data.  They showed the need for the introduction of a $p-p$ force and confirmed the charge independence of all nuclear forces. The second was the formulation by Fermi in 1933  of a theoretical model for the amplitude of neutron $\beta$-decay in which he introduced for the first time the concept of quantised fermion fields. Since that time quantum field theory became the universal language  of  modern theoretical physics. The third important development was Yukawa's introduction of the {\it meson} as an intermediary for the nuclear forces. Initially he thought of a vector field whose three space components would mediate the weak interactions and the zero component the nuclear forces, but it was  soon replaced by a spin zero one. The final synthesis is due to N. Kemmer in 1938\cite{Kemmer}. It took six years, as well as the work of many physicists, for Heisenberg's original suggestion of 1932 to become the full isospin symmetry of hadronic physics we know today. It is a remarkably short time, given the revolutionary nature of the concept. 

Heisenberg's isospin space was three dimensional and the transformations form a group $O(3)$, or $SU(2)$, like our familiar rotations. However the concept has been subsequently enlarged as new particles were discovered and larger internal symmetry groups were brought into evidence. The space of elementary particle physics became a multi-dimensional manifold, with complicated geometrical and topological properties, and only a sub-space of it, the four-dimensional Minkowski space, is directly accessible to our senses.
\vskip 0.3cm

$\bullet$ {\it Non-abelian gauge symmetries.} Naturally, one would expect the $SU(2)$ gauge theory to be constructed following the principles we sketched above: we had the global symmetry and we only needed to make it local. But here history took a totally unexpected route. A historical accident --  the discovery of general relativity -- introduced an unnecessary complication. The fascination which general relativity had exerted to all this generation of physicists was such, that for many decades people were unable to conceive local transformations without diffeomorphisms. Therefore they used the theory of gravitation even in places where it had no business to be there. I call it “the GR-syndrome”. 

The first person who tried to construct the gauge theory for $SU(2)$ is Klein\cite{Klein2} who, in an obscure conference in 1938,  presented a paper with the title: {\it On the theory of charged fields}. The most amazing part of this work is that it follows an incredibly circuitous road: He considers general relativity in a five dimensional space and compactifies {\it \`a la} Kaluza-Klein. Then he takes the limit in which gravitation is decoupled. In spite of some confused notation\footnote{He starts from the discovery of the muon, misinterpreted as the Yukawa meson, in the old Yukawa theory in which the mesons were assumed to be vector particles. This provides the physical motivation. The aim is to write an $SU(2)$ gauge theory unifying electromagnetism and nuclear forces. He has a five-dimensional theory but  he takes the $g_{4\mu}$ components of the metric tensor to be $2\times 2$ matrices. He wants to describe the $SU(2)$ gauge fields but the matrices he is using, although they depend on three fields, are not traceless.  In fact, answering an objection by Møller, he added a fourth vector field, thus promoting his theory to $U(1) \times SU(2)$.}, he finds the correct expression for the field strength tensor of $SU(2)$. He considered massive vector bosons and it is not clear whether he worried about the resulting breaking of gauge invariance. I cannot find out whether this paper has inspired
anybody else's work because the proceedings of this conference are not
included in the citation index. As far as I know, Klein himself did not follow
up on this idea\footnote{He mentioned this work in a 1955 Conference in
Berne\cite{Klein3}.}. 

The second work in the same spirit is due to Pauli\cite{Pauli} who, in 1953,  sent to A. Pais a manuscript, with the title “Meson-nucleon interactions and differential geometry”,  in which he developed precisely this approach: the construction of the $SU(2)$ gauge theory as the flat space limit of a compactified higher dimensional theory of general relativity. He was closer to the approach followed today because he considered a six dimensional theory with the compact space forming an $S_2$.  Although he never published this work, I believe that  he was aware of Klein’s 1938 paper  because the two were in frequent correspondance\footnote{ It seems that Pauli's opinion on higher dimensional theories was fluctuating. He was the one who showed Kaluza's paper to Klein in 1926, but, when in 1931 the latter was about to move to Stockholm, he sent him a typical Pauli letter in which he writes: [In Stockholm]{\it “$\cdots$ you $\cdots$ will not have to worry about $\cdots$ the fifth dimension (or similar topics) $\cdots$
I am not of the opinion that finding new laws of nature and indicating new directions is one of your great strengths, although you have always developed a certain ambition in this direction $\cdots$ I find much more beautiful those of your papers which deal with applications of known theories $\cdots$”} (He refers to Klein's papers on the paradox with the Dirac equation, or the Klein-Nishina formula on the cross section of the electron Compton scattering.) From that I conclude that, at least in 1931, Pauli did not think much of five dimensional theories. However, two years later, he published his own version of such a theory, the “Projective relativity theory”\cite{KK}.}. He had realised that a mass term for the gauge bosons breaks the invariance and  in late 1953 his enthusiasm began to wane. “If one tries to formulate field equations $\cdots$ one will always obtain {\it vector mesons with rest mass zero}” (his italics). In fact, }he had an animated argument on this subject during a seminar by Yang in the Institute for Advanced Studies in Princeton in 1954\cite{Pauli-Y}. What is surprising is that Klein and Pauli, fifteen years apart one from the other, decided to construct the $SU(2)$ gauge theory for strong interactions and both choose to follow this totally counter-intuitive method. Note that, contrary to their earlier five-dimensional theories, they both were considering at the end the flat space limit. General relativity was only used as a mathematical trick in order to compute the $SU(2)$ field strength tensor. Apparently it did not occur to either of them that it would have been much simpler to do it directly. Striking examples of the GR-syndrome.  C.N. Yang and R.L. Mills\cite{YM} were the first to understand that the gauge theory of an internal symmetry takes place in a fixed background space which can be chosen to be flat, in which case general relativity plays no role. Since that time non-abelian gauge theories became part of high energy physics. It is not surprising that they were immediately named {\it Yang-Mills theories.} The extension to other groups including direct products, was done by R. Utiyama\cite{Uti} as well as S.L. Glashow and M. Gell-Mann\cite{GG}.
\vskip 0.3cm

$\bullet$ {\it From Yang-Mills to physics.} Although Yang and Mills wanted to construct a gauge theory of nuclear forces, their work has never been used for that purpose. In fact, the first attempts concerned the weak interactions.  In 1957,  Schwinger had conjectured\cite{Schw} that the Fermi theory should be modified with the introduction of an {\it intermediate vector boson} (IVB) $W_{\mu}^{\pm}$.
\be 
\label{IVB1}
H_{I}=gJ^{\mu}(x)W_{\mu}^-(x)+hc
\ee
with $g$ a new dimensionless coupling constant. This way weak interactions looked pretty much like the electromagnetic ones, a vector boson coupled to a current, but with some very important differences: (i) The photon is massless and the e.m. interactions are long ranged. The weak interactions are known to be short ranged, so the $W$'s must be massive. (ii) The photon is neutral, the $W$'s are charged. (iii) The electromagnetic current is conserved, the weak current is not. It was soon clear that these differences implied that the theory (\ref{IVB1}), with massive vector bosons, was non-renormalisable. In order to turn the IVB model into a Yang-Mills theory, Schwinger assumed the existence of a triplet of intermediate bosons, which he called $Z^{\pm, 0}$, the two charged ones mediating the weak interactions and the neutral one being the photon. A year later, in 1958, S.A. Bludman\cite{Blud} built an $SU(2)$ Yang-Mills theory for weak interactions in which all three gauge bosons were coupled to $V-A$ currents. No connection with electromagnetism was assumed.  

The most important contribution from this period dates from 1961\footnote{It has a submission date of Sept. 9 1960.} and it is due to S.L. Glashow\cite{Glash}. It is the work that introduced  the $SU(2)\times U(1)$  electroweak gauge theory.   
Glashow proposed a unified description for weak and electromagnetic interactions and his model is the one we use today. The only missing element is the spontaneous symmetry breaking mechanism which was invented in 1964, see below.  In the Introduction we read {\it “The mass of the charged intermediaries must be greater than the $K$-meson
mass, but the photon mass is zero -- surely this is the principal stumbling block
in any pursuit of the analogy between hypothetical vector mesons and photons.
It is a stumbling block we must overlook.”} Glashow understands that this is a major problem, he has no solution to offer, but he shows the good physical judgement to overlook it. The theory has two neutral gauge bosons -- one associated to $U(1)$ and a second associated to the third generator of $SU(2)$ -- and introduces the idea of a mixing between the two. The photon field is a linear combination of the two neutral vector fields with an angle which Glashow called $\theta$ (today it is called $\theta_W$).

The Glashow and Gell-Mann paper\cite{GG}, in addition to extending the Yang-Mills theory to more general groups, introduces several new elements some of which have been incorporated into our present gauge theories.   In line with Glashow's remark in his previous paper\cite{Glash}, they notice that the introduction by hand of a mass for the  gauge bosons breaks gauge invariance and they talk about {\it partially gauge invariant theories.}  They extend the Yang-Mills construction to any algebra which can be written as a direct product of simple factors. The well-known result of associating a coupling constant to every  factor in the algebra appeared for the first time in this paper. Even the seed for a grand unified theory was there. In a footnote they say:
{\it “The remarkable universality of the electric charge would be better understood were the photon not merely a singlet, but a member of a family of vector mesons comprising a simple partially gauge invariant theory.”} Finally, they correctly identify the problems related to the absence of strangeness changing neutral currents and the small value of the $K_1^0-K_2^0$ mass difference, problems which will turn out to be crucial for the construction of the Standard Model\footnote{I do not know why this paper has not received the attention it deserves, but this is partly due to the authors themselves, especially Gell-Mann, who rarely referred to it.}.

\subsubsection{The theory of renormalisation} 
\label{Ren}

$\bullet$ {\it Simple to understand, hard to prove.} As we said previously, the theory of renormalisation, at least in the form we use in quantum field theory, dates from the late nineteen-forties. However, its roots are much older. I do not know who was the first to realise that the concept of point particles introduced short distance singularities. For example, the two classical forces, electromagnetism and gravitation,  are both described by a $1/r$ potential which becomes singular for $r=0$. This makes the self energy of an electron infinite and the Rutherford atom unstable. Lorentz  tried to solve the electron problem by smearing the charge distribution over a finite size, an intuitively attractive idea but very hard to implement in practice, especially in a relativistic framework. The development of the quantum theory solved the problem of atomic stability in a very strange way: introducing non-commutativity in phase space, it expressed the problem of the electron energy in an atom as an eigenvalue problem which, for the Coulomb potential, is shown to have a discrete  spectrum bounded from below\footnote{It is worth mentioning here that Schr\"odinger introduced his formulation of quantum mechanics in a series of articles with the title {\it“Quantisierung als Eigenwertproblem”} or, {\it “Quantisation as an Eigenvalue Problem”}\cite{Schrod1}.}. The early attempts to formulate a quantum theory of radiation introduced new forms of divergences, in addition to the electron self energy, and the history in the late twenties and thirties is quite confusing\footnote{In the book  “Inward Bound”  by A. Pais\cite{hist1}, Chapter 16 has the title {\it “Battling the infinite”} and covers this period.}. 

As far as I know, the term “renormalisation” was first used by R. Serber\cite{Serber} in 1936 in an attempt to understand the problem of vacuum polarisation, but I think that the real precursor of the modern theory of renormalisation is probably H.A. Kramers who, in 1938, introduced the idea of a {\it mass renormalisation} for the electron\cite{Kramers}. He was also a discussion leader at the Shelter Island Conference. Although the basic ideas were laid down more than 75 years ago, the detailed principles, both physical and technical, evolved considerably. They grew simpler to understand, but quite complicated to prove. For many years the theory was considered as  a mathematically murky process consisting in adding and subtracting infinities. Even the founding fathers for a long time shared this opinion\footnote{Feynman, in his report at the 12th Solvay Conference on Physics, noting the impossibility to compute mass ratios, said {\it “I still hold this belief,} [that the theory must fail] {\it and do not subscribe to the philosophy of renormalisation.”}\cite{Solvay}. Schwinger, during the nineteen-sixties, lost interest in quantum field theory and attempted to develop a new approach (Source Theory) today forgotten. I remind that in these years we were still deep in the Dark Ages.}. 
Today we understand that nothing is more remote from the truth. The theory of renormalisation offers the only known mathematically consistent way to define the successive terms in the perturbation expansion of a quantum field theory.
I want to emphasise here that, if in a calculation of a physical quantity we encounter a divergent expression, it means that we have made a mathematical mistake. Did we make a mistake in formulating quantum field theory at short distances? Yes we did! Let us take the example of a self interacting scalar field. The Lagrangian density and the canonical commutation relations are 
\be
\label{r3}
{\cal L}=\frac{1}{2}\left (\partial_{\mu}\phi\right )\left (\partial^{\mu}\phi\right )-\frac{1}{2}m^2\phi^2-\frac{\lambda}{4!}\phi^4~;~~~\left [\phi(\bm{x},t), \dot{\phi}(\bm{y},t)\right]=\im \hbar \delta^3(\bm{x}-\bm{y})
\ee
We know that the Dirac $\delta$-function is not a “function” but a special form of what we call “a distribution”. Many properties of well-behaved functions do not apply to it. In particular, the multiplication is not always a well defined operation; $(\delta(x))^2$ is meaningless. The presence of the $\delta$-function in (\ref{r3}) implies that the field $\phi(x)$ is also a distribution, so the product $\phi^2$ is ill defined. Yet, it is precisely expressions of this kind that we write in every single term of our Lagrangian, so  it is not surprising that our calculations yield divergent results. 

A conceptually simple way to solve the problem -- not necessarily the most convenient one for practical calculations -- would be to replace the field products in (\ref{r3}) by splitting the points:
\be
\label{r4}
\phi(x)\phi(x)\rightarrow \lim_{a\rightarrow 0}\phi(x+\frac{a}{2})\phi(x-\frac{a}{2})
\ee
This expression is perfectly well defined for all values of the parameter $a$, except $a=0$. In terms of distributions this means that the product is defined up to an arbitrary distribution ${\cal F}(a)$ which has support (i.e. it is non-zero), only at $a=0$. Such a distribution is a superposition of the $\delta$-function and its derivatives. 
\be
\label{r5}
{\cal F}(a)=\sum_i C _i\delta^{(i)}(a)
\ee
with the $C_i$'s arbitrary real constants.
The moral of the story is that the quantisation rules for a local field theory imply that every term in the Lagrangian contains in fact a set of arbitrary constants which must be determined by experiment. Renormalisation is the mathematical procedure which allows us to do it. How many parameters are needed in order to define a given field theory? The answer involves the distinction between {\it renormalisable} and {\it non-renormalisable} theories. For the first a finite number suffices. For the second we need an infinite number, which means that non-renormalisable theories have no predictive power. A final remark: Renormalisation has a well-deserved reputation of complexity although its principle is rather simple. The complications are technical. For example, in order to define the Lagrangian of eq. (\ref{r3}) we need only three arbitrary constants. Two of them are associated to the two-point function and can be thought as defining the mass $m$ and the normalisation of the field, and the third one is associated to the four-point function and defines the value of the coupling constant $\lambda$. However, the rigorous proof of this statement at any order of perturbation, requires some quite complicated calculations\footnote{The computational rules which are used in practical calculations were first formulated by Feynman and they are still called {\it Feynman rules}. For multi-loop diagrams they involve integrations over some auxiliary parameters and the difficulty consists in proving that the resulting expressions are always integrable. For example, in 1961, T.T. Wu discovered that the na\"ive rules needed to be modified for some very high order diagrams containing sub-diagrams with more than 11 external lines\cite{Wu}. I remember the admiration aroused by this result, since most of us could not even draw such a diagram, let alone study its analyticity properties.}. As we said earlier, the early formulation of the theory to all orders of perturbation is due to Dyson\cite{QED-Rev}. It was further developed by many people, in particular by N.N. Bogoliubov and O. Parasyuk in 1955\cite{Bog-Par}, but the first mathematically sound proof showing that the method works to all orders is due to   K. Hepp in 1966\cite{Hepp}. In 1973 H. Epstein and V. Glaser presented a new proof, formulated directly in position space in the spirit of equations (\ref{r4}) and  (\ref{r5}), which exhibits the simple physical principles in a more transparent way\cite{Epst-Gl}. This story shows that for almost two decades people were using renormalised perturbation theory without having a proof for its mathematical consistency. It is also interesting to notice that during all this time several text books on quantum field theory were published, all presenting the theory of renormalisation, but no one mentioning the fact that a real proof was missing! An example of the rather dilettante attitude of physicists regarding mathematical rigor. 
\vskip 0.3cm
$\bullet$ {\it Renormalisation and gauge invariance -- QED.} The theory of renormalisation was first applied to quantum electrodynamics which is a gauge theory. Gauge invariance introduces some extra complications which were not immediately identified. I will not follow here the chronological order, but I will try to explain the problems which will turn out to be crucial for the renormalisation of the Yang-Mills theories.

The characteristic feature of such theories is the presence of vector fields $A_\mu(x)$ which, from the field theory point of view, imply the introduction of four degrees of freedom. However, Poincaré invariance tells us that a massive spin-one particle has three degrees of polarisation and a massless one two. It is this mismatch between the Lorentz and the Poincaré counting which complicates the study of quantum field theories describing particles with spin higher than 1/2. We must include constraints in order to eliminate the redundant degrees of freedom. For the massless case they are the analogues of the gauge conditions we encountered in classical electrodynamics. 

Let us start with QED. The invariance under gauge transformations  implies that, if we find a solution for the vector potential $A_\mu(x)=\hat{A}_\mu(x)$, we can construct an infinity of others by  writing $A_\mu(x)=\hat{A}_\mu(x)+\p_\mu \theta(x)$ with $\theta(x)$ an arbitrary function. The gauge conditions are supposed to restrict this arbitrariness and, hopefully, choose a unique solution\footnote{V.N. Gribov has shown that,  for Yang-Mills theories, there exists no gauge condition yielding a unique solution\cite{Gribov}. This phenomenon is known as “Gribov ambiguity”. It is important in the attempts to give a global definition of a gauge theory, but it does not affect the perturbation expansion.}.
Since we know that in Nature only transverse photons are physical, the obvious idea is to impose the Coulomb gauge condition ${\bm \nabla}\cdot {\bm A}(x)\!=\!0$. It is easy to show that in this gauge  we are left with the two “physical” degrees of freedom, to wit the transverse photons.  It is the method used by Dirac in the calculation of the spontaneous emission probability, the first physical application of the quantised radiation field\cite{Dirac-SpEm}. 

In high energy physics the use of a covariant gauge, such as the Lorenz gauge $\p^\mu A_\mu=0$,  is much more convenient. In a simple-minded approach, we just plug the condition in the equation of motion which becomes $\Box A_\nu=j_\nu$. The four components of the vector field decouple, so the theory seems to describe four independent degrees of freedom. It is the method which was used in the early calculations\cite{QED-Rev}. Therefore, the price for keeping Lorentz covariance is to work with a theory which includes unphysical degrees of freedom. The mathematical description was proposed in 1950 by S.N. Gupta and K. Bleuler\cite{Gup-Bl} and is known as the {\it Gupta-Bleuler method}. It involves an enlarged Hilbert space, containing longitudinal and scalar photons, in addition to the physical transverse ones. The strange feature -- a consequence of the fact that the Minkowski metric is not positive definite -- is that the states with scalar photons do not have positive norm.

At risk of appearing pedantic, I want to develop this approach in a more formal way, which will make the generalisation to the Yang-Mills theory easier. I will introduce the Lorenz gauge condition using a {\it Lagrange multiplier}, a method invented by Lagrange in 1788. The QED Lagrangian is written as
\be
\label{GT1}
{\cal L}=-\frac{1}{4}F_{\mu \nu}(x)F^{\mu \nu}(x)+b(x)\p^\mu A_\mu(x)-j_\mu(x) A^\mu(x)
\ee
where the current $j_\mu$ is given in terms of the electron fields by $j_\mu=e\bar{\psi}\gamma_\mu \psi$ and it is conserved by virtue of the Dirac equation. $b(x)$ is an auxiliary field, called {\it Lagrange multiplier}. Indeed, varying  independently with respect to $A_{\mu}$ and $b$, we obtain the system
of equations
\be
\label{GT2}
\Box A_{\nu}(x)-\partial_{\nu}b(x)=j_{\nu}(x)  \;\;\; {\rm and} \;\;\; \partial_{\mu}A^{\mu}(x)=0
\ee
which means that the gauge condition has become an equation of motion. Furthermore, applying $\p^\nu$ to the first equation and using the conservation of the current we find $\Box b(x)=0$ which shows that the auxiliary field $b$ is a free field and does not affect the dynamics. We could even generalise (\ref{GT1}) in order to impose a family of gauge conditions by writing
\be
\label{GT3}
{\cal L}=-\frac{1}{4}F_{\mu \nu}(x)F^{\mu \nu}(x)+b(x)\p^\mu A_\mu(x)-\frac{a}{2}b^2(x)-j_\mu(x) A^\mu(x)
\ee
with $a$ an arbitrary real constant. It is straightforward to show that $b$ is still a free field and the Lagrangian takes the form
\be
\label{GT4}
{\cal L}=-\frac{1}{4}F_{\mu \nu}(x)F^{\mu \nu}(x)+\frac{1}{2a}\big (\p^\mu A_\mu(x)\big )^2-A^{\mu}(x)j_{\mu}(x) 
\ee 
We thus obtain a family of gauge fixed Lagrangians depending on the parameter $a$. We can compute the Feynman rules from this Lagrangian and we find for the photon propagator
\be
\label{GT5}
 G^{\mu \nu}_{\rm F}(k)=\frac{-\im}{k^2+\im \epsilon}\left(\eta^{\mu \nu}
  -\frac{ k^{\mu}k^{\nu}}{k^2 (1-a)}\right)
\ee
The choice $a=1$, which yields  $G^{\mu \nu} =-\im \frac{\eta^{\mu \nu}}{k^2+\im \epsilon} $, is known as the Feynman gauge, and that with $a=0$ and $G^{\mu \nu} = \frac{-\im}{k^2+\im \epsilon}\left(\eta^{\mu \nu} -\frac{ k^{\mu}k^{\nu}}{k^2} \right)$, the Landau gauge. 

All these steps appear to be arbitrary and, in some sense, they are. Their final justification is based on the fact that, as we can show for QED, they lead to a mathematically consistent and physically correct theory. We recognise immediately two problems:

The first is that of gauge invariance. For every choice of gauge, for ex. every choice of $a$ in (\ref{GT3}), we obtain a different field theory: the Feynman rules are different and so are the resulting correlation functions. We must prove that all these different field theories yield the same values for all physically measurable quantities. The second is that of unitarity. The theory is formulated in a Hilbert space containing unphysical, negative norm states. We must prove that in a scattering experiment, if the initial state contains only physical, transversely polarised photons, the final state will also be physical with probability equal to one. These problems were addressed in the early formulation of QED, in particular by Dyson and Schwinger\cite{QED-Rev}. In 1949 W. Pauli and F. Villars wrote the first gauge invariant regularisation scheme\cite{Paul-Vill} which guarantees the consequences of gauge invariance at all intermediate steps of the calculation. In 1950 J.C. Ward showed that the conservation of the electromagnetic current yields a set of identities for the correlation functions, known since as {\it Ward identities}\cite{Ward}, in terms of which the gauge invariance of the physical results could be obtained.  
\vskip 0.3cm
Before leaving the subject of QED, I want to mention the results for massive vector fields, obtained by A. Proca\cite{Proca} and E.C.G. Stueckelberg\cite{Stueck1}, much before the formulation of the renormalisation program. 

As we said earlier, the first Yukawa theory assumed that the meson had spin equal to 1. This motivated the study of massive vector fields and in 1936 A. Proca wrote the equation of motion for such a field\footnote{I write here the equation for a real field, although Proca considered a complex one. Incidentally, Proca thought that his equation could describe the electron-positron system. The title of his article is {\it “Sur la théorie ondulatoire des électrons positifs et négatifs.”} or, {\it “On the wave theory of positive and negative electrons.”} I suspect that this misunderstanding was due to a confusion regarding the representations of the Lorentz group, which were not known at the time. The spinorial and the vector representations -- ($0,\1/2$)$\oplus$($\1/2, 0$) and ($\1/2,\1/2$) in modern notation -- are both four dimensional, and people erroneously thought they were equivalent. The Dirac spinor belongs to the first and the vector field to the second}. It takes the simple form
\be
\label{Proca}
\p^\mu F_{\mu \nu} +m^2 A_\nu=j_\nu
\ee
and can be derived from the Lagrangian density
\be
\label{Proca0}
{\cal L}_P=-\frac{1}{4} F_{\mu \nu}F^{\mu \nu}+\1/2 m^2A_\mu A^\mu +ej_\mu A^\mu
\ee
It does not require any gauge fixing because the mass term breaks gauge invariance. If the current is conserved we obtain $\p^\nu A_\nu=0$, which shows that the equation describes only three physical degrees of freedom. It is the equation we use today for particles like $W^\pm$ and $Z^0$. 

Proca did not compute the Green function corresponding to his equation, but had he done so, he would have found the general expression\footnote{I do not know who was the first to write this expression. It is given in the book of G. Wentzel\cite{Wentzel} of 1943, one of the first books on quantum field theory.}
\be
\label{Proca1}
G^{\mu \nu}_{\rm F}(k)=\frac{-\im}{k^2-m^2}\left(\eta^{\mu \nu}
  -\frac{ k^{\mu}k^{\nu}}{m^2}\right)
\ee
The important difference with the photon propagator in (\ref{GT5}) is the behavior for large momenta: it is $k^{-2}$ for the photon and $k^0$ for the massive Proca equation. As a result a field theory with massive vector fields appears to be more singular than QED. In today's language, a field theory defined by (\ref{Proca0}) contains only the three physical degrees of freedom of a massive vector field, but it is non-renormalisable by power counting.

Stueckelberg proposed a different formulation and wrote
\be
\label{Stueckel1}
(\Box +m^2)A_\mu=j_\mu
\ee
This looks very strange because it seems to describe four degrees of freedom -- the four components of the vector field -- and, under the usual quantisation rules, the quanta created from $A_0$ would yield states with negative norm. We conclude that this equation is not appropriate to describe a massive vector field. Furthermore, the condition $\p^\nu A_\nu=0$ does not follow from the equation and we obtain instead a weaker form $(\Box +m^2)\p^\mu A_\mu=\p^\mu j_\mu$ which implies that, for a conserved current, $\p^\mu A_\mu$ is a free field. On the other hand the Green function of this equation is just the $\eta^{\mu \nu}$ part of (\ref{Proca1}), so it behaves at large momenta like $k^{-2}$.

Stueckelberg had the brillant idea to add a massive scalar field $B(x)$ with the same mass $m$, thus increasing the number of degrees of freedom from 4 to 5. $B(x)$ is assumed to be coupled to the same current $j_\mu$ with a derivative coupling proportional to $1/m$. In today's notation the Lagrangian density for the $A_\mu - B$ system could be written as
\be
\label{Stueckel2}
{\cal L}_{S_1}=-\1/2 (\p_\mu A^\nu)^2+\1/2 m^2A_\mu A^\mu +\1/2(\p_\mu B)^2-\1/2m^2B^2+ej_\mu(A^\mu+\frac{1}{m}\p^\mu B)
\ee
With an incredible foresight and intuition, in the absence of precise rules for the perturbation expansion, Stueckelberg understood that the $B$ field in this Lagrangian supplies the missing $k_\mu k_\nu/m^2$ term of the propagator (\ref{Proca1}). As a result, the two Lagrangians (\ref{Proca0}) and (\ref{Stueckel2}) will give identical results for scattering amplitudes as long as the initial and final states contain only the physical degrees of freedom of a massive vector field. I will not present the original argument here because today this result follows easily from the standard Feynman rules. The propagators of both $A_\mu$ and $B$ behave like $k^{-2}$ at large momenta, but the theory is still non-renormalisable because of the derivative coupling of the $B$ field which brings a power of $k$ at every vertex. 

In 1941, W. Pauli\cite{Pauli1941} remarked that Stueckelberg's Lagrangian has a hidden gauge symmetry: It remains invariant under the transformation $A_\mu \rightarrow A_\mu + \p_\mu \Lambda$ and $B \rightarrow B+m\Lambda$, {\it provided} the function $\Lambda(x)$ satisfies the free field equation $(\Box +m^2)\Lambda (x)=0$. We see that, for the physical sector of the theory, the field $B(x)$ could be eliminated by a suitable choice of gauge. We can perform a redefinition of the electron field $\psi$ of the form
\be
\label{Stueckel3}
\psi'(x)= {\mathrm {exp}}\left[ -\frac{\im e B(x)}{m}\right ] \psi(x)
\ee
which looks like a gauge transformation, but with the field $B$ as the gauge function. The Lagrangian (\ref{Stueckel2}), including the fermion fields, becomes
\be
\label{Stueckel4}
{\cal L}_{S_2}=-\1/2 (\p_\mu A^\nu)^2+\frac{m^2}{2}A_\mu A^\mu +\1/2(\p_\mu B)^2-\frac{m^2}{2}B^2+\bar{\psi}'(\im \dsl -m_e)\psi'+e\bar{\psi}'\gamma_\mu \psi'A^\mu
\ee
The $B$-field has decoupled and is now a harmless free field. ${\cal L}_{S_2}$
is still quantised in a space containing unphysical negative norm states, but
it is renormalisable by power counting. Furthermore, since field redefinitions such as (\ref{Stueckel3}) do not change the physical $S$-matrix\footnote{This result, which we can derive heuristically using the reduction formula in quantum field theory, has been proven rigorously and it is known as {\it Borchers theorem}\cite{Borch}.}, the three Lagrangians (\ref{Proca0}), (\ref{Stueckel2}) and (\ref{Stueckel4}) give the same results when computing physical scattering amplitudes. Here is the magic of the Stueckelberg formalism:
the Lagrangian (\ref{Proca0}) which contains only physical degrees of freedom, although it is non-renormalisable by power counting, for all practical purposes behaves effectively like a renormalisable one. It is the phenomenon we shall encounter, in a more complicated version, in the Standard Model. 

\vskip 0.3cm
$\bullet$ {\it Renormalisation and gauge invariance -- Yang-Mills.} The extension of the renormalisation program from QED to Yang-Mills has not been an easy exercise. For many years the physical motivation was missing. There was no evidence that Yang-Mills fields play any role in particle physics. In fact the scarce early studies used Yang-Mills as a laboratory to gravitation because both theories share the common feature of being gauge invariant with self-coupled gauge fields. The Lagrangian density is
\be
\label{YMRen1}
{\cal L}_{YM} =-\frac{1}{4}F^a_{\mu \nu}F^{a \mu \nu} ~~{\mathrm {with}}~~F^a_{\mu \nu}=\p_\mu A^a_\nu-\p_\nu A^a_\mu +g f^{abc} A_\mu^b A_\nu^c
\ee
$A^a_\mu(x)$, $a=1, \cdots, N$  is the multiplet of the vector gauge fields and we sum over $a$. For $SU(2)$ $N=3$, for $SU(3)$ $N=8$ etc. $g$ is the coupling constant and $f^{abc}$ are numbers which characterise the Lie algebra. For $SU(2)$ we have $f^{abc}=\epsilon^{abc}$. To this Lagrangian we can add matter fields, spinor or scalar, but, as we see, Yang-Mills, even without matter fields,  is a fully interacting quantum field theory, contrary to QED where a theory containing only photons is a free field theory. 

In retrospect we can see easily that the na\"ive quantisation rules fail for Yang-Mills theories. Let us try to impose the gauge condition $\p^\mu A^a_\mu(x)=0$ using the method of the Lagrange multiplier we introduced in (\ref{GT1}). We add to (\ref{YMRen1}) a term proportional to $b^a(x)\p^\mu A^a_\mu(x)$. It is an easy exercise to verify that $b^a(x)$ is no more a free field.

The first person who noticed this problem was Feynman. In 1962 he was lecturing in a conference on the theory of gravitation\cite{Feyn-Gr} and attempted to compute graviton one loop diagrams\footnote{The title of his lecture was {\it “Quantum Theory of Gravitation”}.}. He found that the unphysical components did not decouple and checked that the same phenomenon appears also when computing one-loop Yang-Mills diagrams. Feynman knew better that anybody else that the perturbation expansion is {\it defined} by the Feynman rules, so he proceeded to modify the rules by postulating that, at every closed loop, we should add a new set of unphysical degrees of freedom designed to cancel the ones which are present in the covariant propagator of massless vector -- or tensor for gravity -- fields. Feynman discovered this rule at one loop diagrams but conjectured that it should be true at all orders. In 1967, L.D. Faddeev and V.N. Popov\cite{Fad-Pop} re-derived this result in a formal way, using the Feynman path integral method, which shows its general validity. These new unphysical degrees of freedom are known as {\it Faddeev-Popov ghosts.} The question of renormalisability had not been addressed.

\subsubsection{Renormalisation and symmetry} 
\label{Ren-Sym}
The importance of symmetry has been recognised quite early in classical physics, but it is only during the first half of the twentieth century that Amalie Emmy Noether gave a mathematical formulation of this importance by establishing the connection between symmetries and conservation laws. In field theory invariance under a group of transformations is manifest in two ways: first by the field content. Fields are assumed to fill linear  irreducible representations of the symmetry group. An early example of this property was Kemmer's 1938 prediction\cite{Kemmer} of the existence of the neutral pion. Second, by imposing relations among the masses and coupling constants of the theory. An obvious example is the equality of masses for all fields belonging to the same irreducible representation. Another one is the absence of terms in the Lagrangian that violate the symmetry, which implies that the corresponding coupling constants should vanish. As a result of Noether's theorem, we can prove that, for a local field theory, the invariance under a set of continuous transformations yields the existence of currents which are conserved by virtue of the equations of motion. This in turn implies the existence of generalised Ward identities among the correlation functions of the theory\cite{Coleman-sym}.

I do not know who was the first to spot a possible conflict between renormalisation and symmetry. In quantum field theory we have to deal with the divergent expressions of the perturbation expansion and the process of renormalisation consists of choosing certain values for a finite set of constants. Conflict will arise if there is no choice compatible with the requirements of a given symmetry. In our jargon, a symmetry of the classical equations of motion that cannot be implemented in the renormalised quantum theory is called {\it “anomalous”.} Anomalies may appear only in the process of handling divergencies, therefore a simple way to make sure they will not occur is to introduce a regularisation scheme which respects the symmetry in question\footnote{There may exist non-perturbative anomalies, which do not manifest at any finite order of perturbation, but I will not discuss them here. For an example, see E. Witten, ref. \cite{Witten}.}. Since every regularisation scheme involves the introduction of a dimensionful parameter, a good way to guess which symmetries may be anomalous is to look for those that are incompatible with such parameters. In this section I will show the case of chiral symmetry, which requires the presence of massless fermions, and in a following section that of the symmetry under scale transformations.

\begin{figure}
\begin{center}
\includegraphics[height=35mm]{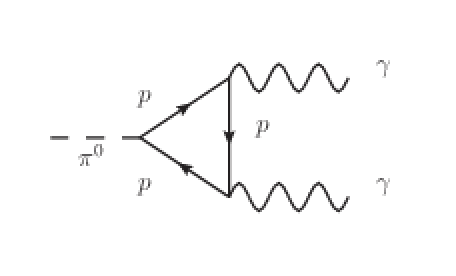} 
\caption{The $\pi^0$ decay diagram via a proton loop.}
\label{Fig-pizero-dec}  
\end{center}    
\end{figure}

The anomalous conservation of the axial current can best be seen in the simple model of quantum electrodynamics. At the classical level we obtain the conservation equations for the two currents, the vector and the axial.
\be
\label{ABJ1}
j_\mu=\bar{\psi}\gamma_\mu\psi~~~~\p^\mu j_\mu=0~~~{\mathrm and}~~~ j_\mu^{(5)}=\bar{\psi}\gamma_\mu \gamma_5 \psi~~~~\p^\mu j_\mu^{(5)}=2\im m\bar{\psi} \gamma_5  \psi
\ee
with $m$ the electron mass.
It was found that in the quantum theory we cannot satisfy both these conservation equations. If we choose to keep the conservation of the vector current, a natural choice since this current is coupled to the photon field, the conservation of the axial current is modified. The discovery of this strange phenomenon has a long history, like a play in several acts.
\vskip 0.2cm
$\bullet$ {\it Act I. The life-time of the neutral pion.} As we said already, in 1938 N. Kemmer conjectured the existence of a neutral Yukawa meson in order to restore isospin invariance of the nuclear forces\cite{Kemmer}. In 1940 S. Sakata and Y. Tanikawa\cite{Sak-Tan} remarked that it could have escaped detection because it could decay into two photons with a very short lifetime. It is precisely what turned out to be the case. The first estimation of the life-time is due to J. Steinberger in 1949\cite{Steinb}. He assumed the pion to decay via the triangle diagram of Figure \ref{Fig-pizero-dec} with a virtual proton running in the loop. He did not know the pion spin and parity and considered all possibilities $0^+$, $0^-$, $1^+$ and $1^-$. The diagram is superficially divergent and Steinberger used the Pauli-Villars regularisation scheme\cite{Paul-Vill}. It was one of the first correct one loop calculations in quantum field theory outside QED because the techniques were still quite new. In the abstract we read: {\it “The results are quite different from those of previous calculations, in all those cases in which divergent and conditionally convergent integrals occur before subtraction,
but identical whenever divergences are absent.”} The “previous calculations” Steinberger refers to did not respect gauge invariance for the final photons. Although Steinberger does not address the question of the conservation of the axial current, his result for the $\pi^0$ decay rate contains what will become later the axial anomaly\footnote{It is interesting to note that $\pi^0$ was discovered the following year by Steinberger, Panofsky and Steller\cite{Steinbetal}. So, Steinberger is the only particle physicist who computed theoretically the decay properties of a particle and subsequently discovered it experimentally.}. 
\vskip 0.2cm
$\bullet$ {\it Act II. The Schwinger model.} In 1962 Julian Schwinger proposed the toy model of quantum electrodynamics with massless fermions in a two-dimensional space-time\cite{Schwing-mod}. The model turns out to be exactly solvable, but the solution describes a physical system completely different from the one we guess  in perturbation theory: instead of massless fermions interacting with the electromagnetic field, we obtain a set of free massive bosons. For many years this model served as  playground for the study of more complicated theories. Schwinger's motivation was the connection between gauge invariance and the mass of the photon and we shall mention it again in section \ref{SSB}. He never used the term “anomaly”, but a byproduct of his analysis was the discovery that the quantum system obeys equations which differ from the ones we obtain classically. In particular, the exact theory has no conserved axial current.
\vskip 0.2cm
$\bullet$ {\it Act III. The Adler-Bell-Jackiw anomaly.} In 1964 M. Gell-Mann proposed an algebraic scheme based on the algebra of $SU(3)\times SU(3)$ as an approximate invariance of strong interactions\cite{CA1}, \cite{CA2}. It assumes the approximate conservation of eight vector and eight axial currents and the generators of the two $SU(3)$ factors are the combinations $V-A$ and $V+A$, hence the name {\it “chiral symmetry”}. This scheme turned out to have far reaching consequences for our understanding of the fundamental interactions and we shall see it again in section \ref{SSB}. It is exemplified by the simple model of three free fermions, for ex. the three quarks $u$, $d$ and $s$. At the limit of massless quarks the classical system is invariant under $U(3)\times U(3)$ and the breaking due to the mass terms was assumed to be weak. This assumption appears to be fully justified for the $SU(2)\times SU(2)$ subgroup -- because of the very small masses of the $u$ and $d$ quarks --  and reasonable for the heavier $s$ quark. 

During the nineteen sixties this current algebra scheme became the starting point of numerous theoretical investigations and gave rise to many important results\cite{CA2}.
Of particular significance for our story is a theorem obtained by the Scot David Sutherland who was fellow at the CERN Theory Division. Sutherland was the first to look at the consequences of current algebra in the presence of the electromagnetic interactions and derived a simple, albeit totally unexpected, result: at the limit of exact chiral symmetry the electromagnetic decays of the neutral $0^-$ mesons, such as the $\pi^0$ or the $\eta$, are forbidden\cite{Sutherl}. The argument is very simple and depends only on the canonical commutation relations among the neutral components of the vector and axial currents. This result was embarassing for two reasons: first, it was in obvious contradiction with experiment. Second, it seemed to be against all na\"ive expectations coming from simple field theory models. This second problem was not immediately recognised and brings me to the celebrated Bell-Jackiw paper on anomalies. 
\begin{figure}
\begin{center}
\includegraphics[height=35mm]{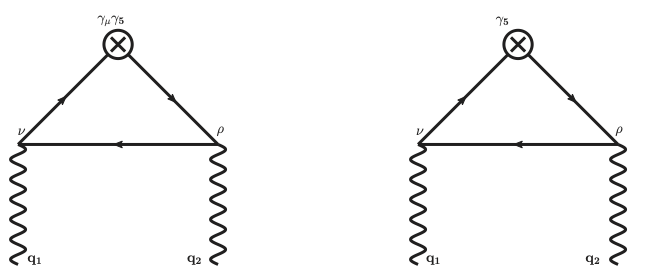} 
\caption{The two diagrams showing the QED axial anomaly.}
\label{Fig-anom}  
\end{center}    
\end{figure}

As far as I know the first person to worry about the compatibility of Sutherland's theorem with field theory and current algebra was Roman Jackiw, an American fellow who was spending a year at CERN on leave from the Harvard Society of Fellows. Jackiw discussed this problem with me and we both asked the advice of Henri Epstein, a French mathematical physicist and a world expert on field theory. Epstein explained to us that na\"ive consequences of classical equations of motion and symmetries in field theory are not necessarily correct at the quantum level. I was satisfied with this explanation, which involved singular products of distributions and all the associated mathematical artillery, but Jackiw was not. Especially when he became aware of Steinberger's 1949 result. Discussing with John Bell, he realised that Steinberger's model for the pion was essentially identical to the models which were supposed to exemplify the algebra of currents. Now the contradiction was clear: the same model  quantised canonically was giving rise to chiral symmetry which, presumably,  included Sutherland's theorem; but  explicit one loop computations gave a contradictory result. Bell and Jackiw wrote a remarkable paper\cite{BJ} which contains two parts: in the first they expose the problem in a clear and unambiguous way. They compute the one-loop diagrams of Figure \ref{Fig-anom} using Pauli-Villars regulators, the only regularisation scheme respecting gauge invariance which was known at the time, and they correctly identify the origin of the problem as being the masses of the auxiliary regulator fields which break the conservation of the axial current. The second part, although incorrect, is also very interesting: they attempt to adjust the regularisation scheme in order to restore the full chiral symmetry. As it turned out, this modified scheme does not lead to a consistent renormalisable quantum field theory. They were the first in a long series of unsuccessful attempts to regularise the anomaly away. A bit later, S. Adler addressed the same problem\cite{Adl-an} but with a different philosophy: instead of trying to eliminate the anomaly, he accepted it as the correct form of the divergence of the axial current including quantum corrections. For QED equation (\ref{ABJ1}) gets modified as
\be
\label{ABJ3}
\p^\mu j_\mu^{(5)}=2\im m\bar{\psi} \gamma_5  \psi+\frac{\alpha}{4\pi}\epsilon_{\mu\nu\rho\sigma}F^{\mu\nu}F^{\rho\sigma}
\ee
with $\alpha$ the fine structure constant. It was further shown\cite{Adl-Bard} that this one-loop result, including the value of the coefficient $\alpha/4\pi$,  gets no higher order corrections. Rare are the examples in quantum field theory in which a one loop effect remains unchanged at all orders. Furthermore, there is no consistent subtraction scheme respecting gauge invariance, which could eliminate the anomaly and restore the classical conservation equation (\ref{ABJ1}). 

For pure QED the anomaly in the conservation of the axial current is a curiosity. This current is not coupled and it is not part of the renormalisation program. However, for more complicated models it plays a crucial role, as we just saw regarding the decay of the neutral pion.  
Looking again at eq. (\ref{ABJ3}) we notice a peculiar feature: the anomaly is independent of the mass of the fermion that goes around the loop. Any fermion, light or heavy, with charge $e$ will give the same contribution.  A few years later, in 1971,  Julius Wess and Bruno Zumino found a set of consistency conditions which determine the form of the possible anomalous terms in any gauge theory\cite{WZ}. All these results have been important in the building of an anomaly-free version of the Standard Model\footnote{With the subsequent introduction of powerful mathematical techniques for the study of gauge theories, the Wess-Zumino consistency conditions became the starting point of a new field of research, that of topological field theories.}. 
\vskip 0.2cm
This is not the end of the play, but it is time for an intermezzo. The final act belongs to the Standard Model and will be presented in section \ref{SM-Anom}.

\subsubsection{The renormalisation group} 
\label{Ren-Gr}
The concept was introduced by Stueckelberg and A. Petermann in 1951\cite{Stueckel-Peter}. In principle, it is quite simple. As we said previously, the theory of renormalisation involves the introduction of a set of arbitrary constants, to be determined by experiment. In general they are the values of parameters such as masses, coupling constants and the normalisation of the fields. Their determination depends on the way we choose to apply the renormalisation program, each one yielding a different quantum field theory. Stueckelberg and Petermann asked the question under which conditions all these theories are physically equivalent. The answer is that they are equivalent provided all the differences can be absorbed in a redefinition of the renormalisation constants. In the process of proving this statement, they introduced a formal way to map one theory to another and proved that these mappings form a group\footnote{The first paper in reference \cite{Stueckel-Peter} is a short report in a Conference. The second is a presentation of the general theory of renormalisation which, for the first time, introduces the mathematical theory of distributions. Incidentally, Stueckelberg and Petermann use the terms “normalisation theory” and “normalisation group”. These papers went mainly unnoticed, first because they seemed to be only formal and, second, because they appeared in Helvetica Physica Acta, a journal not widely read. The second article is written in French.}.

In 1954, Gell-Mann and F. Low\cite{Gell-Low} re-invented the concept of the renormalisation group with a very important new element: they understood that, under certain conditions, the transformations of the group applied to a correlation function can be represented by changes of the energy scale. They used this property to compute the asymptotic behavior of the propagators in QED. It is this remark that made the renormalisation group a powerful tool for the study of quantum field theory at high energies.
Their analysis is close to the one we are doing today because they show that the asymptotic behavior is governed by two functions, $\psi$ and $q$ in their notation, which can be computed order by order in a perturbation expansion and are functionally related to our $\beta$ and $\gamma$ functions. They even noticed the special role of the zeroes of $\psi$, which are the same with the zeroes of our $\beta$-function, and conjectured that the physical value of $\alpha$ in QED could be computed this way. In their work there is no reference to Stueckelberg and Petermann. 

In line with the general disinterest in quantum field theory, renormalisation group ideas were largely ignored by the high energy physics community until the end of the 1960s\footnote{Some important results were obtained by the Landau school, see below. There exists also a dedicated chapter in the book by N. N. Bogoliubov and D. V. Shirkov\cite{Bog-Sh}, first published in Russian in 1957.}. They became fashionable by the end of the decade, following the rebirth of quantum field theory. Some important papers published at that time contributed to this effect. 

Probably the most significant for particle physics were the ones by K. Symanzik and C.G. Callan\cite{Sym-Cal} with the celebrated Callan-Symanzik equation. Strictly speaking, they both address a question which is not directly related to the renormalisation group, although it leads to the same physical results. I will present them shortly. Here I want to mention an important work by K.G. Wilson who first applied the renormalisation group ideas to strong interactions\cite{Wilson1}. Although the concept of asymptotic freedom was not known, Wilson presented a clear analysis of the high energy behavior of coupling constants with the possibility of running into fixed points or even limiting cycles. In a series of subsequent publications\cite{Wilson2} he applied the renormalisation group ideas to the physics of phase transitions and critical phenomena.

Using modern notation, we can write the renormalisation group equation for the simplest scalar field theory described by the Lagrangian density (\ref{r3}) as follows: 
The physical theory, i.e. the one which can be used to compare with the experimental data, is the one in which the value of the mass equals the physically measured value $m_{\mathrm {ph}}$, the field is normalised to one on the mass-shell, and the coupling constant is the value of the four-point function at some physically measurable point, for example the scattering amplitude at threshold. The corresponding three renormalisation conditions take the form\footnote{The renormalisation program is usually formulated in terms of the “amputated, one-particle-irreducible $n$-particle correlation functions” $\Gamma^{(n)}$. They are defined as the sum of all connected diagrams with $n$ external lines which cannot be separated into two disconnected parts by cutting a single internal line (the meaning of “one-particle-irreducible”) and from which we have dropped the propagators of the external lines (the meaning of “amputated”).}
\be
\label{RG1}
\Gamma^{(2)}(p^2=m_{\rm {ph}}^2)=0~;~~\frac{\partial \Gamma^{(2)} (p^2)}{\partial p^2}\Bigg|_{p^2=m_{\rm {ph}}^2}=\im~;~~\Gamma^{(4)}(p_1,\ldots,p_4)\Big|_{m_{\rm {ph}}^2}=\im\lambda
\ee
The concept of the renormalisation group stems from the realisation that, from the purely technical point of view, nothing forces us to choose these physical renormalisation conditions; any other choice would be acceptable, provided it yields well-defined values for the three arbitrary constants. The example which is most commonly used is to introduce a mass parameter $M$ and rewrite the second and third of the equations (\ref{RG1}) as\footnote{We could use even more general conditions, for example not keeping the physical mass in the first of the equations (\ref{RG1}), but no substantially new insight is gained. Examples of such more general renormalisation group equations are given in reference \cite{WeinbRG}.}
\be
\label{RG2}
\frac{\partial \Gamma^{(2)} (p^2)}{\partial p^2}\Bigg|_{p^2=M^2}=\im~;~~\Gamma^{(4)}(p_1,\ldots,p_4)\Big|_{\mathrm {point} \,M}= -\im\lambda^{(M)}
\ee
In the last condition by “point $M$” we mean some point in the space of the four momenta $p_a^2=M^2$, $a=1,\ldots,4$, provided it is a point at which $\Gamma^{(4)}$ is regular. Similarly, $\Gamma^{(2)}(p^2)$ and its derivative should be well defined at $p^2=M^2$. For historical reasons the point $M$ in momentum space is called the {\it subtraction point}. 

Since $M$ has no physical meaning, we can change it by a transformation $M_1 \longrightarrow M_2=\eta M_1$ with $\eta$ a numerical factor, which can take any value in a domain including the value $\eta=1$ and such that all resulting points $M_2$ belong to the domain of analyticity of $\Gamma^{(2)}$ and $\Gamma^{(4)}$. These transformations form a group, called {\it the renormalisation group}. For every value of $M$ we obtain a different set of correlation functions and the condition that all these theories describe the same physics is the well-known equation of the renormalisation group.
\be
\label{RG3}
\left[M\frac{\partial}{M}+\beta\left(\lambda, \frac{m}{M}\right) \frac{\partial}{\partial \lambda} -n\gamma\left(\lambda, \frac{m}{M}\right)\right]\Gamma^{(n)}(p_1,...,p_{n};m,\lambda;M)=0
\ee  
$\beta$ and $\gamma$ can be computed order-by-order in the perturbation expansion and, since they are dimensionless, they depend only on $\lambda$ and the ratio $m/M$\footnote{Here $m$ is the physical mass. In order to simplify the notation we have dropped the subscript ${\mathrm {ph}}$.}. 

The renormalisation group equation (\ref{RG3}) describes the response of the system under  a rescaling of the subtraction point. One could argue that it is of limited interest because it only solves a problem we have created for ourselves by not making the physical choice   $M=m_{\rm {ph}}$. This is undoubtedly correct and we need some more input in order to extract physically interesting information out of this equation. It was the contribution of Gell-Mann and Low and I will come back to it presently.

Let me  come now to the year 1970 and the work of K. Symanzik and C.G. Callan\cite{Sym-Cal}.  The basic remark is that in a four-dimensional renormalisable field theory in the classical approximation, the only terms in the Lagrangian which break  scale invariance are those with dimensionful parameters, such as mass terms. If they are absent, for example if we consider the theory (\ref{r3}) with $m=0$,  we expect the dependance of the correlation functions on the external momenta to be determined by dimensional analysis. The  2-point function  should be given by $\Gamma^{(2)}(p,\lambda)=f(\lambda) p^2$ 
with $f$ some function of the coupling constant. This is indeed correct at the classical level with $f=\im$. We know however that this behaviour is wrong for the quantum theory because the divergent diagrams in the  perturbation expansion will force us to introduce a dimensionful cut-off $\Lambda$ and we find terms proportional to powers of $\log (p^2/\Lambda^2)$. In our previous terminology, we find that the symmetry under scale transformations is anomalous. The Callan-Symanzik equations are the correct Ward identities of broken scale invariance\footnote{A very good pedagogical presentation can be found in S. Coleman, reference \cite{Coleman-Dil}.}. They take the form
\be
\label{RG4}
\left[m\frac{\partial}{m}+\beta(\lambda) \frac{\partial}{\partial \lambda} -n\gamma(\lambda)\right]\Gamma^{(n)}(p_1,...,p_{n};m,\lambda)=m^2\delta(\lambda)\Gamma^{(n)}_{\phi^2}(p_1,...,p_{n};m,\lambda)
\ee
where $m$ denotes the physical mass $m_{\rm {ph}}$. $\Gamma^{(n)}_{\phi^2}$ is called “the $n$-point function with a zero momentum $\phi^2$ insertion”. Its meaning is explained in references \cite{Sym-Cal} and \cite{Coleman-Dil} but we won't need it here. The functions $\beta$, $\gamma$ and $\delta$ can be computed in perturbation theory.  They codify the anomalies of scale invariance. Contrary to what we found in the axial anomaly, they receive non-zero contributions at all orders of the perturbation expansion. The $\gamma$ function is often called “the anomalous dimension of the field”\footnote{This terminology, although it is commonly used, it  is misleading because it sounds as if the quantum theory violates dimensional analysis, which obviously cannot happen. For a discussion see reference \cite{Coleman-Dil}.}.

The two equations, (\ref{RG3}) and (\ref{RG4}), although they look vaguely similar, they describe totally different properties of the theory. Varying the subtraction point $M$ we obtain a set of quantum field theories, but only one of them, obtained for $M=m_{\rm{ph}}$, is the physical one. It follows that the renormalisation group transformations map one unphysical theory to another. Scale transformations on the other hand change the value of the physical mass, so they map one physical theory to another physical theory. 

Now that we explained in what these equations are different, we can turn into
what they have in common. They both describe the response of the theory under a change of a dimensionful parameter. Were it the only such parameter, we could appeal to ordinary dimensional analysis and trade this dependence with that on the external momenta.  We expect that this can be achieved at very high energies when masses  could be neglected. It was the remark of Gell-Mann and Low. This na\"ive expectation turns out to be correct order by order in the perturbation expansion, provided we look at quantities which are free from infrared singularities\cite{IR}. Let us rescale all external momenta $p_i\rightarrow \rho~ p_i$ with a real parameter $\rho$ and take $\rho >>1$. Starting from either one of the two equations (\ref{RG3}) or (\ref{RG4}), we obtain 
\be
\label{RG5}
\left[-\rho\frac{\partial}{\partial \rho}+\beta(\lambda)\frac{\partial}{\partial
  \lambda}-n\gamma(\lambda)\right]\Gamma_{\mathrm {as}}^{(n)}(\rho~p_i, \lambda)=0
  \ee
with $\Gamma_{\mathrm {as}}^{(n)}$ the asymptotic form of the Green function at large momenta in a kinematical region free from infrared singularities. 

The differential equation (\ref{RG5}) was solved around the end of the 18th century by Gaspard Monge, who was studying problems of laminar flow in hydrodynamics. We make a change of variables following the hydrodynamic flow: $(\rho, \lambda) \rightarrow (\rho, \bar{\lambda})$ with $\bar{\lambda}(\lambda, \rho)$ solution of the {\it characteristic equation}
\be
\label{RG6}
\left[-\rho\frac{\partial}{\partial \rho}+\beta(\lambda)\frac{\partial}{\partial \lambda}\right]\bar{\lambda}( \lambda, \rho)=0,~~~\bar{\lambda}(\lambda, \rho=1)=\lambda
\ee
which is equivalent to
\be
\label{RG7}
\rho \frac{\partial \bar{\lambda}}{\partial \rho}=\beta(\bar{\lambda})
\ee
The general solution of (\ref{RG6}) is now given by
\be
\label{RG8}
\Gamma_{\rm as}^{(n)}(\rho p_1,...,\rho p_{n};\lambda)=\Gamma_{\rm as}^{(n)}(
p_1,...,p_{n};\bar{\lambda})\exp\left \{-n\int_{1}^\rho \d\rho'
\gamma[\bar{\lambda}(\lambda ,\rho')]\right \}
\ee
The physical meaning of (\ref{RG8}) is clear. Scaling all momenta of a Green
function by a common factor $\rho$ and taking $\rho$ large, has the
following effects: (i) It multiplies every external line by the scale dependent factor $\exp(-\int_1^\rho \d\rho' \gamma(\bar\lambda(\lambda,\rho')))$\footnote{ We see that the Green functions  get rescaled as if they had dimensions related to $\gamma(\lambda)$. This is the origin of the term “anomalous dimension”.} and (ii) it replaces the physical coupling
constant $\lambda$ by an effective one, $\bar{\lambda}=\bar\lambda(\lambda,\rho)$, which is the solution of
(\ref{RG7}). For this reason $\bar{\lambda}$ is
 often called {\it running} coupling constant.
 
Gell-Mann and Low, in their 1954 paper, recognised the significance of the possible zeros of the $\beta$ function. Equation (\ref{RG7}) shows that, if $\beta(\lambda)=0$ for some $\lambda=\lambda^*$, the effective coupling constant $\bar{\lambda}( \lambda^*, \rho)$ becomes independent of $\rho$. They conjectured that the fine structure constant $\alpha \approx 1/137$ could be the solution of such an eigenvalue equation $\beta_{\rm {QED}}(\alpha)=0$. Obviously, as long as perturbation theory is our only guide, we cannot say anything about
the properties of $\beta(\lambda)$ for arbitrary $\lambda$ and we do not know whether it has
any non-trivial zeros. There have been several attempts to guess the properties of $\beta$ beyond perturbation  -- for ex. using the method of Padé approximants -- 
with some good results in problems of statistical mechanics, but no success in high energy physics. We know that $\beta(0)=0$, because $\lambda=0$ corresponds to a free field theory. We also know that the behavior of the effective coupling constant at high energies depends on the sign of the first non-zero term in the expansion of $\beta$.
\be
\label{RG9}
\beta(\lambda)=b_0\lambda^2+b_1\lambda^3 +\dots ~~,~~b_0=\frac{3}{16\pi^2}
\ee
The first term $b_0$ is positive, so $\bar{\lambda}$ grows with $\rho$. This behavior is common to all renormalisable quantum field theories which were known in the fifties and early sixties and it was often assumed to be universal. L.D. Landau concluded\cite{Landau1} that all these theories were logically inconsistent, unless the  coupling constant was zero, in which case they were trivial. In his article with I. Pomeranchuk they note {\it “We reach the conclusion that within the limits of formal electrodynamics a point interaction is equivalent, for any intensity whatever, to no interaction at all.”}\footnote{This was the famous {\it zero charge problem}. At high energies the effective charge $\bar{\alpha}(\alpha, \rho)$ grows without limits unless the physical charge $\alpha$ vanishes.} It is obvious that statements of this kind, coming from authorities with the status of  Landau, contributed in throwing a total discredit on quantum field theory. 
Let me remark however, that Landau was right to say that {\it “\dots weak coupling electrodynamics is a theory which is, fundamentally, logically incomplete\dots”}\footnote{Today we expect the problem to be solved in some kind of ultraviolet completion of the theory, for example in the framework of a grand unified theory.}. The mistake, hard to avoid with the knowledge of his time, was to generalise it to all quantum field theories and assume that no other behavior was possible. {\it “\dots We are driven to the conclusion that the Hamiltonian method for strong interactions is dead and must be buried, although of course with deserved honor.”}\cite{Landau2} The picture changed with the discovery of asymptotically free quantum field theories, but they belong to the new era and will be presented in section \ref{QCD}.

\subsubsection{Higher internal symmetries -- The quarks}
\label{Quarks}
In section \ref{GT} we presented the construction of the isospin symmetry of nuclear forces by Heisenberg and Kemmer. It is based on the group $SU(2)$ and the pair proton-neutron belongs to the two-dimensional fundamental representation. The discovery of strange particles brought the need to enlarge this symmetry and the “obvious” extension was $SU(2)\rightarrow SU(3)$ by enlarging the  doublet to a triplet. It was the model proposed by S. Sakata in 1956\cite{Sak} in which he added the $\Lambda$-hyperon to the two nucleons to form a fundamental triplet of $SU(3)$. This scheme did not fit the observed hadrons and in the following years many  groups  were considered as candidates for this larger symmetry. Finally it turned out that Nature was more subtle: $SU(3)$  is the right answer, but the fundamental triplet is not formed by particles we see. In 1961 M. Gell-Mann and Y. Ne'eman formulated the model which became known as “The eightfold way”\cite{SU3} in which the low-lying hadrons form octets of $SU(3)$. The peculiar property of this scheme is precisely the absence of physical particles belonging to the fundamental triplet representation. This led Gell-Mann and, independently G. Zweig, to propose in 1964 a composite model in which hadrons are bound states of constituents -- named “quarks” by Gell-Mann and “aces” by Zweig -- which fill the missing triplet\cite{quarks}. What is less known is that similar ideas had been expressed earlier, I believe first by L.B. Okun\cite{Okun}. He started from the Sakata model but he interpreted the members of the fundamental triplet not as the physical $p$, $n$ and $\Lambda$, but as fictitious particles\footnote{He used the term “sakatons”.}. In fact, our ideas on quarks followed a similar trajectory. At first they were considered as just another layer of compositeness and experiments were designed to discover them\footnote{There have been  even several unsubstantiated claims of discovery.}. One of the motivations to built ISR, the first hadronic collider, was to discover quarks. However, the picture became soon more complicated. The need to satisfy the requirements of Fermi statistics, led to the introduction of color\cite{color} and the hypothesis of “color blindness” led to the notion of confinement. Even today we have no real understanding of it and we cannot rigorously derive it from first dynamical principles\footnote{It is one of the unsolved problems endowed with a prize of one million dollars from the Clay Mathematics Institute. For a review of the various ideas around confinement see reference \cite{Sussk}.}.
As a result, in spite of some phenomenological successes of the simple non-relativistic quark model, most people shifted their views to considering quarks as mere mathematical devices to keep track of symmetry properties. Naturally, M. Gell-Mann, who strongly advocated this opinion, played a major role to its general endorsement. 
 The atmosphere changed with the SLAC results on deep inelastic scattering, but I will discuss them in section \ref{QCD}.

\subsubsection{Spontaneously broken symmetries (SSB)} 
\label{SSB}
I do not know who invented this term, which I find rather misleading\footnote{Sidney Coleman, in one of his Erice lectures, wanted to replace it with “secret symmetry”\cite{Coleman-SSB}, but it was not adopted.}, but the phenomenon is well known in many problems of classical and quantum physics, especially those related to phase transitions. Its main features can be summarised as follows: 

If the dynamics of a physical system is invariant under a group of transformations, we expect the equations of motion to admit symmetric solutions. To cite P. Curie\cite{Curie} {\it “ When certain causes produce certain effects, the elements of symmetry of the causes must be found in the effects produced.” } Expressed this way, the statement sounds almost trivial, however it may happen that, in a certain region of the parameter space, the symmetric solution becomes unstable. It is often the case in infinite systems exhibiting the phenomenon of phase transitions. In such cases the minimum energy solutions, which represent the stable states, are degenerate and have less symmetry than the equations of motion. The degree of degeneracy corresponds to the symmetry deficit. The symmetry is not “broken”, but it is “hidden” because it is not manifest in the spectrum of states. For a quantum system the degeneracy of the ground state is translated into the appearance of zero energy excitations. In the following I will present only the applications of this phenomenon to particle physics.
\vskip 0.3cm

$\bullet$ {\it Spontaneous breaking of global internal symmetries.} The phenomenon was first studied by Yoichiro Nambu\cite{Nambu} and Jeffrey Goldstone\cite{Goldst}. Nambu was motivated by a concrete physical problem, that of determining the hadronic axial vector current of weak interactions. Following the experimental discovery of parity violation and the measurement of the neutrino helicity in $\beta$-decay, the Fermi theory of the weak interactions was determined to be of the current$\times$current form. The current $J_\mu$ was the sum of a hadronic and a leptonic piece, each one being a superposition of a vector and an axial part.
\be
\label{Fermi1}
{\cal H}_W=\frac{G_F}{\sqrt{2}}J^\mu J^{\dagger}_\mu~;~~~~J_\mu=J_\mu^{(h)}+J_\mu^{(l)}~;~~~~J_\mu=V_\mu-A_\mu
\ee
In writing (\ref{Fermi1}) we made the assumption of {\it universality} which, in a rather vague sense, implied that the Fermi coupling constant $G_F$ was the same for all currents. It is simple to write the leptonic part in terms of the fields of known leptons\footnote{Today we must add also the $\tau$ piece.}
\be
\label{Fermi2}
J_\mu^{(l)}=\bar{\psi}_e\gamma_\mu(1+\gamma_5)\psi_{\nu_e} + (e\rightarrow \mu, \nu_e \rightarrow \nu_\mu) 
\ee
but there is nothing analogous for the hadronic part for which an expression in terms of free fields is meaningless. Therefore we have to look at particular matrix elements and the first one is neutron $\beta$-decay. Ignoring the small momentum transfer between the neutron and the proton, we can write the matrix element as
\be
\label{Fermi3}
\bra{p}J_\mu^{(h)}\ket{n}=\bar{u}_p\gamma_\mu \big{[} g_V(0) + g_A(0) \gamma_5\big{]}u_n
\ee
In general, $g_V$ and $g_A$ are two form factors which depend on the square of the momentum transfer\footnote{If we relax the zero momentum transfer approximation, we obtain the contribution of  additional form factors.}, but in our approximation they are
 two real numbers to be determined by experiment. They are the vector and the axial weak charges of the nucleon.  If we factor out the Fermi coupling constant, which is determined by the rate of muon decay, the experimental results give us the values
\be 
\label{Fermi4}
g_V(0)\simeq 1~,~~~g_A(0)\simeq 1.25
\ee
The surprising result is that of $g_V$. I remind that, in the absence of strong interactions, we would have obtained $g_V=g_A=1$, therefore the value in (\ref{Fermi4}) means that, for the vector part of the current, all strong interaction corrections manage to sum up to zero! It is the result we obtain for the electromagnetic current, where the electric charges of the proton and the positron are exactly equal -- in spite of the proton having strong interactions -- as a consequence of the conservation of the electromagnetic current. This strongly suggests that the vector hadronic current of the weak interactions must be a conserved current of the strong interactions, which we should identify with the Noether current of a symmetry. S.S. Gershtein and Y.B. Zeldovich and, independently, R.P. Feynman and M. Gell- Mann\cite{CVC}  formulated the {\it Conserved Vector Current} (CVC) hypothesis, according to which {\it the vector part of $J_\mu^{(h)}$ belongs to the triplet of the conserved isospin currents of strong interactions.} Is it possible to make a similar hypothesis for the axial current? In other words, can we assume that $\p^\mu A^{(h)}_\mu =0$, thus promoting  the isospin symmetry of the strong interactions from $SU(2)$ to $SU(2)\times SU(2)$? At first sight the answer is no and there is plenty of evidence against it. First, a conserved axial current would imply $g_A(0) \simeq 1$ in contradiction with (\ref{Fermi4}). Second,    such a hypothesis appeared to be incompatible with the data on leptonic pion decay which had been confirmed by the observation at CERN of the electronic mode with the correct rate. Indeed, since the pion is a pseudoscalar, its leptonic decay amplitude is given by the matrix element of the axial current
\be 
\label{Fermi5}
\bra{0}A^{(h)}_\mu(0)\ket{\pi(q)}=f_{\pi}\, q_\mu~;~~~~\bra{0}\p^{\mu}A^{(h)}_\mu(0)\ket{\pi(q)}=\im f_{\pi}m_{\pi}^2
\ee
with the pion decay constant $f_{\pi} \simeq 130\,  {\rm MeV} \simeq 0.93 \, m_\pi$. Third, and most important, there is no trace of such a chiral symmetry in the spectrum of hadrons. The proton and the neutron form a doublet of isospin but they have no partners with opposite parity.
For all these reasons, the conservation of the axial current seemed to be excluded.

Nambu was the first to observe that all this “evidence” was in fact misleading. Let us start with the value of $g_A(0)$ and the pion decay amplitude. They share a striking feature: they are both numerically weak. Indeed, in the scale of strong interactions  the change from 1 to 1.25 is a rather small renormalisation, and $m_{\pi}^2$ is very small compared to the square of a typical hadron mass, which is of order 1 GeV$^2$. So, the question is: could we assume that the axial current is {\it approximately} conserved? But then, what about the third piece of evidence, the absence of chiral symmetry in the spectrum of hadrons? Here the lessons we learned from the phenomenon of spontaneous symmetry breaking in statistical mechanics come handy. Translated into the language of particle physics, they imply that the invariance of the equations of motion under a group of transformations may be  realised in at least  two different ways. In the first, which we call {\it Wigner realisation}, the eigenstates of the Hamiltonian form irreducible representations of the symmetry group with degenerate eigenvalues.  This is the case of the isospin symmetry.  However, we saw that there exists an alternative way, which in our jargon is  called {\it the Nambu-Goldstone realisation}. The  symmetry is {\it spontaneously broken} and it is not manifest in the spectrum of states. We saw that it contains a massless excitation which, in the language of particle physics, implies the presence of a zero mass state, {\it the Goldstone particle}. Following Nambu, we postulate 
 that this  phenomenon occurs in the strong interactions. To be precise, we assume that the strong interaction Hamiltonian can be written in the form
\be
\label{eq:PCAC5} 
H_{\rm strong}=H_{\rm symmetric}+H_{\rm breaking}
\ee
where $H_{\rm symmetric}$ is invariant under chiral $SU(2)\times SU(2)$ transformations\footnote{The generators of the two $SU(2)$ factors are the combinations $V+A$ and $V-A$ respectively, hence the term “chiral symmetry”.} and $H_{\rm breaking}$ breaks this invariance, but its effects are relatively small. Furthermore, we assume that $SU(2)\times SU(2)$ is spontaneously broken to the diagonal subgroup $SU(2)\times SU(2)\rightarrow SU(2)_V$, with $SU(2)_V$ involving only vector currents, which we  identify with isospin.  In the absence of $H_{\rm breaking}$ this spontaneous breaking  would lead to three massless pseudo-scalar Goldstone bosons. Switching on  $H_{\rm breaking}$ turns these bosons into ``pseudo-Goldstone bosons" with small, non-zero masses. We assume that they can be identified with the pions. This scheme is known as {\it PCAC} for {\it Partial Conservation of the Axial Current}. 

A few months after Nambu, Goldstone published a paper with the title {\it Field Theories with “Superconductor” Solutions}\cite{Goldst}. I will explain the meaning of the word {\it “Superconductor”} in the next section, but the paper contains the first concrete field theory models that exhibit the phenomenon of SSB. In particular, it contains the model  we use today in which the breaking is triggered by a scalar field whose square mass becomes negative. It is the “Standard Model” of spontaneous symmetry breaking. Goldstone shows that all such models contain a massless scalar, hence  the origin of the term {\it Goldstone boson}\footnote{The connection between SSB and massless particles has been proven later under very general assumptions, independent of particular field theory models. See, for example reference \cite{Kastleretal}}. 

Before closing this section I want to mention a paper which played a very important role in all subsequent developments. It is the {\it $\sigma$-model} of Gell-Mann and Maurice Lévy\cite{Gell-Lev}. The title is {\it “The axial vector current in beta decay”} and, although it does not mention the phenomenon of SSB -- it was submitted a few days before the paper by Nambu -- it offers a concrete field theoretic framework for implementing  Nambu's idea. Both papers, by Nambu and Gell-Mann--Lévy,  start by noticing an intriguing relation,  discovered earlier by M.L. Goldberger and S.B. Treiman\cite{Gold-Tr}, between two apparently unrelated quantities, to wit $g_A$ of $\beta$-decay and the pion-nucleon coupling constant. Gell-Mann and Lévy's model contains a doublet of nucleon fields $\Psi(x)$, whose right- and left-handed components form representations (1/2, 0) and (0, 1/2) of chiral $SU(2)\times SU(2)$, respectively, and a quartet of scalar fields transforming like a (1/2, 1/2) representation of the chiral group. Since $SU(2)\times SU(2)$ is locally isomorphic to $O(4)$, the $2\times 2$ matrix of the scalars fields can be written as  $\Phi(x) = [\sigma(x)\I1 + \im {\bm \tau}\cdot {\bm \pi}(x)]/\sqrt{2}$ with ${\bm \tau}$ the three Pauli matrices. 
The Lagrangian density is given by
\bq
\label{sigmalin}
{\cal L}= & \bar{\Psi}_Li\dsl \Psi_L+  \bar{\Psi}_Ri\dsl \Psi_R +g\bar{\Psi}_L \Phi \Psi_R +h.c.  \nn \\
 & +\frac{1}{2}Tr[(\partial^{\mu}\Phi)(\partial_{\mu}\Phi^{ \dag})]-\frac{1}{2}M^2Tr[\Phi\Phi^{ \dag}]-\frac{\lambda}{4!}Tr[\Phi\Phi^{ \dag}]^2
\eq
Note that chiral symmetry forbids a mass term for the fermions. Gell-Mann and Lévy break the symmetry explicitly by adding a term linear in the $\sigma$ field, but we can follow Goldstone and choose $M^2<0$ with the result that the fermions acquire a mass and the three pion fields become massless.  They are the Goldstone bosons.  It is the mechanism  we shall apply to the Standard Model. The quantity, which corresponds to the order parameter of the spontaneous symmetry breaking, is the vacuum expectation value of the $\sigma$ field, which is non-zero in the broken phase. In addition to the three nearly massless pions, the model predicts the existence of a $0^+$ particle, the $\sigma$, of unknown mass\footnote{Notice the analogy with the Standard Model.}. However, as it was already pointed out by Gell-Mann and Lévy,  in the spectrum of $0^+$ particles, there is no obvious candidate for  $\sigma$. Thus we assume that this role is played by a composite operator and the order parameter is the vacuum expectation value of a fermion--anti-fermion pair $<\bar{\Psi}\Psi>_0$. This phenomenon is known as “dynamical symmetry breaking” and Nambu and Giovanni Jona-Lasinio proposed field theory models for it\cite{Nambu}. 

The saga of identifying the axial current of weak interactions is one of the great achievements of theoretical particle physics of the early sixties. It culminated with a series of papers by Gell-Mann who formulated the complete algebraic scheme known as “Current Algebra”\cite{CA1} \cite{CA2} which we mentioned in section \ref{Ren-Sym}. 
\vskip 0.3cm

$\bullet$ {\it Spontaneous symmetry breaking in the presence of gauge interactions.} The picture changes completely if we consider the same mechanism in the presence of gauge interactions\footnote{If the term “spontaneous symmetry breaking” applied to a global symmetry is misleading, that of  “spontaneous breaking of a gauge symmetry”, which we often use, is plainly wrong.}. The history is quite complicated because the phenomenon was in fact discovered twice; so we have two parallel stories with people having different motivations and often ignoring each other. 

Chronologically, the first story is that of the Meissner effect which describes the fact that a magnetic field appears to be screened and does not penetrate inside a superconductor.
A phenomenological description of the effect was proposed by Fritz and Heinz London, who introduced the {\it London penetration length} $\lambda$. L.D. Landau and V.L.  Ginzburg\cite{Land-Ginz}, in their famous 1950 article, wrote the first model describing superconductivity as a phase transition in terms of an order parameter $\Psi$; it is equal to zero in the normal phase and takes a non-zero value in the superconducting phase. In terms of $\Psi$, the equation for the electromagnetic potential takes the form
\be
\label{Landau}
\Delta {\bm A}= \dots +\frac{4\pi e^2}{mc^2}|\Psi|^2 {\bm A}~~~\Rightarrow ~~{\bm A}(x)
\sim {\bm A}(0)~\e^{-x/\lambda}
\ee
where the dots stand for terms which are not important for the discussion. We see in this equation that $\Psi$ acts like a mass term for the photon and, as a result, in the superconducting phase, ${\bm A}$ decays exponentially. 

In 1957, J. Bardeen, L.N. Cooper and J.R. Schrieffer\cite{BCS} formulated the microscopic theory of superconductivity, known as {\it the BCS theory}, in which they give the physical origin of the order parameter $\Psi$ in terms of the Cooper pairs. The following year P.W. Anderson, in a series of two papers\cite{Anderson}, showed that, in the BCS theory, the photon inside a superconductor becomes effectively massive. This is witnessed first by the existence of a mass gap and, second, by the appearance of waves with longitudinal polarisation. The same conclusion was reached also by Nambu in 1960\cite{Nambu1} who examined the BCS theory in the Hartree-Fock approximation. The article, which refers to Anderson, has a good discussion of gauge invariance. This concludes the presentation of the effect in superconductivity and explains the title of Goldstone's paper\cite{Goldst}.

In parallel, and mostly independently, a different story developed in particle physics with the motivation of providing a mass to the Yang-Mills bosons. As far as I know, the first person to worry about this problem was Schwinger who, in 1962, wrote two papers\cite{Schwinger-mass}, both with the title {\it Gauge Invariance and Mass.} The first has a general discussion analysing the theoretical basis for the belief that gauge invariance implies a zero mass for the gauge bosons. In a simplified form the argument can be summarised as follows: let $\Pi_{\mu \nu}(q)$ be the 1P-I two-point function of the vector boson. In a gauge invariant theory it has the general form
\be
\label{Schwing1}
\Pi_{\mu \nu}(q)=\Pi(q^2)\left (g_{\mu \nu}-\frac{q_{\mu}q_{\nu}}{q^2}\right)
\ee
If we respect gauge invariance, at any given order of the perturbation expansion we find  $\Pi(0)=0$ and this implies a zero mass for the vector boson. However, Schwinger notes that, in a non perturbative regime, nothing prevents $\Pi(0)$ to be different from zero, thus yielding a massive boson. In the second paper he studies the two-dimensional  field theory model, known as {\it the Schwinger model,} we introduced in section \ref{Ren-Sym}. As we said, the model is exactly solvable with the surprising result that it describes a free, massive boson! So, Schwinger concludes that the “theorem” that gauge invariance implies a zero boson mass can be proven only in perturbation\footnote{In these papers Schwinger makes no reference to superconductivity.}.

It seems that Schwinger had reached this conclusion earlier. At the International Conference on Elementary Particles, held at Aix-en-Provence in Sept. 14-20 1961, Feynman, in his Summary Talk, says: {\it “\dots Since gauge invariance is usually believed to imply that the mass} [of
  the gauge bosons] {\it is zero, the first prediction of these theories\dots  is
disregarded. Schwinger pointed out to me however, that one can use gauge
invariance to prove that the mass of the real photon is equal to zero, only if
one assumes that in the complete dressed photon, there is a finite amplitude
to find the undressed one.'' } It is precisely what happens in the theories we use today. In a modern language, the two Hilbert spaces, the first built above the vacuum containing the “undressed” photon, and the second above the vacuum of the “dressed” one, are orthogonal. 

Motivated by Schwinger's remarks, people looked again at the gauge invariance -- mass connection. I want to mention two contributions:
In 1963 Anderson wrote a paper\cite{Anderson1} with the title {\it “Plasmons, Gauge invariance and Mass”} in which he presents his previous results on plasma waves as an example of  mass generation. In the Abstract he writes : {\it “Schwinger has pointed out that the Yang-Mills vector boson} [(He only considers the abelian theory)] {\it \dots does not necessarily have zero mass.\dots We show that the theory of plasma oscillations is a simple non-relativistic example exhibiting all of the features of Schwinger’s idea.”} At the same time Maurice Lévy wrote a non-local gauge invariant version of QED with a massive photon\cite{Levy}. So, by the early sixties, the problem of the gauge boson masses was clearly set. We had a concrete example of mass generation in a non-relativistic theory, but no explicit relativistic field theory models. As usually, a well formulated problem is soon solved. In the following years we have an avalanche of papers:

B.W. Lee and collaborators\cite{BenLee} considered models with spontaneous symmetry breaking and came very close to the solution we know today.

W. Gilbert\cite{Gilbert} criticised these attempts and claimed that the superconductivity example was not relevant for a relativistic theory. His argument is based on a new proof of the Goldstone theorem. He argued that in a Lorentz invariant theory, there is always a massless boson associated to every generator of a spontaneously broken symmetry. It was the first model independent proof of this theorem. Very briefly, it goes as follows: Let us consider a relativistic theory invariant under a group of continuous transformations. By Noether's theorem we have a conserved current $J_\mu({\bm x},t)$ and the associated time-independent charge $Q$. We assume that the symmetry is non-trivial and, therefore, there exists at least one local operator $A(x)$ which transforms non trivially.
\be
\label{Gilb1}
\p^\mu J_\mu=0~~;~~~Q=\int \d {\bm x} J_0({\bm x},t)~~;~~~[Q,A]=\delta A~~;~~~\delta A\neq 0
\ee  
We now suppose that the symmetry generated by $Q$ is spontaneously broken. This implies that the vacuum state $\ket{\Omega}$ is degenerate and it is not annihilated by $Q$. Let us consider the following Green function
\be
\label{Gilb2}
{\cal A}_\mu (k)=\int \d^4x \e^{\im kx} \bra{\Omega}[j_{\mu}(x),A(0)] \ket{\Omega}= k_ {\mu} F(k^2)
\ee
with $F$ some function of $k^2$. In a Lorentz invariant theory this is the most general form of ${\cal A}_\mu (k)$. Furthermore, the non-vanishing of $\delta A$, implies that $F$ is not equal to zero. Now we multiply by $k^\mu$ and use the conservation of the current. We find:
\be
\label{Gilb3}
k^{\mu}{\cal A}_{\mu} =0~~\Rightarrow ~~k^2 F(k^2)=0~~\Rightarrow ~F(k^2) \sim \delta(k^2)
\ee
which shows the existence of a zero mass singularity, interpreted as the presence of a zero mass particle. Gilbert concludes that, in a relativistic theory, it is impossible to have spontaneous symmetry breaking without massless particles.

The same year, and independently of Gilbert\footnote{Gilbert's paper has submission date March 30 and Englert's and Brout's June 26. There was no ArXiv or internet at that time.}, F. Englert and R. Brout published their paper\cite{Eng-Br} which contains the complete solution, as we know it today, using elementary scalar fields. They consider both, abelian and non-abelian gauge theories. As most papers of that period, the emphasis was on models for strong interactions. 
They discuss chiral theories and also mention the possibility of a dynamical symmetry breaking. They refer to Nambu for SSB and Schwinger. They have no reference to superconductivity.

Still in 1964, P. Higgs published two papers\cite{Higgs}. In the first he answers Gilbert's objection by considering an abelian gauge theory in the Coulomb gauge. Explicit Lorentz covariance is lost and Higgs shows that Gilbert's argument does not apply. As a result, the photon becomes massive. However he argues that physical quantities, such as the mass spectrum, cannot depend on the gauge and they should be the same even in a covariant gauge. In the second paper he presents a full study of the abelian model and considers extensions to include the breaking of flavor $SU(3)$. He makes the connection between the would-be Goldstone bosons and the longitudinal degrees of freedom of the gauge particles and states explicitly the prediction for a physical scalar particle. He refers to the Englert and Brout paper, but also to Anderson and superconductivity. 

Before going on let us give a first idea of what went wrong with Gilbert's proof. A mathematically more complete explanation will be discussed presently. At first sight Gilbert's assumptions, although not explicitly spelled out, are the standard ones of relativistic quantum field theory, to wit Lorentz invariance, locality etc. He shows the existence of a zero mass singularity in the Green functions of eq. (\ref{Gilb2}). Before going on, let me emphasise that all explicit calculations in perturbation using various Lorentz covariant field theory models, confirm Gilbert's result: the zero mass singularity is there. From that he wants to infer the existence of a zero mass particle. But for this conclusion you need to assume that there exists no other zero mass singularity with the wrong sign which cancels the first one in all $S$-matrix elements. For that you need to assume that all states in the Hilbert space have positive norm. This is part of the usual axioms of quantum field theory, but we all know that it does not apply to a gauge theory. Take QED. We can use a non covariant gauge, like the Coulomb gauge, in which only physical positive norm states appear, but explicit Lorentz covariance is lost. It was Higgs' argument. Alternatively, we can use a covariant gauge, but then we should quantise the theory following the  Gupta-Bleuler method in which we have negative norm states. We can check by explicit calculations that the zero mass singularities do cancel in all physical matrix elements. 

The final stone in the edifice was laid down by G.S. Guralnik, C.R. Hagen and T.W.B. Kibble\cite{Kibble}, again in 1964. They present a detailed analysis of the abelian model with all  aspects we use today. In particular, the counting of degrees of freedom: 2 polarisation states of a massless gauge boson + 1 degree of freedom of the would-be Goldstone scalar = 3 polarisation states of a massive vector boson, appears there. They explain Gilbert's result by pointing out that, for a gauge theory, the integral over all three space giving the charge operator in equation (\ref{Gilb1}) does not converge -- because gauge theories induce long range correlations -- and one should first compute the integral in a finite volume $V$ and study carrefully the limit $V\rightarrow \infty$. Among the references they include Goldstone, Gilbert, Englert and Brout which was already published and Higgs as a preprint. They mention superconductivity with no detailed references. 

\subsection{The synthesis -- A model of leptons}
\label{Weinb}

The Yang-Mills theories and the BEH mechanism were, in principle, known since 1964. I say “in principle” because very few had noticed them and practically no one was expecting  to use them in weak interactions. Therefore, it is not surprising that nobody noticed a short letter with the rather uninspired title {\it A model of leptons}\cite{Weinb1}. I was at CERN at the time and, together with other fellows, we had a study group discussing the papers in the literature. I remember it was the late Bruno Renner who reported on Weinberg's paper and we all decided it was uninteresting! We promptly forgot everything about it. And we were not the only ones. With present knowledge we see immediately that Weinberg in a single stroke solves three fundamental problems: first, and most important, he uses the Brout-Englert-Higgs  (BEH) mechanism to give masses to the intermediate gauge bosons. Second, he shows that the same mechanism generates a mass term for the charged lepton. Indeed, since the right and left components of the Dirac spinors transform differently under the gauge group, a direct mass term is forbidden. Third,  the same mechanism yet again gives rise to Glashow's mixing, and produces  a mass for the $Z^0$ while  leaving the photon massless. Then, why did we all fail to understand the significance of this paper? I believe that the main reason was lack of interest in the community. Field theory was not considered as a serious tool to make progress in particle physics. Out of the three problems the paper solved, none was  recognised as being important. Take, for example, the problem of the gauge boson masses: since they were not known to exist, few people cared about the way they were getting their masses. A second problem with the paper was the title: “A Model of Leptons”. The only leptonic weak interaction known at that time was muon decay. Nobody wanted to consider such a complicated-looking theory for just a single process. It sounded like using heavy artillery to kill a fly. Furthermore, if there was one property of weak interactions which everybody liked and respected, was precisely the property of  universality, so “A Model of Leptons” went totally unnoticed. This lack of attention which this paper has raised is reflected in the citations it received. It was published in 1967 but it has practically no citations until 1972. When we wrote the paper on charm with S.L. Glashow and L. Maiani in 1970\cite{GIM}, none of us had any recollection of Weinberg’s paper and we do not refer to it. Even more unbelievable, we gave a seminar at MIT in which we presented the weak interactions, both leptonic and hadronic, as a Yang Mills theory.  Weinberg, who was in the audience, asked many questions but, even he, did not make the connection with his previous work\cite{Ilio1}. Of course, starting in 1972, the number of citations skyrocketed.

\section{The electroweak theory}

\subsection{The Fermi theory as an effective field theory} 
{\it Where is the cut-off, or, the vital importance of precision measurements.}
\vskip 0.3cm
We have seen the Fermi theory of weak interactions as a most seminal and inspiring model. Many discoveries of fundamental importance -- $CVC, PCAC$, Chiral symmetry, Spontaneous symmetry breaking, Current algebra -- were made as a result of the efforts to understand the properties of the weak current. However, from the mathematical point of view, Fermi's theory is not  renormalisable  and, therefore, it is only a phenomenological model. In practical terms this means that, if we write any physical amplitude as a power series in the Fermi coupling constant $G_F$, every term in the expansion requires the introduction of a cut-off parameter $\Lambda$. In a renormalisable theory, such as quantum electrodynamics, there exists a well-defined prescription to take the limit $\Lambda \rightarrow \infty$ and obtain unambiguous results, but to a non-renormalisable theory the prescription does not apply. The cut-off must remain finite and its value determines the energy scale above which the theory cannot be trusted. This is the definition of {\it an effective theory.}

Can we estimate an order of magnitude for the cut-off? A very simple method is the following: Ordinary dimensional analysis tells us that a physical quantity ${\cal A}$, for example a  decay amplitude, can be written in a series expansion as:

\be
\label{EfTh1}
{\cal A}= A_1G_F\left ( 1+\sum_{n=2}^{\infty}A_n(G_F\Lambda^2)^{n-1}\right )
\ee
where, in every order of the expansion, we have kept only the highest power in $\Lambda$. We see that the expression $g_{\mathrm {eff}}=G_F\Lambda^2$ acts as an effective, dimensionless coupling constant. The expansion will become meaningless when $g_{\mathrm {eff}}\sim 1$, which, for the numerical value of $G_F$, gives $\Lambda \sim $ 300 GeV, a value which, for the accelerators of the 1960's, was essentially infinite\footnote{It is what happens in the Standard Model where the role of $\Lambda$ is played by the vector boson masses.}. 

It was B.L. Ioffe and E.P. Shabalin\cite{Iof-Shab}, from the Soviet Union, who first remarked that, in fact, one can do much better. Let us go back to the expansion (\ref{EfTh1}) and consider also the sub-dominant terms in powers of $\Lambda$. We can rephrase their argument and write any physical quantity as a double expansion in $g_{\mathrm {eff}}$ and $G_F$:

\be
\label{EfTh2}
{\cal A}=\sum_{n=0}^{\infty}A_n^{(0)}g_{\mathrm {eff}}^n~+~G_FM^2\sum_{n=0}^{\infty}A_n^{(1)}g_{\mathrm {eff}}^n~+~(G_FM^2)^2\sum_{n=0}^{\infty}A_n^{(2)}g_{\mathrm {eff}}^n~+~\dots
\ee
where the quantities $A_n^{(i)}$ may contain powers of the logarithm of $\Lambda$.  $M$ is some mass parameter, which, for a typical quantity in particle physics, is of the order of 1 GeV. The first series contains the terms with the maximum power of $\Lambda$ for a given power of $G_F$, they are called {\it the leading divergences.} Similarly, the second series contains all the {\it next-to-leading divergences}, the third the {\it next-to-next-to-leading divergences}, etc. Following Ioffe and Shabalin, let us choose for ${\cal A}$ a quantity in strong interactions, for example the energy levels in a nucleus. The leading divergences represent the weak interaction corrections to this quantity. We can visualise them as resulting from the emission and re-absorption of a virtual $W$. But weak interactions violate parity and/or strangeness, therefore the high precision with which such effects are known to be absent in nuclear physics gives a much more stringent bound for $\Lambda$, of the order of 2-3 GeV. Similarly the next-to-leading divergences contribute to “forbidden” weak interaction processes, such as $\Delta S$=2 transitions (the $K^0_L-K^0_S$ mass difference), or $K^0_L\rightarrow \mu^+ \mu^-$ decays. Again, the precision measurements of such quantities give the same 2-3 GeV limit for $\Lambda$. This was obviously unacceptable because it would mean that the breakdown of Fermi's theory should had been seen already. Why did most people not worry about it? I suspect that the main reason was a widespread mistrust towards field theory in general and higher-order diagrams in particular. Since we have no theory, why bother about its higher-order effects? For most physicists of the nineteen-sixties, studying the higher order effects of weak interactions was useless in the absence of a full theory of strong interactions. There were many speculations about strong interactions providing the cut-off for the weak ones. People had not understood this profound connection between strong and weak interactions, which is manifest by the weak current being a generator of the strong interaction symmetries\footnote{Note that Ioffe and Shabalin\cite{Iof-Shab} had already pointed out that the divergences of weak interactions could not be affected by strong interactions, as long as the latter satisfy current algebra.}.

\subsection{Fighting the infinities}

By the middle sixties a long and painful struggle against the infinities started. Although it was fought by few people, it has been an epic battle given in two fronts: The first, the phenomenology front, or the bottom-up approach, aimed at finding the necessary modifications to the theory in order to eliminate the disastrous leading and next-to-leading divergences. The second, the field theory front, or the top-down approach,  tried to find the conditions under which a quantum field theory involving massive, charged, vector bosons is renormalisable. It took the success in both fronts to solve the problem.

\subsubsection{Early attempts}
{\it Can we determine the Cabibbo angle? Are we ready to sacrifice elegance?}
\vskip0.3cm
In the early attempts the effort was not focused on a particular physical problem, but aimed instead at eliminating the divergences, at least from physically measurable quantities. Some were very ingenious, but lack of space does not allow me to present them in any detail. A very incomplete list contains:
\begin{itemize}
\item
The {\it physical} Hilbert space contains states with negative metric\cite{Lee-W}. The introduction of negative metric states is considered unacceptable because it implies violation of the unitarity condition. However, T.D. Lee and G.C. Wick observed that, if the corresponding “particles”, in this case the weak vector bosons, are very short lived, the resulting unitarity violations could be confined into very short times and be undetectable.
\item
The V-A form of the Fermi theory is an illusion and, in reality, the intermediate bosons mediating weak interactions are scalars\cite{Kum-Seg}. By a Fierz transformation, the effective Lagrangian could look like a vector theory for some processes. This way the theory is renormalisable, but at the price of loosing all insight into the fundamental role of the weak currents.
\item
The theory (\ref{IVB1}) is an approximation and the real theory contains a large number of intermediaries with couplings arranged to cancel the  most dangerous divergences\cite{Geletal}. The idea was simple: divergences arise in perturbation theory because a massive vector boson has a propagator which behaves like a constant at large momenta. This behaviour cannot be improved without violating unitarity. However, for a matrix valued field, we can obtain cancelations for some matrix elements. With a clever arrangement of the couplings, we can hide all bad divergences from the physically relevant quantities. A simple idea whose implementation turned out to be very complicated\footnote{S.L. Glashow's remark: “Few would concede so much sacrifice of elegance to expediency."}. 
\item
The weak interaction divergences and the value of the Cabibbo angle\cite{GST}\cite{CM}. The idea was to compute the coefficient of the divergent term, for example in a loop expansion, for both the weak and the electromagnetic contributions. Setting this coefficient equal to zero gives an equation for the Cabibbo angle. The work by itself has today only a historical interest, but, as by-products, two interesting results emerged, summarised in the following two relations: 

\be
\label{CM1}
\tan \theta =\sqrt{\frac{m_d}{m_s}}~~~;~~~\frac{|m_d-m_u|}{m_d+m_u} \sim {\cal O}(1)
\ee
where the masses are those of the three quarks. The first is in good agreement with experiment and relates the Cabibbo angle with the medium strong interactions which break $SU(3)$. The second, obtained by Cabibbo and Maiani, is more subtle: The prevailing philosophy was that isospin is an exact symmetry for strong interactions broken only by electromagnetic effects. In this case one would expect the mass difference in a doublet to be much smaller than the masses themselves. The second relation of (\ref{CM1}) shows instead that isospin is badly broken in the quark masses and the approximate isospin symmetry in hadron physics is accidental, due to the very small values, in the hadronic mass scale, of $m_u$ and $m_d$. 
\end{itemize}
\subsubsection{The bottom-up approach}
$\bullet$ {\it The leading divergences.}
{\it The breaking of $SU(3)\times SU(3)$.}
\vskip 0.3cm

The leading divergences in the series (\ref{EfTh2}) raised the spectrum of strangeness and parity violation in strong interactions. The first step was to find the conditions under which this disaster could be avoided\cite{BIP}. The argument is based on the following observation: at the limit of exact $SU(3)\times SU(3)$ one can perform independent right- and left-handed rotations in flavour space and diagonalise whichever matrix would multiply the leading divergent term. As a result, any net effect should depend on the part of the interaction which breaks $SU(3)\times SU(3)$. In particular, one can prove that, under the assumption that the chiral $SU(3)\times SU(3)$ symmetry breaking term transforms as a member of the (3, $\bar{3}$)$\oplus$($\bar{3}$, 3) representation, the matrix multiplying the leading divergent term is diagonal in flavour space,  i.e. it does not connect states with different quantum numbers, strangeness and/or parity. Therefore, all its effects could be absorbed in a redefinition of the parameters of the strong interactions and no strangeness or parity violation would be induced. This was first found for the one loop diagrams and then extended to all orders. This particular form of the symmetry breaking term has a simple interpretation in the formalism of the quark model: it corresponds to an explicit quark mass term and it was the favourite one to most theorists, so it was considered a welcome result. 
\vskip 0.3cm
\noindent $\bullet$ {\it The next-to-leading divergences.}
{\it Lepton-hadron symmetry - Charm.} 
\vskip 0.3cm

The solution of the leading divergence problem was found in the framework of the commonly accepted
theory at that time. On the contrary, the next to leading divergences required
a drastic modification, although, in retrospect, it is a quite natural one\cite{GIM}. 

Let us first state the problem. A firmly established experimental fact is that flavour changing weak processes obey certain selection rules: One of them, known as the $\Delta S =1$ rule, states that the flavour number, in this case strangeness $S$, changes by at most one unit. A second rule is that the allowed $\Delta$Flavour = 1 processes involve only charged currents. It follows that $\Delta S =2$ transitions, as well as  Flavour Changing Neutral Current  processes (FCNC), must be severely suppressed. The best experimental evidence for the first is the measured $K_L-K_S$ mass difference which equals 3.48 $10^{-12}$ MeV and, for the second, the branching ratio $B_{\mu^+ \mu^-}= \Gamma (K_L\rightarrow \mu^+ \mu^-)/\Gamma (K_L\rightarrow all)$ which equals 6.87 $10^{-9}$. It was this kind of tiny effects which led to the small value of the cut-off we mentioned earlier. In fact, this problem can be addressed at two levels. They are both easier to visualise in the framework of the quark model. At the
limit of exact flavour symmetry, quark quantum numbers, such as strangeness,
are not well defined. Any basis in quark space is as good as any other. By
breaking this symmetry the medium strong interactions choose a particular basis,
which becomes the privileged one. Weak interactions, however, define a
different direction, which forms an angle $\theta_C$ with respect to the first
one. Having only three quarks to play with, one can form only one charged current  of the form postulated by Cabibbo\cite{Cabib}: 

\be
\label{Cab1}
J_{\mu}^C(x)=\bar{u}(x)\gamma_{\mu}(1+\gamma_5)[\cos \theta_C~d(x)+\sin \theta_C~s(x)]
\ee

The expression (\ref{Cab1}) can be interpreted as saying that the $u$ quark is coupled to a certain linear combination of the $d$ and $s$ quarks, $d_C=\cos \theta_C~d+\sin \theta_C~s$. The orthogonal combination, namely $s_C=-\sin \theta_C~d+\cos \theta_C~s$ remains uncoupled. Notice the difference with the leptonic current. We have four leptons, two neutrals, the $\nu_{(e)}$ and the $\nu_{(\mu)}$ and two negatively charged ones, the electron and the muon. They are all coupled and the weak current (\ref{Fermi2}) has two pieces.

The first level of the problem is to consider a theory 
satisfying a current algebra.  The neutral component of the current will be related to the commutator of $J_0^C$ and $J_0^{C\dagger}$ and will contain terms like $\bar{d}_Cd_C$, thus having flavour changing pieces. Notice again that this does not happen with the leptonic current. The commutator of the current (\ref{Fermi2}) with its hermitian adjoint has no terms violating the two lepton flavor numbers. Phrased this way, the solution is almost obvious: we must use the $s_C$ combination, but, in order to do so, we must have a second up-type quark\cite{GIM}. If we call it $c$, for {\it charm}, the charged weak current (\ref{Cab1}) will have a second piece:

\be
\label{GIM1}
J_{\mu}(x)=\bar{u}(x)\gamma_{\mu}(1+\gamma_5)d_C(x)+\bar{c}(x)\gamma_{\mu}(1+\gamma_5)s_C(x)
\ee
or, in a matrix notation,

\be
\label{GIM2}
J_{\mu}(x)=\bar{U}(x)\gamma_{\mu}(1+\gamma_5)CD(x)
\ee
with
\be
\label{GIM3}
U=\left ( \begin{array}{c} u\\c  \end{array} \right )~~~;~~~D=\left ( \begin{array}{c} d\\s  \end{array} \right )~~~;~~~C=\left ( \begin{array}{cc}\cos \theta &\sin \theta \\-\sin \theta & \cos \theta \end{array} \right ) 
\ee

The important point is that, now, a current $J_3$, given by the commutator of $J$ and $J^{\dagger}$, is diagonal in flavour space.

\begin{figure}
\begin{center}
\includegraphics[height=25mm]{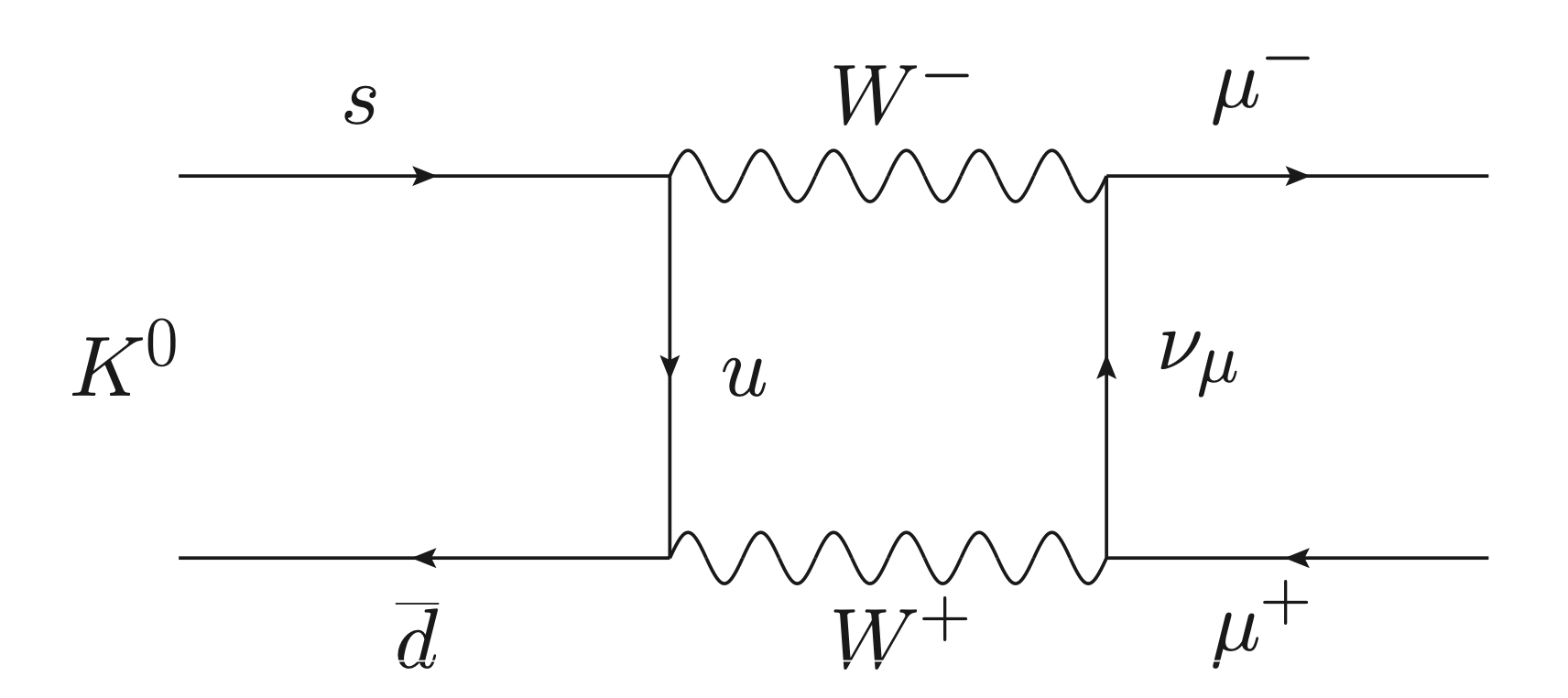} \includegraphics[height=25mm]{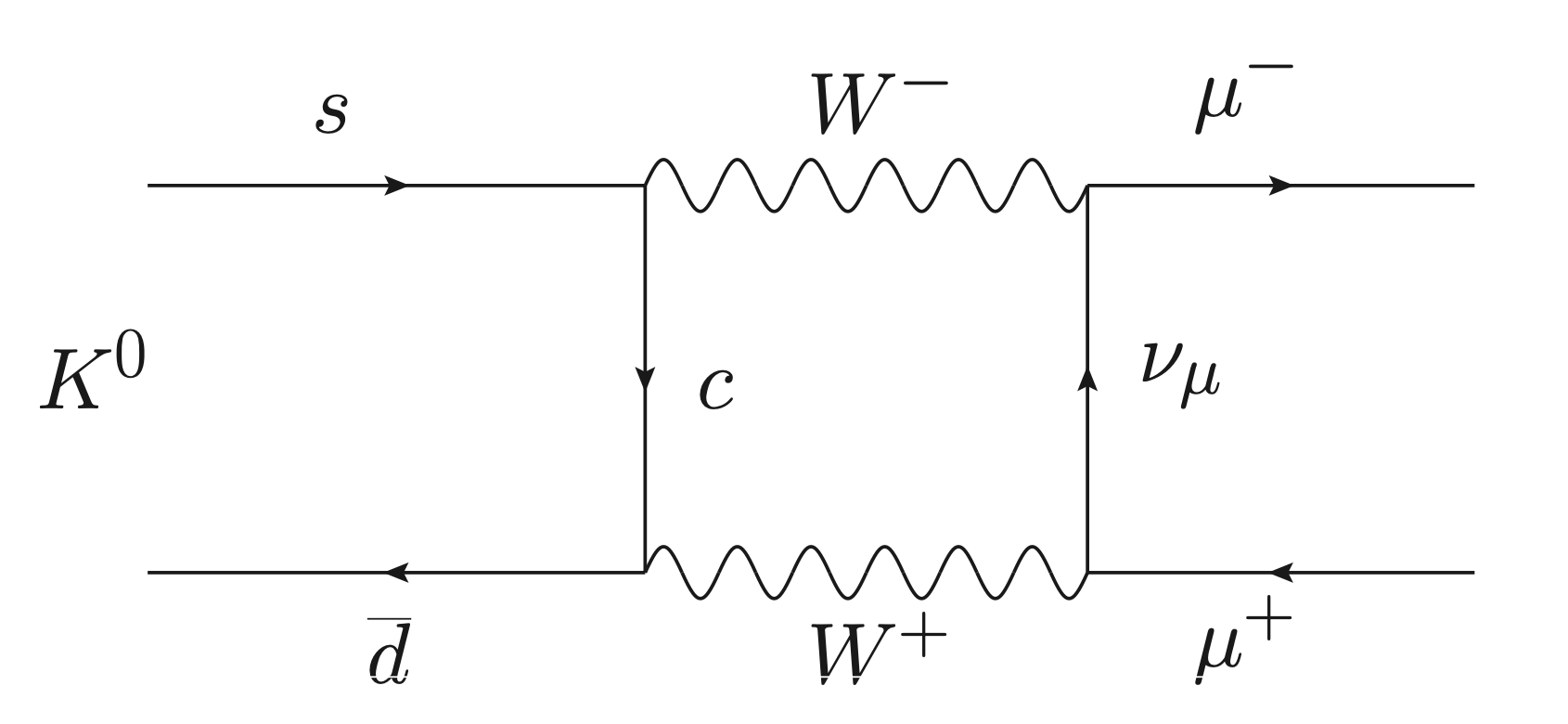}
\caption{The one-loop contribution to $K^0 \rightarrow \mu^+ + \mu^-$.  In a three quark theory (left), the charm contribution (right).}
\label{FigKtomu1}  
\end{center}    
\end{figure}

This solves the first level of the problem, but it is not enough to explain the observed rates. For example, the
$K_L\rightarrow \mu^+ \mu^-$ decay can be generated by the box diagram of
Figure \ref{FigKtomu1} (left), which, although of higher order in the weak interactions, it is quadratically divergent and contributes a term proportional to $G_Fg_{\mathrm {eff}}$.

Here comes the second ingredient of the mechanism. With a
 fourth quark, there is a second diagram, with $c$ replacing $u$, 
Figure~\ref{FigKtomu1} (right). 
In the limit of exact flavour symmetry the two diagrams cancel. The breaking of flavour symmetry induces a mass difference between the quarks, so the sum of the two diagrams is of order $g^4 (m_c^2-m_u^2)/m_W^2 \sim G_F(G_Fm_c^2)$. Therefore, Ioffe and
Shabalin's estimations can be translated into a limit for the new quark mass
and yield an upper bound of a few GeV for the masses of the new hadrons\cite{GIM}. This
fact is very important. A prediction for the existence of new particles is
interesting only if they cannot be arbitrarily heavy\footnote{Theoretical proposals involving new quark species had been made earlier\cite{Newq}, but they had no particular motivation and no connection to weak interactions.}. 

In the early days of the Fermi theory there was a kind of symmetry between hadrons and leptons: proton-neutron {\it vs} neutrino-electron. The first discovery of heavy flavours appeared to break this symmetry: we had two new leptons, the muon and its associated neutrino, but only one new hadron, the strange quark. The introduction of the charmed quark restored this symmetry. By doing so, the mechanism obtained two important results: (i) It solved the technical problem of the low value of the Ioffe and Shabalin cut-off by replacing it with the masses of new hadrons. (ii) It opened the way to a formulation of the theory in terms of current operators which satisfy algebraic properties. It is this second result that made possible the use of Yang-Mills theories for the entire weak interactions, leptonic as well as hadronic. The price was the prediction for an entire new sector of hadronic physics, the charmed particles, with masses which were bounded from above by a few GeV. In particular we predicted the existence of a new $1^-$ boson.

The idea of introducing new  quark flavors in order to explain weak interaction phenomena was brilliantly extended in 1973 by M. Kobayashi and T. Maskawa\cite{Kob-Mask} who noticed that a model with three quark families offers a natural mechanism for $CP$ violation.

{\it Some personal reminiscences:} I mentioned already an informal seminar we gave at MIT immediately after we discovered the charm mechanism. It took place in F. Low's office and gathered most of the MIT theorists. For us it was a good test and we passed it successfully. Low praised our solution and Weinberg, who asked many questions,  although he did not make the connection with his earlier work, he liked the mechanism very much. The same evening at dinner Pucci Maiani (Luciano had been married a few months before) remarked that we looked very happy. Shelly told her that we expected our work to be part of future physics text books. 

We submitted the paper to the Physical Review and, a couple of weeks later, we received the report. The referee had read our paper very carefully, he said it was interesting and worth publishing, but he raised one objection: a na\"ive power counting for a massive Yang-Mills interaction gives $(G\Lambda^6)^n$ while we were implicitly assuming it to be $(G\Lambda^2)^n$. He then went on to remark : “The na\"ive behavior can undoubtedly be improved, but the assertion that it can go down to $(G\Lambda^2)^n$ must be either proven or deleted.” He did not say it was wrong, which proves that he knew the problem very well. We only changed a few words in the paper and resubmitted, this time successfully. 
The epilogue is that it took Shelly and myself -- Luciano had left already for Italy -- some lengthy and hard work to prove that our initial estimate was right. We published two papers\cite{GI} which, although I consider them to be among the most intelligent papers I have ever written, turned out to be irrelevant for the subsequent developments. 

\subsubsection{The top-down approach: The Yang-Mills quantum field theory.}
\vskip 0.3cm
I believe that the first person who decided to look at the Yang-Mills quantum field theory as the dynamical theory of weak interactions was M.J.G. Veltman. Around 1966, he was trying to understand the deeper origin of the conservation, or near conservation, of the weak currents. In particular, he tried to throw some light on the general confusion which prevailed at that time concerning the so-called “Schwinger terms” in the commutators of two current components. While he was on a visit from CERN to Brookhaven, he wrote a paper in which he suggested a set of divergence equations which generalised the notion of the covariant derivative of quantum electrodynamics. This fundamental idea was taken up next year and developed further by J.S. Bell\cite{Velt-Bell}. These equations looked like the covariant derivatives of  non-abelian gauge theories, so Veltman decided to study their field theory properties. The electrodynamics of charged vector bosons had been formulated already by T.D. Lee and C.N. Yang in 1962\cite{Lee-Yang}. They had shown that electromagnetic gauge invariance allows to express the vector boson's charge $e$, magnetic moment $\mu$ and quadrupole moment $Q$ in terms of only two parameters $e$ and $\kappa$, as $\mu=e(1+\kappa)/2m_W$ and $Q=-e\kappa/m_W^2$. The resulting theory is highly divergent but Veltman noticed that many divergences cancel for the particular value $\kappa=1$\footnote{The fact that the theory with $\kappa=1$ has better convergence properties had been noticed also by Glashow\cite{Glashow-EMVect}.}. It is the value predicted by a theory in which $W^{\pm}$ and the photon form a Yang-Mills triplet. For Veltman this was a clear signal that the theory of weak and electromagnetic interactions must obey a Yang-Mills gauge invariance. 

Studying the renormalisation properties of Yang-Mills theories turned out to be a very hard and complex problem, both conceptually and practically. On the conceptual side, neither the results of Feynman\cite{Feyn-Gr} and Faddeev and Popov\cite{Fad-Pop} nor those on spontaneous symmetry breaking were widely known and Veltman had to rediscover even the basic Feynman rules. On the practical side, the number of terms grew very fast and Veltman had to develop a computer program to handle them\cite{Scoon}. He called it “Schoonschip”,\footnote{“Clean ship” in dutch} and it was the first program of symbolic manipulations applied to theoretical high energy physics. 

A few years earlier, while working for his thesis, Veltman had become familiar with a method to compute Feynman diagrams using the so-called “cutting rules” which were proposed in 1960 by R.E. Cutkosky\cite{Cut}. They consist in cutting the diagram in all possible ways and replacing the cut propagators by the mass-shell delta functions. The method gives the imaginary part of the diagram and we can recover the real part by writing a dispersion relation. Veltman developed this approach and turned it into a very powerful tool particularly adapted to explore theories for which the entire field content is not a priori known\footnote{In doing so, he was concentrated in the calculation of scattering amplitudes. For Veltman all quantum field theories that give the same $S$-matrix were considered as identical. This purely semantic question was the source of endless disputes in conferences and meetings between him and many other field theorists.}. A personal exposition of his approach can be found in  reference\cite{Velt1}. He first studied massive Yang-Mills theories at the one loop level\cite{Velt2} and claimed cancellations of divergences, although the papers, especially the first one, were not easy to follow. In retrospect we see that Veltman allowed himself to change the Feynman rules as if the theory was gauge invariant although the mass terms break the symmetry. In fact, with a remarkable intuition, he was rediscovering by trial and error the one loop effects of the  Brout-Englert-Higgs mechanism. In 1970, in collaboration with H. Van Dam, he studied the zero mass limit of gauge fields, both for Yang-Mills and gravity\cite{Velt3} and proved, by explicit calculation, that the limit was not continuous because the longitudinal degrees of freedom of the massive fields do not decouple when the mass goes to zero. Similar results had been obtained also by A.A. Slavnov and L.D. Faddeev\cite{Slav-Fad}, although they were not widely known. 
In 1968 Veltman took a sabbatical leave and spent the 68-69 year at Orsay. He lectured on Yang-Mills fields and there were some lecture notes in the form of a preprint which, as far as I know, has not been published. Some of us, including myself, found in these notes our first introduction to Yang-Mills and path integrals. 

In addition to Veltman and the Utrecht school, there were some other isolated and rather confidential contributions which addressed the question of the renormalisation properties of gauge theories\cite{Ren-YM}. There is a great confusion in the early literature and one finds contradictory statements regarding the renormalisability of massive Yang-Mills. In most cases the problem is semantic, but it took some time before the situation was cleared up. The correct statement, which I will explain in section \ref{Ren-mYM}, is that a massive Yang-Mills theory is non-renormalisable, unless the gauge boson masses are generated through a Brout-Englert-Higgs mechanism of spontaneous symmetry breaking. As we saw in section \ref{SSB}, this mechanism leaves at least one physical scalar field. The mass of this scalar particle $m_S$ is an arbitrary parameter of the theory, but we cannot get rid of it by sending it to infinity since, in this case, we recover the massive Yang-Mills theory without the phenomenon of spontaneous symmetry breaking, which is non-renormalisable. Through the radiative corrections, all measurable quantities in the Standard Model depend on $m_S$ and, in the late seventies, Veltman computed the one loop effects in an effort to restrict the value of the Higgs mass\cite{Velt4}. As it turned out, no severe restrictions could be obtained because the dependance was only logarithmic. He called it {\it “the screening theorem”.} As a result, no precise predictions on $m_S$ were known until the actual discovery of the particle at the LHC.

In 1969 Veltman was joined in Utrecht by Gerard 't Hooft, a graduate student with whom he would share the Nobel Prize 30 years later\footnote{Initially Veltman did not want to assign to this young student the renormalisability of Yang-Mills as a thesis project because he thought it was too risky. He proposed instead the problem of $A_2$ splitting which was quite popular at that time. 't Hooft came back a bit later declaring he did not like it, so Veltman gave up and proposed Yang-Mills. In fact, 't Hooft showed good taste because the problem of $A_2$ splitting  turned out to be a fake.}. In 1970 't Hooft followed the Cargèse summer school -- his application to the Les Houches summer school had been turned down -- in which he heard B.W. Lee lecturing about the renormalisation of the $\sigma$-model with spontaneous symmetry breaking. According to 't Hooft, this was a revelation and shed new light to the renormalisation properties of Yang-Mills theories\cite{tHooft1}. Rare are the examples of PhD students whose results had such a profound and lasting influence in physics. The year 1971 was the {\it annus mirabilis}. 't Hooft published two papers, landmarks in the road to the Standard Model\cite{tHooft2}. The first has the title {\it “Renormalisation of massless Yang-Mills Fields”} and presents the first complete formulation of the quantum gauge theory. In this work 't Hooft generalised the Faddeev-Popov method to arbitrary gauges and derived the corresponding Feynman rules. Furthermore, he invented a gauge invariant regularisation scheme for one-loop diagrams by introducing a fictitious fifth dimension, a precursor of the dimensional regularisation method. With these tools he gave the first complete proof of the renormalisability of the theory. The second paper, with the title {\it “Renormalisable Lagrangians for Massive Yang-Mills Fields”}, incorporates the BEH mechanism and shows that the resulting theory is renormalisable. He uses his previous results in order to compute the Feynman rules for a family of gauges and,  in particular, the one known as “the 't Hooft gauge” which is most convenient for calculations. A limiting case in this family is “the unitary gauge” in which the Lagrangian contains only the physical degrees of freedom. It is non-renormalisable by power counting but, by gauge invariance, has the same $S$-matrix as the one we can compute using, for example, the 't Hooft gauge. It generalises to the theory of the weak and electromagnetic interactions the Stueckelberg effect we saw in section \ref{Ren}. The complete program of the Yang-Mills renormalisation has been published by 't Hooft and Veltman in 1972\cite{Hoof-Velt}. The first of these publications contains also the dimensional regularisation scheme which, in addition to preserve gauge invariance, has the advantage of simplifying enormously all perturbation calculations. It extends previous work by C.G. Bollini and J.J. Giambiagi\cite{Bol-Giam}.

G. ’t Hooft and Veltman gave the first detailed presentation of their results in a small meeting at Orsay in January 1972. 't Hooft was the main speaker and I still remember his marathon lectures. This meeting was remarkable in many respects. First it offered the first complete picture of the renormalisation properties of Yang-Mills theories, including the method of dimensional regularisation. Second it triggered stimulating discussions among the participants. In particular, the vital importance of the axial current anomaly cancellation was stressed in this meeting. Third it initiated a long series of meetings which became known as “triangular meetings” (Paris-Rome-Utrecht).  Subsequently enlarged with the addition of other European centres, they played an important role in the development of a European network in theoretical physics.

In the meantime, the renormalisation program of Yang-Mills theories started attracting wider attention. A.A. Slavnov and J.C. Taylor derived the Ward identities for Yang-Mills theories\cite{Slav-Tayl} and simplified considerably the analysis. B.W. Lee and J. Zinn-Justin reformulated the proof of renormalisability and gauge invariance of spontaneously broken Yang-Mills theories, using  path integrals and Ward identities\cite{Lee-Zinn}. The most novel contribution came from C. Becchi, A. Rouet and R. Stora from Marseille\cite{BRS} (see also I.V. Tyutin from Moscow\cite{Tyutin})  who invented an ingenious method to handle the consequences of gauge invariance. Usually we obtain the Ward identities using the invariance properties of the Lagrangian. With gauge theories however, the Lagrangian we are using for actual calculations is not invariant under gauge transformations because we must make a gauge choice. The problem is accentuated in Yang-Mills theories, or quantum gravity, because, in addition, we need to introduce the Faddeev-Popov ghosts whose form depends on the gauge. B.R.S. and T.  found that the resulting Lagrangian is nevertheless invariant under a new symmetry, called “the BRST symmetry”. The strange property of this symmetry is that the parameters which determine the  transformations are not ordinary numbers but  anti-commuting quantities. Since that time, the BRST symmetry has become an essential tool in the study of all quantum gauge theories.

\subsubsection{Are there any other theories?}
\label{Ren-mYM}
\begin{figure}
\centering
\includegraphics[height=25mm]{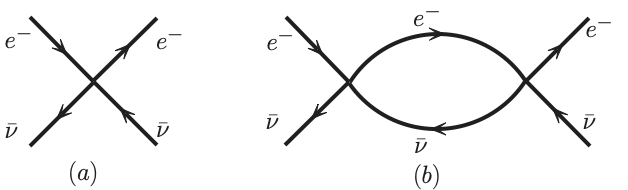}
\caption {The elastic electron--anti-neutrino scattering at first order in
the Fermi coupling constant (a) and at second order (b).} \label{SM4}
\end{figure}

\vskip 0.3cm 
It took a long time to derive a consistent quantum field theory for weak interactions and the final picture was not at all obvious. Its two basic ingredients, Yang-Mills gauge invariance and the phenomenon of spontaneous symmetry breaking, each one considered separately, predicted the appearance of massless particles -- vector gauge bosons and scalar Goldstone bosons -- in obvious contradiction with experiment. It was only by combination that the two  diseases cured each other. In 1973 C.H. Llewellyn-Smith asked the question whether such a behavior could have been anticipated\cite{Llew}. He introduced the concept of {\it tree unitarity} as follows.
Let us consider the Fermi theory and compute the electron--anti-neutrino elastic scattering. At lowest order we have the diagram of Figure \ref{SM4} (a). 
The Fermi coupling constant has dimensions $[M]^{-2}$. At high energies we can
neglect the
fermion masses and the cross section will behave as $\sigma_{\mathrm{el}}\sim G_F^2 s$ where $s$ is the total energy in the centre of mass,
$s=(p_1+p_2)^2$. Remember, however, 
that this cross section is proportional to the imaginary part of the
corresponding one-loop amplitude of Figure \ref{SM4} (b). It follows therefore that, if we write a dispersion
relation for the latter of the form:
\be
\label{elastic3}
M_{\mathrm {el}}(s) \sim \int \frac{\sigma_{\mathrm{el}}(s')}{s'-s}ds'
\ee
it will require two subtractions. Of course, this is in agreement with the
fact that the diagram of Figure \ref{SM4} (b) is quadratically divergent. Therefore, if we want to embed the Fermi theory into a larger, renormalisable field theory, we must improve the high-energy behavior of the tree diagram. This is the concept of tree unitarity. Llewellyn-Smith in successive steps found that such improvement requires: (i) the introduction of intermediate vector bosons, (ii) Yang-Mills couplings among them, (iii) an extra scalar boson, and (iv) relations among masses and couplings of all these fields. In other words, he reconstructed the Standard Model in the unitary gauge. 

One can object that in our search we succeeded because we knew the answer and
this is certainly true. However, now that we have the machinery, we can ask
the more general question: consider a set of fields, spin zero, spin one-half
and spin one. Among the latter, some are assumed massless and some are
massive. Write the most general hermitian, Lorentz invariant, interaction
Lagrangian among them with terms of dimension smaller or equal to four. Our theory so
far is characterised by a large set of parameters, masses and coupling
constants. Impose ``tree unitarity'' for all amplitudes {\it i.e.} no cross section computed in the tree approximation should grow at infinity. This gives
relations among the parameters. The resulting theory is equivalent to a
Yang-Mills gauge theory. The relations among the parameters simply express the
BRST symmetry in the unitary gauge. If the BRST identities are not satisfied the theory cannot be renormalisable. All massive vector bosons correspond to spontaneously
broken generators via a BEH mechanism with two possible exceptions: (i) the trivial one of free
fields, (ii) massive neutral vector bosons coupled to conserved currents
following the Stueckelberg formalism. In other words, we can prove that the road that led to the Standard Model is unique\cite{Tikt}.

\subsection{With, or without, neutral currents}
\label{NC-Y-N}
\vskip 0.3cm
The Glashow-Weinberg-Salam model contains a massive neutral gauge boson, therefore it implies the existence of weak neutral currents. Such currents had not been detected, but only the strangeness changing ones were really forbidden by experiment. The incorporation of the GIM mechanism\cite{GIM} ensured their absence, so the prediction was limited to strangeness conserving neutral currents. In practice they could manifest in two ways:  neutrino induced or electron induced reactions. The first could be either neutrino-electron elastic scattering $\nu_{\mu} +e^- \rightarrow \nu_{\mu} +e^-$ or inelastic neutrino nucleus scattering of the general form $\nu_{\mu} +N \rightarrow \nu_{\mu} +X$. The second could induce parity violating effects in either atomic physics\cite{Bouchiat2}, or electron-nucleus scattering. Any of these reactions was an experimental challenge, so in the early years some gauge models were proposed trying to avoid neutral currents, either partly\cite{NC1}, or totally\cite{G-G}. In 1973  a European collaboration operating the heavy liquid bubble chamber “Gargamelle” at CERN, settled the issue by observing neutral current events in $\nu_{\mu}$ and ${\bar{\nu}}_{\mu}$ beams\cite{Garg}. Yet, it took some time before the community was convinced\footnote{I won some bottles of very fine wine by betting for neutral currents, in particular against Jack Steinberger.}. There were several reasons for that. First, Gargamelle was the first “large” international collaboration (7 labs, 55 physicists!) and this made people suspicious. I still remember some sarcastic comments: “Can anything right come out from such a crowd?”. Second, initially there was only one “gold plated” event of elastic anti-neutrino--electron scattering. Third, the chamber was 4.8m long but only 2m in diameter. Therefore, the collaboration had to prove that they had full control of the background consisting of neutrons produced in the steel surrounding the chamber and entering from the sides, thus mimicking  hadronic neutral current events. For that, they developed a Monte-Carlo program simulating all neutrino interactions in the entire environment of the chamber. It was the first time that the behavior of such a large and complex detector had been analysed numerically. Finally, there were some contradictory claims coming from the first very high energy neutrino experiments in the United States. 

It took much longer to detect neutral current effects in electron induced reactions. The first evidence came from SLAC observing parity violation in deep inelastic scattering of polarised electrons\cite{SLAC-PV}, followed by the discovery of parity violation in atomic physics at the Ecole Normale Supérieure\cite{ENS-PV}.

\subsection{The axial anomaly: The final act}
\label{SM-Anom}
I came back to Europe in October 1971 and joined the Department in Orsay where I had studied. These were  times of great expectations. We had been through so many lean years in particle physics that most physicists had lost hope to ever experience the excitement of great discoveries. There were so many new things to learn, so many questions to answer. Gauge theories had revolutionised our way of thinking. While we were trying to understand the proof of renormalisability in various gauges, we realised the vital importance of the Ward identities. A change of gauge produces a completely new theory. All these theories that look, and in many respects are, so different, are linked together through the Ward identities. For example, the renormalisable gauges contain unphysical degrees of freedom and it is only by virtue of the Ward identities that they decouple from physical quantities. But then the obvious problem appeared. Weak interactions involve both vector and axial currents. We have seen that in many cases one cannot enforce the conservation of both because of the  triangle anomalies. With Cl. Bouchiat and Ph. Meyer we checked easily that in the $SU(2)\times U(1)$ model the Ward identities were indeed broken and the renormalisability and unitarity proofs did not apply. I felt like a child from whom someone stole his most wonderful toy.  This brings me to the final act of the axial anomaly play which we left unfinished in section \ref{Ren-Sym}. The solution was first found by Cl. Bouchiat, Ph.  Meyer and myself
in 1971\cite{BIM}. The key ingredient is the remark that the anomaly does not depend on the mass of the fermion that goes around the triangular loop. Therefore,  we must add all contributions in order to get the right answer. For the electroweak theory this means that we need both the leptons and the quarks.  A simple calculation shows that the total anomaly produced by the fermions of each family is proportional to  ${\cal A}=\sum_i Q_i$, where the sum extends over all fermions in a given family and $Q_i$ is the
electric charge of the $i$th fermion. Since ${\cal A}=0$ is a necessary
condition for the mathematical consistency of the theory, we conclude that
each family must contain the right amount of leptons and quarks to make the
anomaly vanish. This condition is satisfied by the three colour model with
charges 2/3 and -1/3, but also by other models such as the Han-Nambu model with integer charge quarks. In fact, the anomaly cancellation condition  has a
wider application. The Standard Model could have been invented after the
Yang-Mills theory was written, much before the discovery of the quarks. At
that time the “elementary” particles were thought to be the electron and its
neutrino, the proton and the neutron, so we would have used one lepton and one
hadron doublet. The condition ${\cal A}=0$ is satisfied. When quarks were
discovered we changed from nucleons to quarks. The condition is again
satisfied. If tomorrow we find that our known leptons and/or quarks are
composite, the new building blocks will be required to satisfy this condition
again\cite{tHooft3}. 

The moral of the story is that families must be complete. The title of the GIM
paper\cite{GIM} was “Weak Interactions with Lepton-Hadron Symmetry” and with Bouchiat and Meyer we proved that it was indeed correct. Thus, the discovery of a new lepton, the tau, implied the existence of two new quarks, the $b$ and the $t$, prediction which was again verified experimentally.

\section{QCD}
\label{QCD}
The emergence of quantum chromodynamics as the theory of strong interactions has been presented in several reviews, see for example \cite{Gross1} and \cite{Ellis}. The first one, by David Gross, describes also the international scene which preceded QCD. The second, by R.K. Ellis, offers a view of the technical aspects of the computations which are used today. My presentation will be brief. 

\subsection{Strong interactions are complicated}
Strong interactions entered the  physics scene in the first half of the twentieth century with the development of nuclear physics. It was immediately evident that they are short ranged, no macroscopic effects  related to the strong force have ever been observed, and they are very strong,  since they overcome the electrostatic repulsion among protons  and lead to the formation of stable nuclei. As we said already in section \ref{Orig}, the effective coupling constant is large and the perturbation expansion useless. The Particle Data Group lists more than 30 pion-nucleon resonances with masses between 1 and 2.5 GeV, with spins as high as 11/2, and it is clear that no simple quantum field theory could accommodate all of them. Strong interactions appeared to be very complicated. 

We know today that this complexity shows only a superficial part of the picture. The interactions appear very complicated because the objects we were trying to study, the hadrons, are themselves complex. It is as if we were trying to discover quantum electrodynamics by looking at the interactions among complicated macromolecules\footnote{R.P. Feynman has given a nice analogy which accurately describes the efforts to understand strong interactions: imagine you want to study the mechanism of a fine Swiss watch and, to do that, you take two of them and smash them one against the other.}. By changing the perspective, we discovered a completely different picture.

\subsection{Strong interactions are simple -- Deep inelastic scattering and the parton model}
\begin{figure}
\centering
\includegraphics[height=25mm]{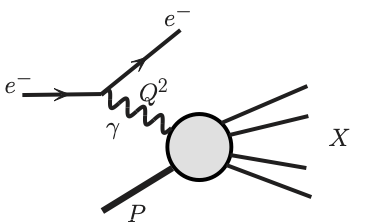}
\caption {Deep inelastic electron-nucleon scattering with the
production of a final state $\ket{X}$. } \label{fig:deepin1}
\end{figure}

The modern  theory of strong interactions was developed as a response to certain intriguing experimental results. In 1966 a high energy  linear electron  accelerator was commissioned at Stanford in the Laboratory which became known as {\it  Stanford Linear Accelerator Center} (SLAC). By the late 1960s and early 1970s a series of experiments  were  carried out at SLAC studying the high energy and large momentum transfer scattering, often called {\it deep inelastic
scattering,} of electrons off nucleons. The process is shown schematically in Figure \ref{fig:deepin1}
 in the one photon exchange approximation. It is an {\it inclusive process,}  meaning that, in the final state, only the electron\index{inclusive process} 
was measured. The  state denoted by $X$ was not. 
This choice was dictated partly by some theoretical considerations\footnote{The story goes back to the Algebra of Currents\cite{CA1}. A good collection of original articles can be found in \cite{CA2}. Particularly interesting for this discussion is the one by J.D. Bjorken\cite{Bjork}.}, but also by experimental constraints: (i) The detectors were not able to distinguish between a $\pi^+$ and a proton if their kinetic energy was much higher than  1 GeV. (ii) The forward and backward regions were not covered, so a complete identification of all particles in $X$ was not possible. (iii) The electrons in the beam were coming in bunches but the detectors were not able to resolve multiple collisions.  Therefore, what was measured corresponds to a sum over all possible $X$. It is easy to show that this differential cross section is given by the matrix element of the commutator of two current operators:
\be
\label{eqDIS1}
\frac{\d\sigma}{\d\Omega' \d E'} \sim \frac{e^4}{Q^4}L_{\mu \nu}W^{\mu \nu} ~~~{\mathrm{with}}~~~W_{\mu \nu} = \frac{1}{2} \int \d^4y ~\e^{\im Q\cdot y} ~\bra {P} [J_\mu^{h\dagger}(y)J_\nu^h (0)] \ket{P}
\ee
where $L_{\mu \nu}$ is the leptonic part which, summed over electron polarisations and in the limit of vanishing electron mass, is given by $L_{\mu\nu} = 2\left(k_{\mu}k'_{\nu}+k_{\nu}k'_{\mu} -\eta_{\mu \nu}k\cdot k'\right)$. $k$ and $k'$ are the momenta of the initial and final electrons respectively, and $Q=k-k'$.  $J_\mu^h$ is the hadronic part of the electromagnetic current and $ \ket{P}$ the one-nucleon state with momentum $P$.
We can show that this inclusive cross section can be expressed in terms  of two functions, called “structure functions”, $F_1$ and $F_2$. Kinematically they can depend only on two dimensionless variables which are chosen traditionally as $M^2_N/q^2$ and $x=q^2/2\nu$ with $\nu=2P\cdot Q$ and $q^2=-Q^2$. The surprising result
of the SLAC experiment was that, when both $\nu$ and $q^2$ become large with  their ratio $x$ kept fixed, these structure functions were  approximately functions only of
$x$. Here “large” means
large compared to the nucleon mass. This property became known as {\it scale invariance};  varying the energy scale of the experiment, $q^2$ and $\nu$ with $x$ fixed, does not affect the cross section. It is easy to see that when $q^2\rightarrow \infty$ with $x$ fixed, the dominant region in the integral of equation (\ref{eqDIS1}) is the one with $y^2\sim 0$, in other words in the deep inelastic region we probe the vicinity of the light cone. 

The property of scale invariance is very interesting, because it is very easy to  “understand”  it using
a na\"{\i}ve and wrong reasoning. When $q^2\rightarrow \infty$ with fixed $x$,
the second variable $M_N^2/q^2$ goes to zero. Na\"{\i}vely, a function
$f(x,M_N^2/q^2)$ can be approximated by 
$f(x,M_N^2/q^2)=f(x,0)+\frac{M_N^2}{q^2}f^{(1)}(x,0)+\dots $
where $f^{(1)}$ is the first derivative of $f$ with respect to $M_N^2/q^2$
keeping $x$ fixed. 
So, when $M_N^2/q^2$ goes to zero, we expect to be left with
only the $x$ dependence. This argument, however, is technically wrong because it assumes that the structure functions $F_{1,2}$ are analytic in the variable $M_N^2/q^2$ around the point $M_N=0$ and can be expanded in a power series. This assumption is false in quantum field theory, because, as we know, we often encounter  infrared divergences when we attempt to set a mass parameter equal to zero.

Feynman had built a simple model, which  implemented this scale behaviour\cite{Partons}. Let us assume that the target nucleon is made out of elementary constituents, which interact with the incident photon as point
 particles. We shall call  these constituents collectively {\it partons}. If we neglect all
 interactions among the partons we can easily reproduce this scaling
 property. Feynman used the following simple picture: at high energies the virtual photon of Figure \ref{fig:deepin1} sees the target nucleon as a thin disk. The parton model amounts to neglect the interactions among the partons in the plane of the disc. H. Fritzsch and M. Gell-Mann\cite{bi-loc} proposed an algebraic framework for this model by extending  the equal-time commutators of current algebra to a light cone algebra of bi-local operators. The basic assumption was that everywhere at the light cone, one could use free field commutation relations\footnote{Gell-Mann was saying that “$\cdots$ at the light cone  Nature reads only free field theory books.”}. 
 
 The trouble, of course, is that the assumption of no interaction among the partons does not seem to make sense.  The partons cannot be free and, at the same time, bind strongly to form a nucleon. Nevertheless, it was such a schizophrenic behaviour that was
  implied by the data. The partons were almost free when probed by a virtual
 photon in the deep inelastic region and still very strongly bound in ordinary
 hadronic experiments. Strong interactions had this dual behaviour: very complicated with no perturbative expansion at the level of  hadrons and very simple, approximated by a free field theory, at the level of  partons. This picture rings a bell. When discussing the applications of the renormalisation group in section \ref{Ren-Gr}, we found that the effective coupling constant of a renormalisable quantum field theory depends on the energy scale of the experiment. We conclude that if we want to understand SLAC's results on deep inelastic electron-nucleon scattering in terms of a quantum field theory, we must look for one in which the effective coupling constant has the opposite behaviour than the one we found in equation (\ref{RG9}): it should become smaller and smaller as we go to higher and higher energies. We call such theories “asymptotically free” and equation (\ref{RG9}) tells us that they must have $b_0<0$.

\subsection{The road to asymptotic freedom}
Like most of the ideas that built the Standard Model, the road to asymptotic freedom resembles a random walk. I will mention some steps, but I do not guarantee completeness. 

As we said in section \ref{Ren-Gr}, all early calculations showed that the running coupling constant of renormalisable quantum field theories decreased with distance and this behaviour was considered to be universal. The first results which indicated the opposite behaviour were either dismissed or ignored\cite{Grozin}. A short list includes:

$\bullet$ In 1965 V.S. Vanyashin and M.V. Terentev\cite{AF1} computed the first term in the expansion of the $\beta$-function for a massive Yang-Mills theory and found a negative sign. However, since this theory is not renormalisable, nobody paid any attention to the result. 

$\bullet$ As far as I know, the first correct answer for the massless Yang-Mills theory is due to I.B. Khriplovich in 1969\cite{AF2}. He used the Coulomb gauge which is ghost free. Both the sign and the magnitude are correct, but the paper was simply ignored. It is not mentioned in any of the standard reviews. I heard about it only recently from G. Parisi. 

$\bullet$ In July 1972 G. 't Hooft reported in a small meeting in Marseille the negative sign of the  $\beta$-function, again  for the  massless Yang-Mills theory. I was present in this meeting, together with many other theorists. I remember, in particular, T. Appelquist and K. Symanzik. I do not remember any enthusiastic reaction and, to my shame,  I did not make any attempt to  relate this result to the parton model and deep inelastic scattering. I cannot explain this mistake, especially because at that time I was trying to understand scaling using the light-cone algebra of bilocal operators\cite{Ilio-Pasc}. Probably, one of the reasons was that introducing scalar fields to induce the BEH phenomenon destroys asymptotic freedom and the idea of having massless gauge bosons floating around was not taken seriously.

$\bullet$ As far as I can see, the aforementioned three calculations had all a technical motivation: compute the $\beta$-function of a Yang-Mills theory. There was no explicit reference to a theory of strong interactions. On the other hand, the papers by D.J. Gross and F. Wilczek, as well as H.D. Politzer\cite{Gr-Wil-Pol}, had a well-defined goal: an asymptotically free theory of strong interactions. 

$\bullet$ In 1973 S. Coleman and D.J. Gross proved that, in four dimensions, the only asymptotically free theories are unbroken, non-abelian, Yang-Mills theories\cite{Col-Gr}

\subsection{Quantum chromodynamics}
During the 1960s quantum field theory had played a peculiar role in the description of strong interactions. It was totally rejected as a fundamental dynamical theory, but it was very often used, explicitly or implicitly, as a playground in order to guess general properties which were postulated for the scattering amplitudes. For example, the basic postulates of analyticity were extracted from the study of Feynman diagrams. The Mandelstam representation was derived from the square diagram of $\phi^3$. Gell-Mann formulated the algebra of currents by considering a simple quark model, although he did not want to advocate the existence of physical quarks. One of the earliest field theoretic quark models described spin-1/2 quarks interacting with a single neutral vector field, called {\it gluon}, although no one believed in the existence of such a spin-1 particle and no one used this model to compute any physical quantity. The quarks and the gluon were not considered as physical dynamical variables. In 1972 H. Fritzsch and M. Gell-Mann proposed to enlarge the singlet gluon model to a color octet of gluons endowed with Yang-Mills interactions\cite{Gell-Mann-QCD1}, \cite{Gell-Mann-QCD2}. It was the theory of QCD but still, it was not meant to be a dynamical quantum field theory. In the abstract of reference \cite{Gell-Mann-QCD2} we read: {\it“It is pointed out that there are several advantages in abstracting
properties of hadrons and their currents from a Yang-Mills gauge model based
on colored quarks and color octet gluons.”} This insistence to refuse the existence of quarks as “hard grains” inside the hadrons had also influenced the experimental program, in particular that of ISR\cite{ISR}. As late as 1973, proposals to build total absorption detectors completely surrounding the interaction region were rejected. The prevailing philosophy among leading theorists and experimentalists was to deny any interest to the study of scattering at large angles. {\it “Let us forget 90°. Nothing happens at 90°”}. Proposals by young experimentalists to look for events at large angles were scornfully dismissed. 

It was only during the late 1960s that quantitative quantum field theory was re-introduced in high energy physics by a new generation of theorists. I have mentioned already several of them and I want to add here K. Wilson who played a prominent role in advocating the use of quantum field theory to understand the scaling behavior found at  SLAC. His ideas have influenced profoundly our present views and have bridged the gap between particle physics and statistical mechanics. He developed a novel approach to the renormalisation group\cite{Wilson3}  and, in 1968, he   
 proposed a general method to study the product of current operators at short distances, called  {\it the Operator Product Expansion” (OPE)}. This work proved to be extremely powerful and its importance goes beyond the application to deep inelastic scattering\cite{Wilson4}.  We noticed already that in the product of the two currents of eq. (\ref{eqDIS1}) $y^2\approx 0$. If we continue to Euclidean space, $y^2=0$ implies $y_\mu=0$, in other words the product is taken at short distance.  Let $A(x)$ and $B(y)$ be two local operators. Wilson postulated that at short distances, the product $A\cdot B$ can be expanded as:
\be
\label{sd1}
A(x)B(0)|_{x_{\mu}\rightarrow 0} \sim \sum_i C_i(x)O^i(0)
\ee
where the sum extends over the infinite set of all local operators $O^i$ which have the same quantum numbers as the product in the left. $C_i(x)$ are $c$-number coefficient functions which may be singular when $x$ goes to zero. Two points are important in this expansion: First, it is assumed to be an operator equation. The coefficient functions $C_i$ do not depend on the particular matrix element one may consider. Second, the behaviour of $C_i(x)$ at the origin can be determined by dimensional analysis. If $d_A$, $d_B$ and $d_i$ are the dimensions of the operators $A$, $B$ and $O^i$ respectively, the coefficient functions $C_i$ are assumed to behave at short distances, up to logarithmic corrections, as: 
\be
\label{sd2}
C_i(x)|_{x_{\mu}\rightarrow 0} \sim |x|^{d_i-d_A-d_B}
\ee
where $|x|$ denotes the modulus of $x_{\mu}$. It follows that the dominant contribution comes from the operators $O^i$ with the lowest dimension $d_i$. In a four dimensional field theory, if we consider only operators which are monomials in the fields and their derivatives, there is always a finite number of operators with a given $d_i$ and this implies  that any product (\ref{sd1}) will be approximated, with any desired accuracy, by a finite number of terms. This short distance OPE was later extended to a light cone OPE in accordance with equation (\ref{eqDIS1}) in Minkowski space\cite{OPE-lc}.

Quantum chromodynamics, as a dynamical quantum field theory of strong interactions, was proposed, following the discovery of asymptotic freedom, by H.D. Politzer as well as D.J. Gross and F. Wilsczek\cite{Gr-Wil-Pol}. It is an unbroken gauge theory based on color $SU(3)$. It describes the dynamics among $N$ color triplets of spin-1/2 quarks interacting with a color octet of massless spin-1 gluons. At present we have $N=6$. The theory is not restricted to short distances. It covers all scales, including the formation of hadrons. What depends on the scale are the computational tools we use to extract physical information: perturbation expansion techniques at short distances and non-perturbative methods, such as lattice simulations, in the strong coupling regime. QCD has been very successful in both regimes. 

$\bullet$ {\it $e^+ e^-$ annihilation to hadrons.} Perturbative methods are directly applicable when computing correlation functions $G(p_1,\cdots p_n)$ in the kinematical region in which all scalar products $p_i\cdot p_j$ are large compared to mass parameters. The first obvious application is the total hadronic production cross section in $e^+ e^-$ annihilation, at the one-photon exchange approximation. By the optical theorem, it is given by the imaginary part of the photon propagator, as shown in Fig. \ref{QCD9}. The result, normalised by the $\mu^+ \mu^-$ production cross section, is\cite{E-E-tot}:
\begin{figure}
\centering
\includegraphics[height=65mm]{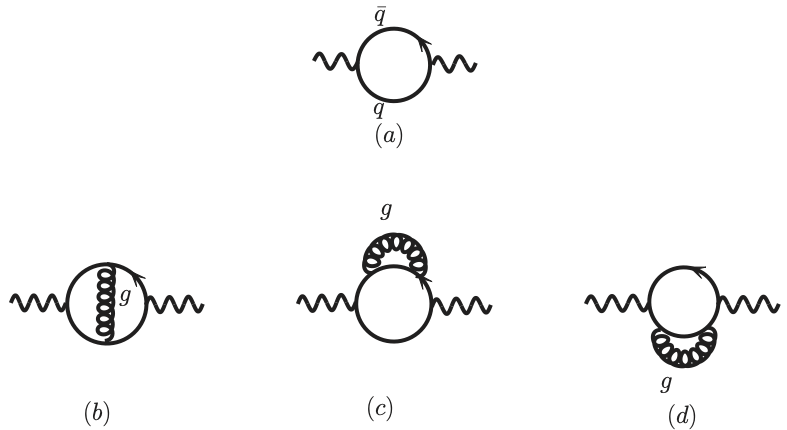}
\caption {The hadronic contributions to the photon propagator at zero
order in $\alpha_s$ (a), and the one-gluon corrections (b), (c) and (d). In our case the last two diagrams give equal contributions.} \label{QCD9}
\end{figure}
\be
\label{eq:R1}
R(Q^2)=\frac{\sigma(e^++e^-\rightarrow{\rm hadrons})}{\sigma(e^++e^-\rightarrow \mu^++\mu^-)}
\ee
\be
\label{eq:R2}
R(Q^2)=\sum_i e_i^2\left(1+\frac{\alpha_s(Q^2)}{\pi} +{\cal O}(\alpha_s^2)\right)~~~~\alpha_s(Q^2)=\frac{1}{4\pi b_0 \ln(Q^2/\Lambda^2)}
\ee
with $b_0=(11-2N_f/3)/4\pi^2$, $N_f$ is the number of quark flavors, $e_i$ is the electric charge of the $i$th quark that can be produced with a photon energy $Q^2$,  and $\Lambda \sim {\cal O}$(200 MeV).

$\bullet$ {\it Deep inelastic scattering.} The $e^++e^-$ total hadronic cross section was easy to analyse because, by applying the optical theorem, we expressed it in terms of a quantity, the photon 2-point function, which depends only on $Q^2$. The property of asymptotic freedom implies that $\alpha_s$ becomes small at large $Q^2$ and we can use perturbation theory. Our next example will be more complicated. It is the calculation of the QCD contributions to the deep inelastic electron-nucleon scattering shown in Figure \ref{fig:deepin1}.  In equation (\ref{eqDIS1}) we expressed the cross section in terms of the matrix element of a product of two currents taken between one-proton states whose momentum is always $P^2=M_N^2$. The dependence on $M_N$ brings a second scale which involves large distances, outside the QCD weak coupling regime. Therefore the deep inelastic structure functions are not directly calculable in perturbation. The first QCD applications used the Wilson OPE in order to factor the large distance part, given by the matrix element of the operators $O_i$ between proton states, from the short distance part given by the functions $C_i$\cite{QCD-early}. The former is $Q^2$ independent and is determined by experiment at some fixed $Q_0^2$. The latter is calculable in a QCD perturbation expansion and predicts violations of exact scaling. This result was not immediately accepted from people who considered QCD as a mere model to extract the light cone current algebra and not as a real dynamical theory\footnote{It seems that when K. Wilson attempted to present to his former PhD supervisor M. Gell-Mann the QCD prediction of scaling violations, the latter did not want to listen. In my report at the London Conference\cite{Il-Lon} I made an analogy with politics and remarked that most people misunderstood the meaning of asymptotic freedom by dropping “asymptotic” and keeping only “freedom”: {\it “As it is often the case, whenever someone talks about freedom, it invariably turns out that he really means something else.”}}. In later years the effects of scaling violations were very well measured, in particular by the HERA collaborations, in full agreement with QCD predictions.

$\bullet$ {\it Hadronic processes -- The APDGL equations.} Applying QCD technology to hadronic collisions was even more complicated because any particular scattering amplitude involves on-mass-shell hadrons and the OPE method is not immediately applicable. One tried instead to isolate infrared-free inclusive observables, such as the average energy $E(\Omega)$ carried into a given solid angular region $\Omega$\cite{Tikto}. An important step, both conceptual and practical, was the combination of QCD with the simple picture of the parton model. They are the Altarelli, Parisi, Dokshitzer, Gribov, Lipatov equations\cite{APDGL} which made possible the use of the asymptotic freedom techniques to a large class of hadronic processes\cite{Ellis}. The equivalence of the two approaches was shown in 1978\cite{Baul-Koun}. 

$\bullet$ {\it QCD in strong coupling.} At energy scales $E \leq $ 1GeV we enter the strong coupling regime and perturbation theory does not apply. Over the years several approximation schemes had been proposed with no significant success. I do not know who was the first to consider the direct numerical estimation of the path integral which defines a quantum field theory by truncation in a finite space-time lattice\footnote{Feynman, in his original 1948 paper which introduced the path integral quantisation method, used a time-lattice but as a means to define the integral, not to compute it numerically.} but, not surprisingly, this method was developed following the increasing power of computing systems. Although the idea sounds simple, its practical application for QCD is not. Many important problems of principle had to be solved, before facing the formidable numerical problems of the simulations. I will mention only two: that of gauge invariance and that of the presence of chiral fermions. K. Wilson was instrumental in the solution of both problems. In physics we use to treat matter fields and gauge fields in a similar way -- we choose a gauge, quantise canonically, compute the corresponding propagators and obtain the Feynman rules -- and this obscures the fact that, mathematically, they are different objects. In the Wilsonian lattice formulation\cite{Wilson5} this difference appears clearly. Let us truncate the four-dimensional Euclidean space by a finite lattice of $N$ points arranged in a hypercube with lattice spacing $a$. The point $x_\mu$ is replaced by an integer $n$, $1\leq n \leq N$ which labels the lattice point. For a field we get $\phi(x)\rightarrow \phi_n$, i.e. a field lives on a lattice point. In contrast, Wilson showed that gauge fields live on oriented lattice links, which explains why mathematicians call gauge fields {\it connections}. In fact, studying lattice gauge theories is a poor man's way to understand differential geometry. The second problem is that of chiral fermions. It is easy to understand its origin. The introduction of a finite lattice spacing $a$ provides an ultraviolet cut-off and Wilson showed that it respects gauge invariance. When we discussed the chiral anomaly we said that there is no cut-off respecting both gauge and chiral symmetry. It follows that chiral fermions cannot be directly introduced on a lattice\cite{Nielsen}. Wilson suggested the method of fermion doubling which is widely used. Lattice calculations of strong interaction effects have become a major research field in recent years with many important, and very precise, results\cite{Lat-rec}. They have contributed in establishing QCD as the theory of strong interactions.

\section{The London Conference}
The theoretical scheme, which became the Standard Theory of particle physics, was fully written in 1973. Yet, it was not generally accepted. For most physicists it was a wild theoretical speculation with no connection to the real world. The main reason for this negative attitude was again the mistrust towards quantum field theory, but it is also true that the model seemed to make many strange predictions with little, if any, experimental support. Let me mention some of them.

$\bullet$ The model predicted the existence of 12 vector bosons. But only one, the photon, was known! Three, ($W^\pm ,~Z^0$) were predicted to be very heavy\footnote{For the 1973 physicists a mass of 100 GeV was essentially infinite!} and the other eight, the gluons, were declared {\it unobservable} by a strange property of {\it confinement.}

$\bullet$ The model predicted the existence of a scalar boson, the BEH, with unknown mass. For many physicists it was a heresy coming after the assumed triumph of the $V-A$ theory of weak interactions.

$\bullet$ Neutral currents were predicted but the obvious ones, the $K^0\rightarrow \mu^++\mu^-$ decay, were excluded. Gargamelle had established the existence of strangeness conserving neutral currents but, as I remarked in section \ref{NC-Y-N}, not everybody was convinced. Furthermore, the possible existence of weak neutral currents had been envisaged before the formulation of gauge theories, so not everybody considered them as a decisive proof of the Standard Model.

$\bullet$ Probably the most “extravagant” prediction was that of the charmed quark, implying the existence of  an entire new hadronic world of charmed particles. For most people the arguments were not considered serious. I still remember the objections: some obscure higher order effects -- triangle diagrams for the anomalies, or square diagrams for the absence of flavor violating neutral currents -- would dictate the structure of the world? Totally absurd! In retrospect, I think that the large majority of physicists rejected this particular prediction  because it went against the prevailing philosophy of compartmentalisation of high energy physics. Theoretical arguments motivated by properties of weak interactions were not admissible to make predictions on hadronic physics, a domain reserved exclusively to strong interactions\footnote{{\it Sutor, ne supra crepidam.} or, {\it Let the cobbler stick to his last.}}.

$\bullet$ The QCD prediction for the ratio $R$ of eq. (\ref{eq:R2}) seemed to be in violent contradiction with experiment. 

This brings me to  the 17th International Conference on High Energy Physics held in London in July 1974. It can be viewed as the last Conference of the Dark Ages, but also the one which announced the New Era. If we look at the program of the previous conferences we find the traditional scheme of sessions on weak, electromagnetic and strong interactions. The man behind the organisation of the London Conference was Abdus Salam who succeeded to inject a small part of a new vision: there were still sessions on strong interactions, resonant physics etc, but we find also a session with a report on gauge theories which included the advances in all three interactions, in other words a report on what was going to become the Standard Model. 
Several old and new results were announced in this Conference. I will mention only those which are important for our story.

$\bullet$ D.C. Cundy gave the plenary report on neutrino physics. Obviously, the Gargamelle results were the central point. I remember him saying: {\it “Those who have bet on neutral currents, now is the time to pay and collect.”}

\begin{figure}
\centering
\includegraphics[height=55mm]{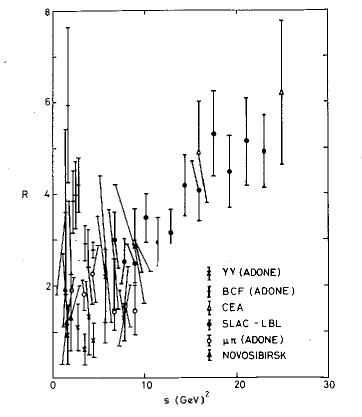}
\caption {A compilation of all early measurements of the ratio $R$, as presented in the 1974 London International Conference on High Energy Physics by Burton Richter.}
\label{R-had1}
\end{figure}

$\bullet$  B. Richter gave the plenary report on $e^+e^- \rightarrow $ hadrons. In Figure \ref{R-had1} I show the graph he presented.
I remind you that the QCD prediction of eq. (\ref{eq:R2}) was that $R$ should approach the value $R=2$ (the sum of the electric charges of $u$, $d$ and $s$ quarks) from above. No such behavior is visible in Figure \ref{R-had1}. 

$\bullet$ In my report on gauge theories\cite{Il-Lon} I said: {\it “\dots the hadron production cross section, which absolutely refuses to fall, creates a serious problem. The best 
explanation may be that we are observing the opening of the charmed thresholds, in which case everything fits together very nicely.”} The addition of a charmed quark would add an extra 4/3 to $R$\footnote{By a numerical accident, the data of Figure \ref{R-had1} contain also the production of the $\tau$ lepton which was not known at the time.}. Salam had ready a bottle of a fine Bordeaux red wine to reward speakers who finish on time. Naturally I was late and I said something like: {\it “I know I am about to loose my bottle, but I am ready to bet now a whole case that, if the weak interaction sessions of this Conference were dominated by the discovery of the neutral currents, the entire next 
Conference will be dominated by the discovery of the charmed particles.”} This convinced Salam to give me the bottle which I opened immediately, poured myself a generous libation, offered the rest to those sitting in the first row, and drank “to charm!”.

\section{The Charming Theory of the New Particles} 
\begin{figure}
\centering
\includegraphics[height=70mm]{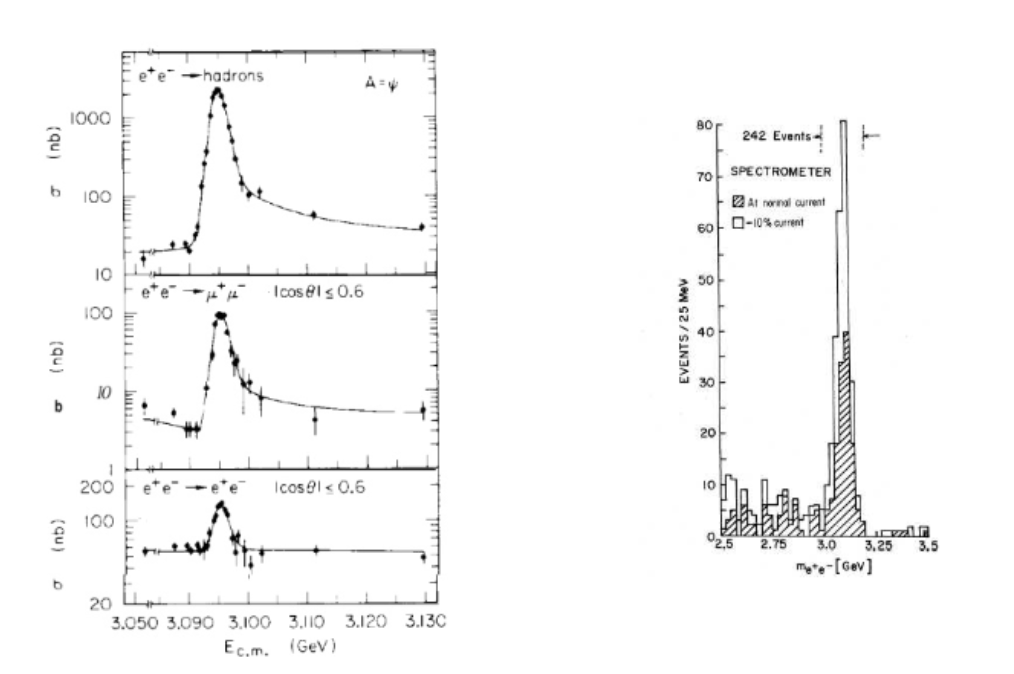}
\caption{The discovery of the $J/\Psi$ meson in Nov. 1974 independently by SPEAR (left) and AGS (right). Both exhibit peaks in the oppositely charged dielectron mass spectrum consistent with the $J/\Psi$ mass at 3.1 GeV. This result was also confirmed by the Frascati group.} \label{R-had2}
\end{figure}

In November 1974 both Brookhaven and Stanford published their results\cite{J/Psidisc}. SPEAR  decided to sweep the region above  3 GeV in fine steps of 1 MeV. To their great surprise they obtained a totally different picture, Figure \ref{R-had2}. 

 The results were announced by both groups on Monday, November 11 1974. They are published at the same issue of Physical Review Letters. The Brookhaven article has submission date November 12 and the SPEAR one November 13. In November 12 M. Greco, a theorist from Rome, arrived at SLAC to give a seminar and found the theory group at a great excitement. He asked S. Drell permission to call Frascati and gave to G. Bellettini the news with the exact position of the resonance\cite{Greco1}. Although 3.1 GeV was at the edge of ADONE's energy range, the Frascati group could confirm the American results\cite{Bacci} and their paper was published in the same issue of PRL\footnote{It seems the G. Salvini, transmitted the paper to PRL over the phone.}.

I was in Paris at the end of 1974 when I received a telephone call from A. Lagarrigue inviting me to an informal meeting to discuss important results from SPEAR. G. 't Hooft was visiting Ecole Normale and I took him along. In those days news did not travel with the speed of internet. B. Jean-Marie had come from Stanford with the results. They were indeed very impressive. A 3 GeV hadron, decaying into pions, with a width of less than 100 keV? Incredible! Lagarrigue asked me what I thought of it and I confess I was bewildered. It was 't Hooft who first gave me the explanation. It is simple to understand in QCD. Let us consider the series of $1^-$ mesons: $\rho$ is a bound state of a quark-antiquark pair of the first family. Its mass lays well above the $2\pi$ mass and its  width is very large $\sim$ 147 MeV. The $\phi$ is an $s\bar{s}$ bound state with a mass of 1020 MeV. It lays barely above the threshold of a $K\bar{K}$ pair, yet its branching ratio to $K\bar{K}$ is 83\% despite the fact that the phase space is tiny. The pure pionic partial width is only 650 keV. In the old days we had a rule, called the OZI (Okubo-Zweig-Iizuka) rule, one of those empirical rules of the dark ages with no real theoretical justification. It stated that in a quark-antiquark bound state, the decay modes requiring the annihilation of the initial $q\bar{q}$ pair, were highly suppressed\footnote{It was called “the re-arrangement model” by H. Rubinstein.}. Let us come now to $J/\Psi$ and assume it is a $c\bar{c}$ bound state. The $0^-$ mesons are supposed to be the pseudo-Goldstone bosons of spontaneously broken chiral symmetry. The latter is very good for the first family, questionable for the $s$ quark, and very poor for charm. Therefore we expect the charmed $0^-$ mesons $D$ to be quite heavy and the mass of $J/\Psi$ to lay below the $D\bar{D}$ threshold. As a result $J/\Psi$ decays mainly into pions. The decay amplitude for a $1^-$ meson goes through three gluons, so the width is proportional to $\alpha_s^3$. Between 1 and 3 GeV $\alpha_s$ has dropped by a factor of two, so we expect the $J/\Psi$ width to be 8 times smaller than the $\phi$ pionic width. It is precisely what is found experimentally. As I said, I first heard this argument from 't Hooft, but later I found it in papers by Appelquist and Politzer as well as  De Rujula and Glashow \cite{Ap-Pol}. The first has a submission date of November 19  and the second November 27. This makes me believe the asymptotic freedom argument was found  earlier.

As expected, the announcement of the experimental discovery triggered a large number of theoretical papers trying to interpret the appearance of these strange resonances with such a narrow width. Many proposed exotic explanations but, going through the early literature, I found three with the correct identification as  $\bar{c} c$ bound states. They are the ones of reference \cite{Ap-Pol} as well as one by C.A. Dominguez and M. Greco\cite{Grecoetal}. It has a submission date of December 30 but, apparently, it is mentioned as preprint in the CERN preprint list with date November 18. It uses duality sum rules to estimate the value of $R$ and considers both $\Psi$ and $\Psi'$ as $\bar{c} c$ bound states\footnote{However, in this paper the narrow widths are erroneously attributed to weak interactions.}.

\begin{figure}
\centering
\includegraphics[height=50mm]{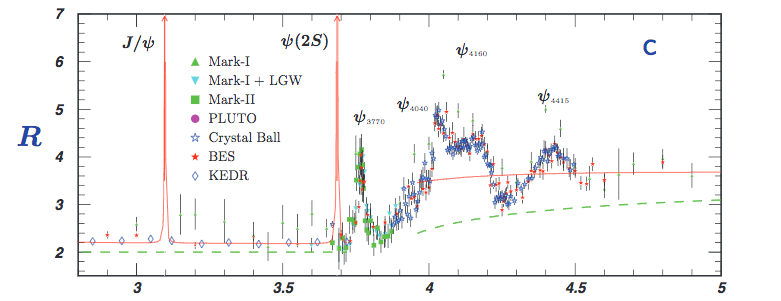}
\caption{The value of $R$ for energies between 3 and 5 GeV.} \label{R-had3}
\end{figure}

Within a year the entire region between 3 and 5 GeV was studied in detail, see Fig. \ref{R-had3}.
As expected, there are broader resonances with masses $\geq$ 4GeV which are those lying above the $D\bar{D}$ threshold. Lo and behold, the particles with naked charm were found among the decay products of these resonances. It was in 1976. In the meantime a rich charmonium spectroscopy\footnote{I believe that the term “charmonium” was  coined by A. De Rujula.} was discovered in full agreement with the theoretical predictions\cite{Alvaro}. Although nobody paid the bet I offered in the 1974 Conference, the entire 1976 one was indeed dominated by charmed particles and gauge theories. The report in this Conference by A. De Rujula had the title: {\it “Theoretical basis of the new particles.”} The first sentence is {\it “I review the four-quark standard gauge field theory of weak, electromagnetic and strong interactions.”} Now we talk about “the four-quark standard gauge field theory”.  The phase transition from Many Models to One Theory was complete. The order parameter has been the fraction of physicists who changed their views: a small minority before 1974 to the large majority after 1976. The complete verification of the theory took many more years and many great discoveries, but the mood of the community had changed. The following discoveries of the vector bosons, the $b$ and $t$ quarks which complete the third family of the $\tau$ lepton, the gluon jets and the BEH scalar as well as the very good general fit using all available data,  were no more great surprises, they were expected. THE STANDARD MODEL had become THE STANDARD THEORY.

\section{From Dream to Expectation}
Feynman has said that progress in physics is to prove yourself wrong as soon as possible. For half a century now we have not been able to prove the Standard Theory is wrong. It has passed successfully all tests and all its predictions have been brilliantly verified. What comes next? 

\begin{figure}
\centering
\includegraphics[height=60mm]{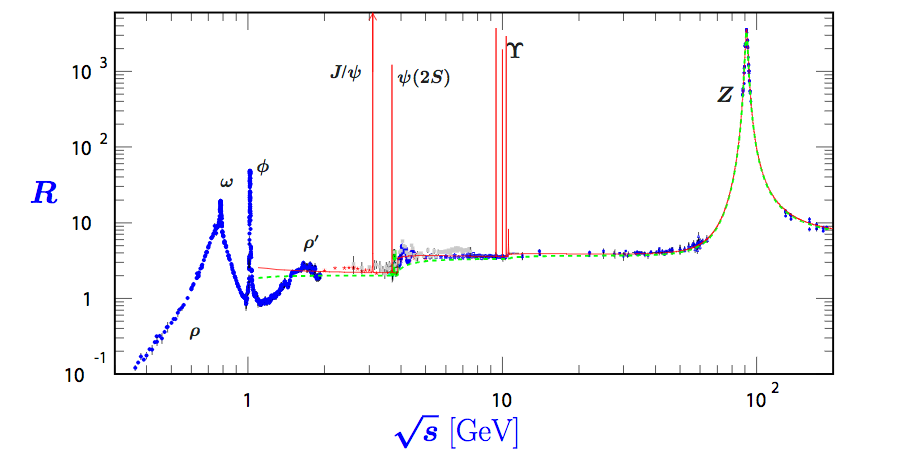}
\caption{The ratio $R$ from low energies, up to and above the $Z$ mass. The green curve is the parton model prediction and the red one includes QCD corrections. Remarkable agreement.} \label{R-had4}
\end{figure}

In 2011 the European Physical Society awarded to Glashow, Maiani and myself the High Energy Physics Prize. At this occasion we were invited to speak at the European Conference in Grenoble. The title of  my talk was “Following the Path of Charm” and I tried to argue that precision measurements at a certain energy scale allow us to make predictions at some higher scale. Let me 
start from an expanded version of a plot showing the value of $R$ from low energies up to and above the $Z^0$ mass, Figure \ref{R-had4}.
What is most remarkable is the precision with which theoretical predictions fit the experimental data in most of the energy interval. The only regions in which there is no agreement are the one at low energy below 1 GeV, in which the QCD is in the strong coupling regime, and small, very localised regions signaling the thresholds for the production of new species of hadrons, charm or $b$. Outside those regions formula (\ref{eq:R2}) gives an excellent fit, although it is just the result of one loop perturbation theory. I concluded that, for reasons that are not fully understood, perturbation theory is reliable outside the regions of strong interactions. 

I tried to use this fact in order to make predictions regarding the multi-hundred GeV scale which was expected to be explored by LHC. The argument was based on the low energy data which favored a Higgs particle of relatively low mass, $\leq$ 200 GeV. It went roughly as follows: The existing data are compatible with the Standard Theory only if the Higgs is light. Therefore, if we find a very heavy Higgs we must also find new interactions which invalidate the Standard Theory calculations. If on the other hand we find a light Higgs, we must find new interactions which stabilise its mass at this low value. I thought that both possibilities were good news for LHC. In my talk I said:

{\it “I want to exploit this experimental fact} [the validity of perturbation theory] {\it and argue that the available
precision tests of the Standard Model allow us to claim with confidence that
new physics is present at the TeV scale and the LHC can, probably, discover it.
The argument assumes the
validity of perturbation theory and it will fail if the latter fails. But, as
we just saw, perturbation theory breaks down only when strong interactions
become important. But new strong interactions imply new physics.”}

My conclusion was that, for LHC, which was about to start operating, new physics was around the corner!

Today we know that LHC found no corner! 

But I secretly believe the argument is correct, only the corner is a bit further down.

Although I will not see it, I am confident some among our young colleagues will find it.

\vskip 2cm

{\bf Acknowledgements:} I have benefitted from the advice and help of many colleagues. Here I mention only those who read an earlier version of these notes and sent me their remarks. They include O. Darrigol, M. Dris, S.L. Glashow, M. Greco, E. Kiritsis, E. Rabinovici and L. Resvanis.

\vskip 2cm

\small{

These notes have touched so many subjects that a complete list of references is impossible. The  selection is necessarily arbitrary and I apologise for the numerous omissions.

}

\begin{thebibliography}{99}
\bibitem{Lat-rec} There is a review on lattice calculations in practically every major high energy physics conference. See also C.T. Sachrajda, contribution in reference \cite{SM-hist2}.
\end{thebibliography}
\end{document}